\documentclass{aa}
\usepackage{subfig}

\usepackage{natbib}
\usepackage{graphicx}
\usepackage[varg]{txfonts}
\usepackage{hyperref}

\begin{document}

\title{Milky Way: New Galactic mass model for orbit computations}

   \author{J. Kla\v{c}ka
          \inst{1}
          \and
           M. \v{S}turc \inst{1}
          \and
          E. Puha \inst{1}
          }

   \institute{Faculty of Mathematics, Physics, and Informatics, Comenius University, 
	Mlynsk\'{a} dolina, 842 48 Bratislava, Slovak Republic\\
              \email{jozef.klacka@fmph.uniba.sk,  michal.sturc@fmph.uniba.sk, emil.puha@fmph.uniba.sk}
             }

   \date{Received 2024}

 \abstract
   {Accurate observational data on the rotation curve of the Milky Way galaxy (MW) are very well understood using the gravitational potentials of the baryonic matter (Pouliasis et al., 2017) and the interpolating function presented by McGaugh et al. (2016). In this way we couple the spherically symmetric mass distribution of the dark matter (DM) to the mass distribution of the baryonic matter (BM).
   A commonly used model of the dark matter distribution can be understood as an approximation of our model.}
   {Creation of a new Galactic mass model for orbit computations based on
   the coupling of the DM with an improved BM model  respecting the theoretical fit of the Galactic rotation curve.  Several applications are given, e.g., values of the Oort constants and mass density in the region of the Sun are found.}
   {The theoretical approach is based on a slight improvement of the BM model by Pouliasis et al. (2017) and on replacement of their DM model by the model based on the interpolating function by McGaugh et al. (2016). The theoretical results are validated by the comparison with observational data.}
   {New Galactic mass model for orbit computations is created. DM distribution is given and the rotation curve for the MW is consistent with observational data. The Tully-Fisher relation holds and the values of several characteristics of the MW for the region of the Sun are given, e.g., rotation speed $v_0$ = ($228.8 \pm 0.2$) ~ \mbox{km} \mbox{s}$^{-1}$, Oort constants $A=(+14.73 \pm 0.03) ~\mbox{km} ~\mbox{s}^{-1} ~\mbox{kpc}^{-1} $, $B=(-13.01 \pm 0.03) ~\mbox{km} ~\mbox{s}^{-1} ~\mbox{kpc}^{-1}$ and the mass density of the baryonic matter $(0.087 \pm 0.001) ~\mbox{M}_{\odot} ~\mbox{pc}^{-3} $. Important MW characteristics are the total BM mass $M_{BM}$ $=$ (8.43 $\pm$ 0.01) $\times$ 10$^{10}$ $\mbox{M}_{\odot}$, the total DM mass $M_{DM}$ $\doteq$ (1.34 $\pm$ 0.01) $\times$ 10$^{12}$ $\mbox{M}_{\odot}$, virial radius and virial mass of the MW, $M_{vir}$ $\doteq$ $M_{DM}$.}
{}


   \maketitle

\section{Introduction}\label{introduction-N}

Rotation curves of galaxies are considered to be an important source of information on mass distribution within the galaxies. The distribution of mass determines the orbital evolution of individual objects moving in the galaxies. A special case is the Milky Way galaxy (hereafter MW), our home galaxy. 
As for the recent publications on the baryonic and dark matter mass distribution, or, equivalently, on the gravitational potentials of the MW or its part, we refer to, e.g., 
{\citet{Allen1991}, \citet{McMillan2011}, \citet{Irrgang2013}, \citet{Kafle},
\citet{Smith2015}, \citet{McKee}, \citet{Barros2016}, \citet{2017A&A...598A..66P}, \citet{2017MNRAS.465...76M}, \citet{2019ApJ...871..120E}, \citet{2023ApJ...945....3S}, \citet{Ou} and \citet{Jiao}. The models of the baryonic matter (BM) are based on observations of the visible matter in the MW and the previous papers considered various contributions of individual Galactic components. Only recently \citet{Snaith2014,Snaith2015} found that the thick disc appears to be as massive as the thin disc, while the scale lengths are discussed in \citet{Bensby2011}, \citet{Bovy2012,Bovy2016}. The Galactic baryonic mass model based on these results is developed by \citet{2017A&A...598A..66P}.  
The distribution of the dark matter (DM) is inferred from the rotation curve.
The equation of motion based on the gravitational accelerations of the BM and the DM can also determine important quantities such as the Oort constants. Thus, improvement of the equation of motion can also improve the values of the Oort constants and the accuracy of their values in comparison with e.g. \citet{bovy17}.

The potential model of \citet{2017A&A...598A..66P} summarizes the older models and presents compact and simple form of the gravitational model of the MW. The gravitational potentials of the model are fully analytical and differentiable at all points. This is the reason why the model by 
 \citet{2017A&A...598A..66P} is widely used since its publication. The parameters of this model such as the enclosed mass, scale lengths and scale heights of the Galactic components are frequently used in recent publications on Galactic mass models, e.g., \citet{2021A&A...654A..25J}, \citet{2022ApJ...926..189N}, \citet{2023ApJ...945....3S}.
 
It is important to better understand the BM and DM distributions in the MW and to better understand and improve the BM and DM models of our galaxy. It is expected that more realistic models fit the observational data.
The BM and DM mass distribution models must respect measurable quantities such as the baryonic mass density in the region of the Sun, Galactocentric velocity of the Sun, the Tully-Fisher relation \citep{1977A&A....54..661T} and other observational results. As a consequence, we can take the baryonic matter potentials from \citet{2017A&A...598A..66P}, the Miyamoto-Nagai profile \citep{Miyamoto}, as a starting point and make only slight changes of the baryonic model parameters to fit the most recent data.
The argument for small changes of the baryonic model parameters is reasoned by the observations on the visible matter, while the DM model may be subjected to larger changes. This enables us to avoid any inconsistencies generated by older models, such as \citet{Allen1991}, \citet{McMillan2011}, \citet{Smith2015}, \citet{2017MNRAS.465...76M}, \citet{2017A&A...598A..66P} or \citet{Ou}, where the fitted parameters and the understanding of the underlying physics were inconsistent with current observations (e.g. more precise rotation curve or respecting the baryonic Tully-Fisher relation). E. g., the model by \citet{2017A&A...598A..66P} yields the rotation speed 243.9 $\mbox{km}~\mbox{s}^{-1}$ and the derivative of the speed $-~0.64$ $\mbox{km}~\mbox{s}^{-1} ~\mbox{kpc}^{-1}$ for the Sun (the Oort constants $A$ $=$ 15.11 $\mbox{km}~\mbox{s}^{-1}~\mbox{kpc}^{-1}$, $B$ $=$ -14.47 $\mbox{km}~\mbox{s}^{-1}~\mbox{kpc}^{-1}$), instead of the values 228.8 $\mbox{km}$ $\mbox{s}^{-1}$ and $-~1.72$ $\mbox{km}~\mbox{s}^{-1} ~\mbox{kpc}^{-1}$ consistent with observations, see Secs. \ref{sec-DM-P} and \ref{dif-vel} below (the Oort constants $A$ $=$ (14.73 $\pm$ 0.03) $\mbox{km}~\mbox{s}^{-1}~\mbox{kpc}^{-1}$, $B$ $=$ (-13.01 $\pm$ 0.03) $\mbox{km}~\mbox{s}^{-1}~\mbox{kpc}^{-1}$). Similarly, if the model by \citet{2017A&A...598A..66P} is corrected using \citet{Allen1991}, then
the rotation speed is 222.0 $\mbox{km}~\mbox{s}^{-1}$ and the derivative of the speed $+~0.042$ $\mbox{km}~\mbox{s}^{-1} ~\mbox{kpc}^{-1}$ for the Sun (the Oort constants $A$ $=$ 13.45 $\mbox{km}~\mbox{s}^{-1}~\mbox{kpc}^{-1}$, $B$ $=$ -13.50 $\mbox{km}~\mbox{s}^{-1}~\mbox{kpc}^{-1}$). However, the corrected DM model of the MW by \citet{2017A&A...598A..66P} and \citet{Allen1991} can be understood as a mathematical approximation of a more realistic model developed in our paper.

In this paper, we are interested in creating a new MW mass distribution model for orbit computation. The BM components are inspired by \citet{2017A&A...598A..66P} and the DM model is developed to be consistent with observations on the rotation curve or 
the mass-speed relation at large Galactocentric distance, 
the baryonic Tully-Fisher relation \citep{1977A&A....54..661T,2000ApJ...533L..99M,2016PhRvL.117t1101M}.

The conventional approaches assume several free parameters for the DM halo model. However, using the recently published empirical results of the observations of the SPARC extragalactic galaxies by \citet{2016PhRvL.117t1101M}, we can develop a new MW mass model that contains no free parameters of the DM halo. 
In this model, the DM distribution is coupled to the BM distribution via an empirical formula from \citet{2016PhRvL.117t1101M}. When using the observational results by \citet{abuter2020detection}, the value of the velocity of the Sun is consistent with \citet{2019ApJ...871..120E}. The resulting rotation curve is also consistent with \citet{2019ApJ...871..120E} and with the baryonic Tully-Fisher relation \citep{1977A&A....54..661T}.

Recent papers show decrease of the rotation curve of the MW for galactocentric distances greater than about 15 $\mbox{kpc}$ e.g. \citet{2023ApJ...945....3S,Ou,Jiao}.
We do not consider this result in our paper. First, the result implies that the MW galaxy behaves in significantly different way than other spiral galaxies characterized by flat rotation curves. 
Second, if the papers enable to calculate the mass density in the region of the Sun, then the models of the papers do not lead to the result consistent with the value
0.10 $\mbox{M}_{\odot}$ $\mbox{pc}^{-3}$ used in the modelling of the Oort cloud of comets in the 21st century,  see, e.g., \citet{Levison2001}, \citet{Korchagin2003}, \citet{Brasser_Morbidelli}. As an example we can mention that the model by
\citet{Ou} yields  0.135 $\mbox{M}_{\odot}$ $\mbox{pc}^{-3}$, the model B2 for the BM disc presented by \citet{Jiao} leads to the value 0.145 $\mbox{M}_{\odot}$ $\mbox{pc}^{-3}$. We can also mention that the DM models presented by \citet{Jiao} give the DM mass density 0.03 $\mbox{M}_{\odot}$ $\mbox{pc}^{-3}$ for the region of the Sun, which is about 3-times greater than the expected value. As for the circular velocity at the Galactocentric distance of the Sun, $v_{c}(R_{0})$, the models by \citet{2023ApJ...945....3S}, \citet{Ou} and \citet{Jiao} yield 210 km/s $<$ $v_{c}(R_{0})$ $<$ 235 km/s.
The model B2 presented by \citet{Jiao} gives $v_{c}(R_{0})$ $=$ 215 km/s.
As for comparison, we can mention that the speed about 229 km s$^{-1}$ holds according to \citet{2019ApJ...871..120E} and \citet{2023ApJ...945....3S}.

{Several applications of the equation of motion of a body moving in the MW are given. Quantities important for the region of the Sun are calculated. The values of the Oort constants are found and compared with the published results, see, e.g., 
\citet{kerr}, \citet{feast1997galactic}, \citet{bovy17}. The mass densities of the BM and the DM are obtained. The oscillation periods of the motion perpendicular to the Galactic equatorial plane are calculated and compared with the published results, see. e.g.,
\citet[p. 5]{2014pmdb.book.....M}. The baryonic Tully-Fisher relation is derived.
Potential improvements of the baryonic model will not change important results of this paper.}

\textit{Section 2} summarizes relevant equations for gravitational accelerations of the BM and the DM within Newtonian gravity approach, in application to the rotation curve.
\textit{Section 3} summarizes relevant gravitational potentials of Galactic baryonic components, taken from \citep{2017A&A...598A..66P}.
\textit{Section 4} finds theoretical rotation curve which follows from the empirical fit to the observations of SPARC extragalactic galaxies applied to the total BM potential of the MW.  
Fig. \ref{Fig1} shows the coincidence between the theoretical result and observations. Consequences of the DM approach by \citet{2017A&A...598A..66P}
are discussed and presented in Fig. \ref{Fig2}. 
The corrected DM model of the MW by \citet{2017A&A...598A..66P} and \citet{Allen1991} can be understood as an approximation of a more realistic model (Secs. \ref{sec-DM-AS-P} and \ref{C-simple-ideas}), see also Eqs. (\ref{eqDM-AS-app}), (\ref{three-req-a-halo}), (\ref{three-req-M-halo}) and (\ref{third-req-res12-sum-R-DM}) in Appendix \ref{appendix-fixed-model}. 
\textit{Section 5} presents a new Galactic mass model, an improved baryonic model of the MW. Figure \ref{Fig3}
shows the coincidence between the theoretical rotation curve based on the model and the observational data, also consistency 
between the theoretical approach and the baryonic Tully-Fisher relation holds.
\textit{Section 6} derives the baryonic Tully-Fisher relation.
\textit{Section 7} discusses relevant quantities for the region of the Sun, speed, rotation period, values of the Oort constants and the baryonic mass density are calculated.
\textit{Section 8} presents vector equation of motion for a Galactic body. This allows the orbit computations.
The conventional approach based on the existence of the DM is used. \textit{Section 9} presents several properties of the DM distribution, including physical understanding of the baryonic Tully-Fisher relation, calculation of the virial radius and the virial mass of the MW. \textit{Section 10} is the discussion focused on the 
comparison of the found rotation curve, mass densities for the Galactocentric distances of the Sun and
mass densities for large values of the Galactocentric distances with the previous results. Gravitational acceleration generated by the DM halo of the MW is presented for spherical DM mass distribution.
Values of the relevant Galactic quantities are summarized in the \textit{Conclusion}.
\textit{Appendix A} summarizes the values of the Oort constants. \textit{Appendix B} offers applications of the Galactic mass model for orbit computations. Motions perpendicular to the Galactic equatorial plane are calculated.
\textit{Appendix \ref{appendix-fixed-model}} discusses and improves
the DM halo model of the MW presented by \citet{2017A&A...598A..66P} and \citet{Allen1991}. 
\textit{Appendix \ref{3-components-disc}} applies
the BM model of the MW 3-component discs presented by \citet{Halle2018}.



\section{Theoretical background}\label{basis-eq}
{The main aim of the paper is to create a new Galactic mass model for orbit computations in the MW. The model is based on 
theoretical understanding of the rotation curve of the MW. We need to find an equation of motion of a body moving in the MW.

Observational results on the rotation curves of galaxies suggest centripetal acceleration relation
\begin{eqnarray}\label{obs-r-0}
\frac{v_{c}^{2}}{R} &=& g_{obs}  
\end{eqnarray}
for the rotation speed $v_{c}$ at the galactocentric distance $R$.

We can take the empirical fit to the observations of SPARC extragalactic galaxies \citep{2016PhRvL.117t1101M}
\begin{eqnarray}\label{obs-r-1}
g_{obs} &=& \frac{g_{bar}}{1 - \exp{(-\sqrt{g_{bar}/ g_{+}})}} ~,
\nonumber \\
g_{+} &=& 1.2 \times 10^{-10} ~\mbox{m} ~\mbox{s}^{-2} °,
\end{eqnarray}	
or,
\begin{eqnarray}\label{obs-Gaugh}
\frac{v_{c}^{2}}{R} &=& g_{total} ~,
\nonumber \\
g_{total} &=& g_{bar} + g_{DM} ~,  
\nonumber \\
g_{DM} &=& \frac{g_{bar}}{\exp{(\sqrt{g_{bar}/ g_{+}})} - 1}  ~,
\nonumber \\
g_{bar} &=& -~\vec{g}_{BM} (R, z = 0) \cdot \hat{\vec{R}} ~,
\end{eqnarray}
where the quantity $g_{bar}$ is the radial component of the
BM gravitational acceleration $-~\vec{g}_{BM}$ in the galactic equatorial plane. Similarly, $g_{DM}$ is the radial component of the DM 
gravitational acceleration $-~\vec{g}_{DM}$  in the galactic equatorial plane, see \citet[Eq. 5]{2016PhRvL.117t1101M}, compare also \citet{Famaey2012}.
The second of Eqs. (\ref{obs-Gaugh}) considers the superposition principle for the total radial component of the gravitational acceleration. It consists both from the BM and the DM accelerations. We stress that the above presented equations hold for the galactic equatorial planes ($z$ $=$ 0). The 3-D form of 
Eqs. (\ref{obs-Gaugh}) is $\vec{\dot{v}}$ $=$ $\vec{g}_{BM}$ $+$ $\vec{g}_{DM}$
and $\nabla \times \left ( \vec{g}_{BM} + \vec{g}_{DM} \right )$ $=$ 0. This statement follows from $\vec{\dot{v}}$ $=$ $-~\nabla \Phi_{BM}$ $+$ 
$\left ( -~\nabla \Phi_{DM} \right )$ $=$ 
$-~ \nabla \left ( \Phi_{BM} + \Phi_{DM} \right )$,
using gravitational potentials of the BM and the DM. 

This paper will consider Eqs. (\ref{obs-r-1}) in the sense of 
Eqs. (\ref{obs-Gaugh}). Thus, the vector generalization considered in this paper will be 
$\vec{\dot{v}}$ $=$ $\vec{g}_{BM}$ $+$ $\vec{g}_{DM}$ (see Eq. \ref{eq-motion-DM} below). 


\section{Baryonic matter gravitational acceleration - \citet{2017A&A...598A..66P}}\label{BM-a}		
We will start with the recent results obtained by \citet{2017A&A...598A..66P}. 

We take gravitational potentials for the baryonic matter components of the MW in the forms:
{
\begin{eqnarray}\label{pot-bulge}
\tilde{\Phi}_{i} (R,z) &=& -~ \frac{G \tilde{M}_{i}}{\sqrt{R^{2} + \left ( \tilde{a}_{i} + \sqrt{z^{2} + \tilde{b}_{i}^{2}} \right )^{2} }} ~, ~i~=~1,~2,~3~,
\end{eqnarray}
where $i$ $=$ 1, 2 and 3 correspond to the thin disc, thick disc and the central bulge. 
Values of the parameters, masses and length scales, are summarized in Table \ref{Table1}.
We remind that the central bulge is spherically symmetric,
the Galactocentric distance is $r$ $=$ $\sqrt{R^{2} + z^{2}}$. 
The total mass of the baryonic matter is ${\tilde{M}}_{BM}$ $=$ $\sum_{i=1}^{3} \tilde{M}_{i}$
$=$ 8.9552 $\times$ 10$^{10}$ $\mbox{M}_{\odot}$.}

\begin{table}[h]
\centering
\def\arraystretch{1.3}
\begin{tabular}{ |c|c|c|c|c| } 
 \hline
 structure & index $i$ & $\tilde{M}_{i}~[10^{10} \mbox{M}_{\odot}]$ & $\tilde{a}_{i}~[\mbox{kpc}]$ & $\tilde{b}_{i}~[\mbox{kpc}]$ \\
 \hline
 thin disc & 1 & 3.9440 & 5.30 & 0.25\\
 thick disc & 2 & 3.9440 & 2.60 & 0.80\\
 bulge & 3 & 1.0672 & 0 & 0.30\\
 \hline
\end{tabular}
\caption{{Values of the parameters for the model by 
\citet{2017A&A...598A..66P}, see Eqs. (\ref{pot-bulge}).}}
\label{Table1}
\end{table}

The radial component of the baryonic matter gravitational acceleration {acting on a body moving in the Galactic equatorial plane} ($z$ $=$ 0) is
\begin{eqnarray}\label{g-bar-rad}
g_{bar} &=& \left [ \frac{\mbox{d} \tilde{\Phi}_{BM} \left ( R, z \right )}{\mbox{d} R} \right ]_{z=0} ~, 
\nonumber \\
\tilde{\Phi}_{BM} \left ( R, z \right ) &=& \sum_{i = 1}^{3} \tilde{\Phi}_{i} 
\left ( R, z \right ) ~,
\end{eqnarray}	
where $\tilde{\Phi}_{i}$, $i$ $=$ 1, 2, and 3 are given by Eqs. (\ref{pot-bulge}).

\section{DM gravitational acceleration}
Sec. \ref{BM-a} presents method for obtaining baryonic matter gravitational acceleration of the MW. {We need to find DM gravitational acceleration of the MW, in order to make theoretical predictions and to obtain theoretical results which may be compared with observational data.}

\subsection{DM acceleration - \citet{2016PhRvL.117t1101M} approach}\label{sec-DM-G}
The third of Eqs. (\ref{obs-Gaugh}) and Eqs. (\ref{g-bar-rad}) hold. This approach is based on the results presented by \citet{2016PhRvL.117t1101M} and also the results by \citet{2017A&A...598A..66P} are used.

We can use Eqs. (\ref{obs-Gaugh}) (or, Eqs. \ref{obs-r-0} and \ref{obs-r-1}) and (\ref{g-bar-rad}) for theoretical construction of the relation $v_{c}$ $=$ $v_{c}(R)$ for the Galactic equatorial plane, $z$ $=$ 0. We obtain {for the rotation speed
$v_{c} \left ( R \right )$}
\begin{eqnarray}\label{rot-curve-theory-old}
v_{c} \left ( R \right ) &=& \sqrt{R ~\frac{\partial \tilde{\Phi}_{BM}}{\partial R} \left [ 1 - 
	\exp{ \left ( -\sqrt{g_{+}^{-1} \frac{\partial \tilde{\Phi}_{BM}}{\partial R}} \right )} \right ]^{-1} } ~,
\nonumber \\
g_{+} &=& 1.2 \times 10^{-10} ~\mbox{m} ~\mbox{s}^{-2} ~.
\end{eqnarray}	
Eqs. (\ref{rot-curve-theory-old}) yield the theoretical rotation curve of the MW. The theoretical rotation curve of the MW is consistent with the observed rotation curve, see Fig. 1.
Fig. 1 compares our theoretical results with the observational data \citep{2019ApJ...871..120E}. 

\begin{figure}[h]
\centering

\subfloat{
	\label{subfig:correct}
	\includegraphics[width=0.95\linewidth]{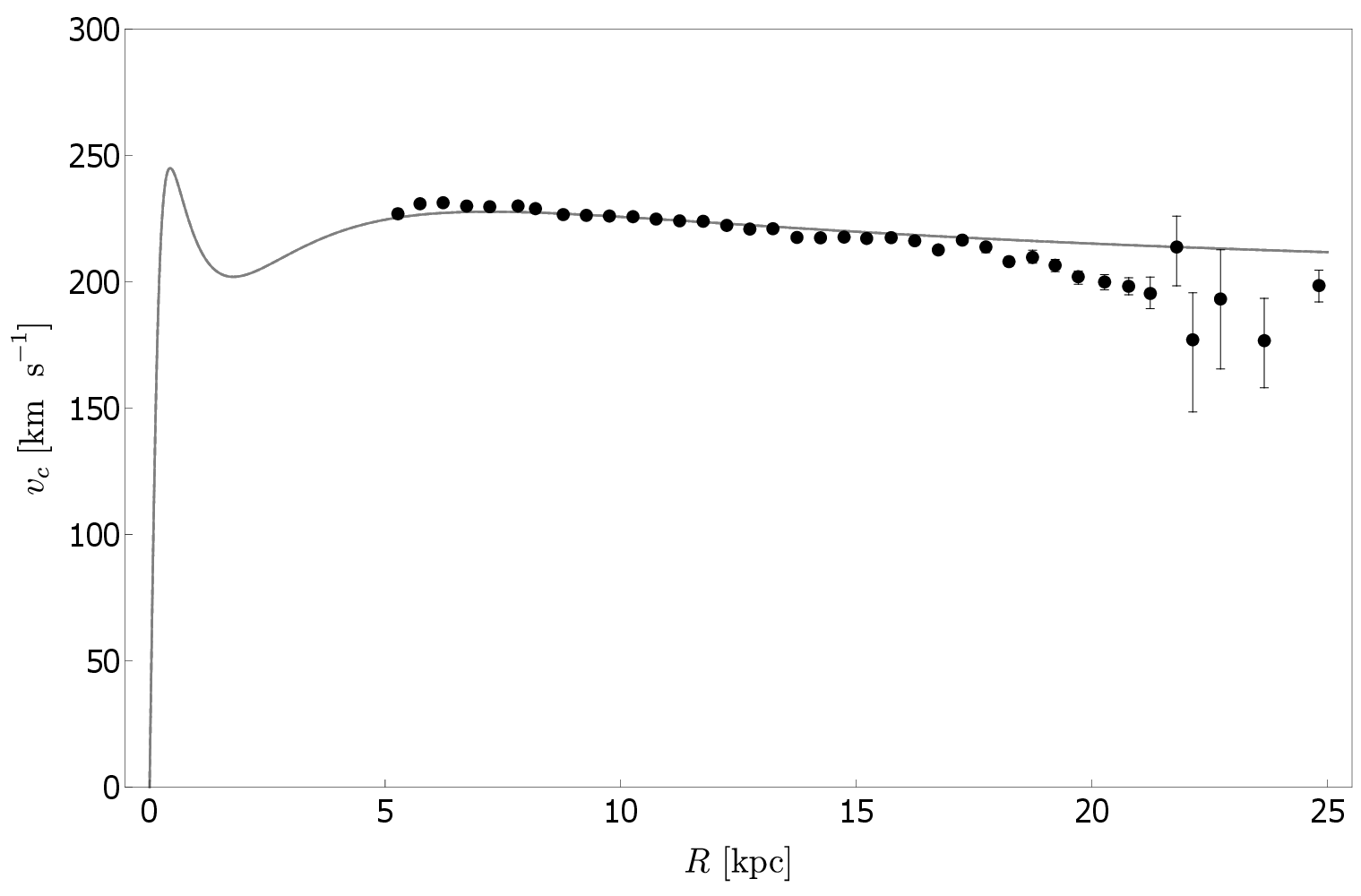} }

\caption{The plot depicts both the theoretical and observational rotation curves of the MW. The circular \textbf{velocity} $v \equiv v_c$ is a function of the Galactocentric distance $R$. The solid curve follows Eqs. (\ref{rot-curve-theory-old}) for the global Galactic field, while the scatter represent the data of \citet{2019ApJ...871..120E}. The solid curve corresponds to the baryonic Tully-Fisher relation for large $R$. The observed lower values of $v_c$ for greater $R$ can be understood as an action of the local gravitational fields of “different tracer populations that have been used for the analyses” (Eilers et al. 2019, Sec. 6).}
\label{Fig1}
\end{figure}

The third of Eqs. (\ref{obs-Gaugh}) yields for the DM rotation speed $v_{cDM}$ $\equiv$ $v_{cDM} (R)$ {the relation}
\begin{eqnarray}\label{v-DM}
v_{cDM}^{2} &=& \left ( \sum_{j=1}^{3} v_{cj}^{2} \right ) 
\left \{ \exp{\left [ \sqrt{\sum_{j=1}^{3} v_{cj}^{2} / \left ( g_{+} R \right )} ~ \right ] - 1} \right \}^{-1} ~,
\end{eqnarray}
where $v_{cj}^{2} / R$ $=$ $\left ( g_{bar} \right )_{j}$ and $j$ $=$ 1 and 2 correspond to the thin and thick discs, $j$ $=$ 3 corresponds to the bulge. Eqs.  (\ref{obs-Gaugh}) and (\ref{v-DM}) yield {for the final theoretical circular velocity $v_{c}$}
\begin{eqnarray}\label{v-total-DM}
v_{c}^{2} &=& \sum_{j=1}^{3} v_{cj}^{2} + v_{cDM}^{2} 
\end{eqnarray}
{and $v_{c}$ should correspond to the observational values}.

\subsection{DM acceleration - \citet{2017A&A...598A..66P} approach}\label{sec-DM-P}
As for the gravitational potential for the DM of the MW, \citet{2017A&A...598A..66P} present, in their Eq. (4),
\begin{eqnarray}\label{eqDM-Poul}
\tilde{\Phi}_{halo}(r) &=& -~ \frac{G M_{halo}}{r} - \frac{G M_{halo}}{1.02 a_{halo}} \times
\nonumber \\
& & \left [ \frac{-1.02}{1 + ( \xi / a_{halo} )^{1.02}} + \ln \left ( 1 + 
\left ( \frac{\xi}{a_{halo}} \right )^{1.02} \right ) \right ]_{\xi = r}^{100 ~\tiny{\mbox{kpc}}}  ,
\nonumber \\
M_{halo} &=& 1.392 ~\times 10^{11} ~\mbox{M}_{\odot} ~,
\nonumber \\
a_{halo} &=& 14 ~\mbox{kpc} ~.
\end{eqnarray}

We can mention that Eqs. (\ref{eqDM-Poul}) does not fulfill the condition of continuity of the potential, $\lim_{r \rightarrow R_{DM}^{-}} \tilde{\Phi}_{halo}(r)$ $=$
$\lim_{r \rightarrow R_{DM}^{+}} \tilde{\Phi}_{halo}(r)$ $=$ $-~G M_{DM}/R_{DM}$. Thus,
Eqs. (\ref{eqDM-Poul}) must contain $-~G M_{DM}/R_{DM}$ instead of $-~ G M_{halo}/r$, where $M_{DM}$ and $R_{DM}$ are the mass and the radius of the halo DM, $M_{halo}$ $\ne$ $M_{DM}$. The real values, $M_{DM}$ $=$ 134 $\times$ 10$^{10}$ $M_{\odot}$ and 
$R_{DM}$ $=$ 158 kpc, are presented by Eqs. (\ref{M_DM}) and (\ref{R_DM}) below. Thus, we will not consider the first term on the right-hand side of Eqs. (\ref{eqDM-Poul}) in calculating 
physical quantities such as accelerations and velocities, i.e., derivatives of $\tilde{\Phi}_{halo}(r)$. Moreover, 
Eqs. (\ref{eqDM-Poul}) pose another physical problem. When taking both $M_{halo}$ and $a_{halo}$ as free parameters to be determined from the rotation curve, Eqs. (\ref{eqDM-Poul}) do not respect the baryonic Tully-Fisher relation.

The radial component of the DM gravitational acceleration ($z$ $=$ 0) is
\begin{eqnarray}\label{g-DM-rad}
g_{DM} &=& \frac{\mbox{d} \tilde{\Phi}_{DM}}{\mbox{d} R} ~,
\nonumber \\
\tilde{\Phi}_{DM} &=& \tilde{\Phi}_{halo}(R)  ~.
\end{eqnarray}	

The radial component of the total gravitational acceleration ($z$ $=$ 0) is given by the second of Eqs. (\ref{obs-Gaugh}), where Eqs. (\ref{g-bar-rad}) and 
(\ref{g-DM-rad}) hold.

As depicted in Fig. 2, the theoretical rotation curve based on the model \citep{2017A&A...598A..66P} exhibits a systematic shift from the observational data.

\begin{figure}[h]
\centering
	\includegraphics[width=0.95\linewidth]{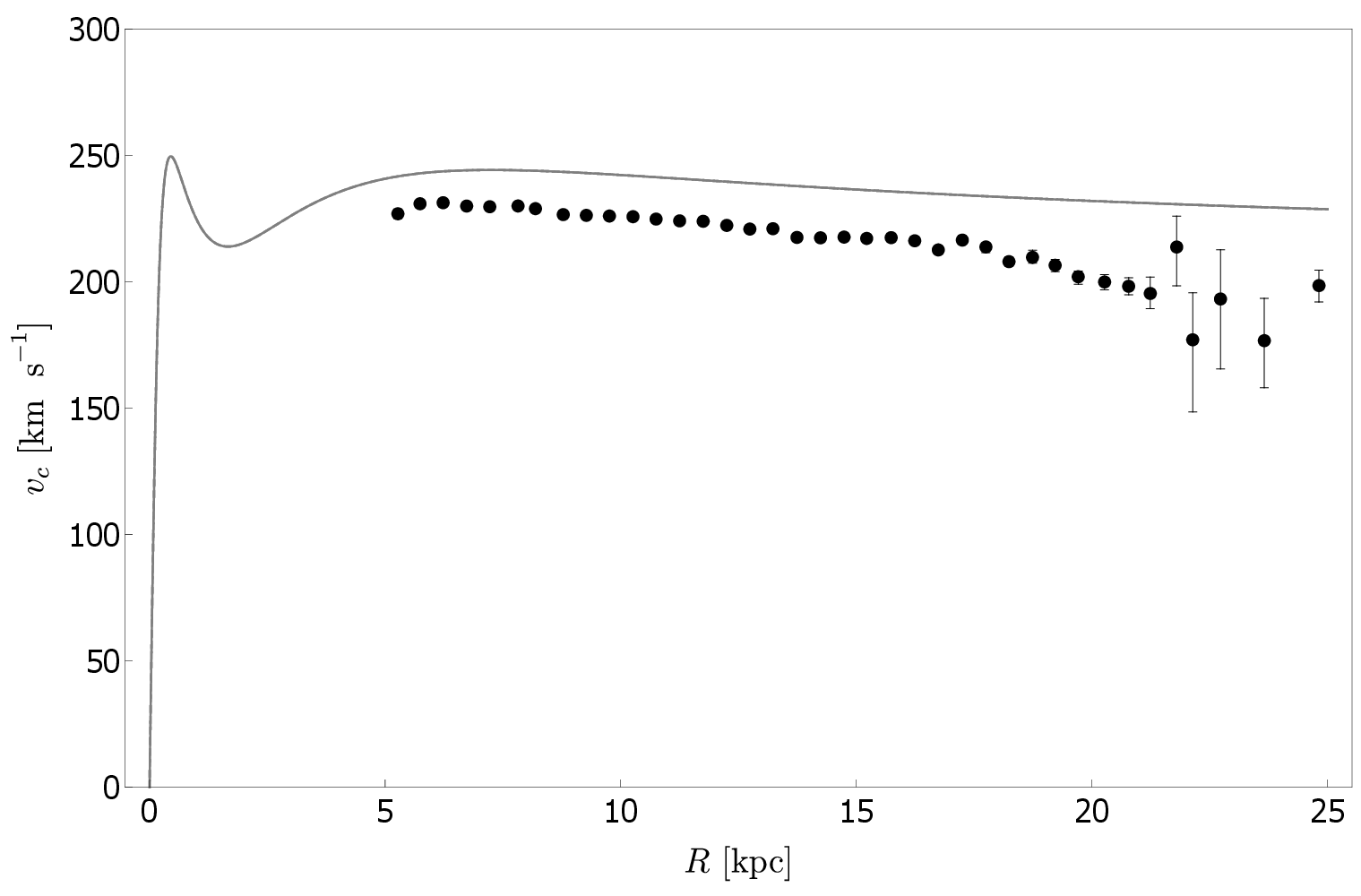}
	\caption{The solid curve is based on the assumption of the existence of the dark matter, see Model I in \citep{2017A&A...598A..66P}. }\label{f1}
	\label{Fig2}
\end{figure}

{Really, we can show the conclusion on the systematic shift explicitly, deriving and calculating the circular velocity $v_{c}$. The} circular velocity $v_{c}$ obtained from Eqs. (\ref{obs-Gaugh})-(\ref{g-bar-rad}) and (\ref{eqDM-Poul})-(\ref{g-DM-rad}) is
\begin{eqnarray}\label{30}
v_{c}^{2} &=& \sum_{j=1}^{3} v_{cj}^{2} + v_{cDM-P}^{2} ~,
\end{eqnarray}
where
\begin{eqnarray}\label{31}
v_{cj}^{2} &=& G \tilde{M}_{j}~ R^{2} / 
[ R^{2} + ( \tilde{a}_{j} + \tilde{b}_{j} )^{2} ]^{3/2} ~, 
~j = 1,~2,~3~,
\end{eqnarray}
the values in Table \ref{Table1} hold for the baryonic matter, and
\begin{eqnarray}\label{32}
v_{cDM-P}^{2} &=&  \left ( G M_{halo} / a_{halo} \right )  \zeta
                                 \left ( 2.02 + \zeta \right ) / \left ( 1 + \zeta \right )^2 ~,
\nonumber \\
\zeta &=& \left ( R/a_{halo} \right )^{1.02} 
\end{eqnarray}
holds for the DM. If $R$ $=$ 8.247 kpc, then $v_{c}$ $=$ 243.9 km s$^{-1}$.
If $R$ $=$ 8.5 kpc, then $v_{c}$ $=$ 243.7 km s$^{-1}$, which is not consistent either
with the circular velocity 221.4 km s$^{-1}$ presented by \citet{2017A&A...598A..66P},
Sec. 2.5.1, p. 7), nor with the value 220 km s$^{-1}$ recommended by IAU. The model represented by Eqs. 
(\ref{pot-bulge}), Table \ref{Table1} and  Eqs. (\ref{eqDM-Poul}) yields the circular velocity 
$v_{c} (R=8.122 ~\mbox{kpc})$ $=$ 244.6 km s$^{-1}$, which is not consistent with the observational data on circular rotation speeds - the speed (229.0 $\pm$ 0.2) km s$^{-1}$ holds for the galactocentric distance $R$ $=$ 8.122 kpc according to
\citet{2019ApJ...871..120E}. 

\subsection{DM acceleration - corrected \citet{2017A&A...598A..66P} approach}\label{sec-DM-AS-P}
As for the gravitational potential for the DM of the MW, it seems that  
\citet{2017A&A...598A..66P} present, in their Eq. (4), incorrect formulae and instead of Eqs. (\ref{eqDM-Poul})
we should write, using \citet{Allen1991},
\begin{eqnarray}\label{eqDM-AS}
\tilde{\Phi}_{halo}(r) &=& -~ \frac{G M_{halo}}{r}  
\frac{\left ( r/a_{halo} \right )^{2.02}}{1 + \left ( r/a_{halo} \right )^{1.02}}
- \frac{G M_{halo}}{1.02 a_{halo}} \times
\nonumber \\
& & \left [ \frac{-1.02}{1 + ( \xi / a_{halo} )^{1.02}} + \ln \left ( 1 + 
\left ( \frac{\xi}{a_{halo}} \right )^{1.02} \right ) \right ]_{\xi = r}^{R_{DM}}  ,
\nonumber \\
M_{halo} &=& 1.392 ~\times 10^{11} ~\mbox{M}_{\odot} ~,
\nonumber \\
a_{halo} &=& 14 ~\mbox{kpc} ~,
\nonumber \\
R_{DM} &=& 100 ~\mbox{kpc} ~.
\end{eqnarray}

Eqs. (\ref{eqDM-AS}) are consistent with the condition
$\lim_{r\to R_{D}^{-}} \tilde{\Phi}_{halo}(r) = \lim_{r\to R_{D}^{+}} (- G M_{DM}/r)$, where
$M_{DM}$ $=$ $M_{halo}$ $\times$ $\left ( R_{DM}/a_{halo} \right )^{2.02} / \left [ 1 + \left ( R_{DM}/a_{halo} \right )^{1.02} \right ]$ and $M_{DM}$ and $R_{DM}$ are the mass and the radius of the DM halo. However, if we take the current values of the Hubble constant, 
then the virial radii of the DM halo are 176.8 kpc for 
$H_{0}$ $=$ (67.4 $\pm$ 0.5) $\mbox{km} ~\mbox{s}^{-1} ~\mbox{Mpc}^{-1}$
\citep{Planck2018} or 163.1 kpc for
$H_0$ $=$ (73.04 $\pm$ 1.04) $\mbox{km} ~\mbox{s}^{-1} ~\mbox{Mpc}^{-1}$ \citep{RiessHubble}. 
The values 176.8 kpc and 163.1 kpc do not match the value $R_{DM}$ $=$ 100 kpc presented by Eqs. 
(\ref{eqDM-Poul}) and (\ref{eqDM-AS}).

The third of Eqs. (\ref{obs-Gaugh}) contains the DM halo characteristic $g_{+}$. The approach used by 
\citet{2017A&A...598A..66P} considers other characteristics of the DM halo. We can consider the approach by \citet{2017A&A...598A..66P} as an approximation to the third of Eqs. (\ref{obs-Gaugh}). Really, if we consider the relation $g_{+}$ $=$ $v_{c}^{2}/r$ and the value $v_{c}$ $=$
(230 $\pm$ 5) km/s, then the result is $r$ $=$ (14.3 $\pm$ 0.6) kpc corresponding to $a_{halo}$ $=$ 14 kpc in 
Eqs. (\ref{eqDM-Poul})  and (\ref{eqDM-AS}). Similarly, for large galactocentric distance: \\
i) the third of Eqs. (\ref{obs-Gaugh}) leads to $v_{c}^{2}$ $=$ $G {\tilde{M}}_{BM} / \tilde{L}$, $\tilde{L}$ $=$ $\sqrt{G {\tilde{M}}_{BM}/g_{+}}$, \\
ii) Eqs. (\ref{eqDM-Poul}) or (\ref{eqDM-AS}) lead to $v_{c}^{2}$ $=$ $GM_{halo}/a_{halo}$. \\
As a consequence, $M_{halo}$ $=$
${\tilde{M}}_{BM} a_{halo}/ \tilde{L}$ $=$ (1.3 $\pm$ 0.1) $\times$ 10$^{11}$ $\mbox{M}_{\odot}$. Thus, one characteristic of the DM halo $g_{+}$ in the third of Eqs. (\ref{obs-Gaugh}) determines two characteristics of the DM halo in Eqs. (\ref{eqDM-Poul})  and (\ref{eqDM-AS}), the scale-length $a_{halo}$ and 
the mass $M_{halo}$.

The radial component of the DM gravitational acceleration ($z$ $=$ 0) is
\begin{eqnarray}\label{g-DM-rad-AS}
g_{DM} &=& \frac{\mbox{d} \tilde{\Phi}_{DM}}{\mbox{d} R} ~,
\nonumber \\
\tilde{\Phi}_{DM} &=& \tilde{\Phi}_{\mbox{\tiny{halo}}}(R)  ~.
\end{eqnarray}	

The radial component of the total gravitational acceleration ($z$ $=$ 0) is given by the second of Eqs. (\ref{obs-Gaugh}), where Eqs. (\ref{g-bar-rad}) and (\ref{g-DM-rad-AS}) hold.

The circular velocity $v_{c}$ obtained from Eqs. (\ref{pot-bulge})-(\ref{g-bar-rad}) and (\ref{eqDM-AS})-(\ref{g-DM-rad-AS}) is
\begin{eqnarray}\label{30-AS}
v_{c}^{2} &=& \sum_{j=1}^{3} v_{cj}^{2} + v_{cDM-AS}^{2} ~,
\end{eqnarray}
where Eqs. (\ref{31}) hold for the baryonic matter and
\begin{eqnarray}\label{32}
v_{cDM-AS}^{2} &=&  \left ( G M_{halo} / a_{halo} \right ) \zeta / \left ( 1 + \zeta \right ) ~,
\nonumber \\
\zeta &=& \left ( R/a_{halo} \right )^{1.02} ~,
\nonumber \\
a_{halo} &=& 14~ \mbox{kpc} 
\end{eqnarray}
holds for the DM. 

If $R$ $=$ 8.247 kpc, then $v_{c}$ $=$ 222.0 km s$^{-1}$.
If $R$ $=$ 8.5 kpc, then $v_{c}$ $=$ 222.3 km s$^{-1}$, while
the circular velocity 221.4 km s$^{-1}$ is presented by \citet{2017A&A...598A..66P}, Sec. 2.5.1, p. 7).  

The model represented by Eqs. (\ref{pot-bulge}), Table \ref{Table1} and Eqs. (\ref{eqDM-AS}) yields the circular velocity 
$v_{c} (R=8.122 \mbox{kpc})$ $=$ 222.9 km s$^{-1}$, which is not consistent with the observational data on circular rotation speeds - the speed (229.0 $\pm$ 0.2) km s$^{-1}$ holds for the galactocentric distance $R$ $=$ 8.122 kpc according to
\citet{2019ApJ...871..120E}. 

\subsection{Comparison of the DM models}\label{Comp-DM-models}
Sec. \ref{sec-DM-G} yields results in a very good agreement with the observational data {on rotation curve of the MW}. An important result is that there is no need to consider a potential with several fitted free parameters, as it is done by \citet{2017A&A...598A..66P}.

Sec. \ref{sec-DM-P} is based on Eq. (\ref{eqDM-Poul}). The potential is not consistent with observations neither for the region of the Sun nor for other regions of the MW. Results of Sec. \ref{sec-DM-P}
are not consistent with the observational data. 
The DM potential by \citet{2017A&A...598A..66P} produces DM rotation curve corresponding neither to observations, nor the baryonic Tully-Fisher relation. 

Another approach to the DM model of the MW halo is presented by \citet{Barros2016}, 
$\Phi_{DM} (r)$ $=$ $(v_{h}^{2} / 2)$ $\ln{\left ( r^{2} + r_{h}^{2} \right )}$ for spherical DM mass distribution. If we require $v_{c}^{2}$ $=$ $\sqrt{G g_{+} M_{BM}}$ for large galactocentric distance $r$ (see Eq. (\ref{large-R-speed-final} below) and $v_{c} (R_{0})$ $\equiv$ $v_{0}$ $=$ 228.8 km/s (see Eq. \ref{speed-sun} below), then $v_{h}$ $=$ $\left (G g_{+} M_{BM} \right )^{1/4}$ $=$ 191.4 km/s (consistent with the value of the Model II in \citet{Barros2016}) and $r_{h}$ $=$ 1.0065 $R_{0}$, where $R_{0}$ is the Galactocentric distance of the Sun. Finally, $\Phi_{DM} (r)$ $=$ $(v_{h}^{2} / 2)$ $\ln{ \left [ \left ( r^{2} + r_{h}^{2} \right )/ \left ( R_{DM}^{2} + r_{h}^{2} \right ) \right ]}$ $-$ $GM_{DM}/R_{DM}$, $M_{DM}$ and $R_{DM}$ are mass and radius of the DM halo,
$\lim_{r \rightarrow R_{DM}^{-}} \Phi_{DM}(r)$ $=$
$\lim_{r \rightarrow R_{DM}^{+}} \Phi_{DM}(r)$ $=$ $-~G M_{DM}/R_{DM}$. We obtain $R_{DM}$ $\doteq$ $r_{vir}$ $=$ 164.2 kpc or 151.6 kpc for
$H_{0}$ $=$ 67.4 $\mbox{km} ~\mbox{s}^{-1} ~\mbox{Mpc}^{-1}$
\citep{Planck2018} or $H_0$ $=$ 73.04 $\mbox{km} ~\mbox{s}^{-1} ~\mbox{Mpc}^{-1}$ \citep{RiessHubble}; analogously, $M_{DM}$ $=$ $(v_{h}^{2}/G)$ $R_{DM}^{3}/(R_{DM}^{2} + r_{h}^{2})$ $=$ 1.40 $\times$ 10$^{12}$ $\mbox{M}_{\odot}$ or 1.29 $\times$ 10$^{12}$ $\mbox{M}_{\odot}$. The mass density of the DM halo for the region of the Sun is $\varrho_{DM} \left ( R_{0} \right )$ $=$ 9.92 $\times$ 10$^{-3}$
$\mbox{M}_{\odot} ~\mbox{pc}^{-3}$.
Surprisingly, the scale-length of the DM halo $r_{h}$ $=$ 8.3 kpc is smaller than the scale-length of the BM of the MW $L$ $=$ 9.895 kpc, see Eq. (\ref{L-value}) below.

\subsection{Improvement of the DM model by  \citet{2017A&A...598A..66P} and \citet{Allen1991}}\label{fixed-model}
The improvement of the DM model by \citet{2017A&A...598A..66P} considers also the approach by \citet{Allen1991}. Detailed derivation is given in Appendix \ref{appendix-fixed-model} and we can summarize
\begin{eqnarray}\label{eqDM-AS-imp}
\tilde{\Phi}_{halo} \left ( r \right ) &=& -~ \frac{G M_{halo}}{r}  
\frac{\left ( r/a_{halo} \right )^{2.02}}{1 + \left ( r/a_{halo} \right )^{1.02}}
- \frac{G M_{halo}}{1.02 a_{halo}} \times
\nonumber \\
& & \left [ \frac{-1.02}{1 + ( \xi / a_{halo} )^{1.02}} + \ln \left ( 1 + 
\left ( \frac{\xi}{a_{halo}} \right )^{1.02} \right ) \right ]_{\xi = r}^{R_{DM}}  ,
\nonumber \\
M_{halo} &=& 6.696 \times 10^{10} ~M_{\odot} ~,
\nonumber \\
a_{halo} &=& 7.863 ~\mbox{kpc} ~,
\nonumber \\
R_{DM} &=& (157.5 \pm 6.4 ) ~\mbox{kpc} ~.
\end{eqnarray}
The values of the model parameters $a_{halo}$, $M_{halo}$ and $R_{DM}$ differ from the values considered by \citet{2017A&A...598A..66P}, compare Eqs. (\ref{eqDM-AS}). 

$\tilde{\Phi}_{halo}(r)$ is a good mathematical approximation to the third equation in Eqs. (\ref{obs-Gaugh}) for large galactocentric distances $r$. The system of Eqs. (\ref{eqDM-AS-imp}) is a good mathematical fit to rotation curve of the MW. The value 1.02 and the small values of $M_{halo}$ and $a_{halo}$ are not physical characteristics of the DM of the MW. While $R_{DM}/a_{halo}$ $\doteq$ 7 for the model presented by \citet{2017A&A...598A..66P}, the improved model given by Eqs. (\ref{eqDM-AS-imp}) yields $R_{DM}/a_{halo}$ $\doteq$ 20. 

Appendix \ref{C-simple-ideas} shows that Eqs. (\ref{eqDM-AS-imp}) can be treated as an approximation to the model developed in Sec. \ref{vec-eq-motion}. 

\section{Baryonic matter gravitational acceleration - improved model}\label{BM-improved} 

\citet{2017A&A...598A..66P} describes distribution of the baryonic matter in the MW and their results are summarized in Eqs. (\ref{pot-bulge}) and Table \ref{Table1}. 
Fig. 1 shows that the baryonic model very well describes rotation curve of the MW if Eqs. (\ref{rot-curve-theory-old}) are used.

\subsection{Motivation}
The baryonic model by\citet{2017A&A...598A..66P} and the observational results of \citet{2016PhRvL.117t1101M} well describe the observed rotation curve \citep{2019ApJ...871..120E}. However,
the detailed calculation leads to the rotation speed of the Sun $v_{0}$ $\equiv$ $v_{c}(R = 8.122 \mbox{kpc})$ $=$ 227.4 km s$^{-1}$, while the observational data lead to $v_{c}(R = 8.122 \mbox{kpc})$ $=$ 229.0 km s$^{-1}$ \citep{2019ApJ...871..120E}. 
We want to remove this small discrepancy. We will improve the Galactic mass model suggested by \citet{2017A&A...598A..66P}. We will obtain a new Galactic mass model for orbit computations.

\subsection{A new Galactic mass model}
We do not want to make great changes of the BM potentials given by \citet{2017A&A...598A..66P} since the authors state that the potentials very well describe mass distributions of the baryonic matter in the MW. The forms of the BM potentials lead to rotation curve of the MW which is consistent with the observed rotation curve (see Fig. 1). Although the rotation curve holds for the equatorial plane, $z$ $=$ 0, we will admit reasonable small changes of the values $\tilde{M}_{i}$, $i$ $=$ 1, 2, 3 and also of the values of the length scales $\tilde{a}_{j}$ and $\tilde{b}_{j}$, $j$ $=$ 1, 2 and 3
in Eqs. (\ref{pot-bulge}), in comparison with \citet{2017A&A...598A..66P}. Thus, we admit only small changes in values of the parameters in the baryonic gravitational potentials. We consider the following forms of the 
baryonic potentials 
\begin{eqnarray}\label{pot-bulge-new}
\Phi_{i} (R,z) &=& -~ \frac{G M_{i}}{\sqrt{R^{2} + \left ( a_{i} + \sqrt{z^{2} + b_{i}^{2}} \right )^{2} }} ~, ~i~=~1,~2,~3~.
\end{eqnarray}

The values $M_{i}$ $=$ $\tilde{M}_{i}$, $a_{i}$ $=$ $\tilde{a}_{i}$ 
and $b_{i}$ $=$ $\tilde{b}_{i}$, $i$ $=$ 1, 2, 3, 
correspond to \citet{2017A&A...598A..66P}. We will determine 
the values of $M_{i} /\tilde{M}_{i}$, $a_{i} / \tilde{a}_{i}$,
$b_{i} / \tilde{b}_{i}$,  $i$ $=$ 1, 2, 3, from the least square method applied to the rotation curve \citep{2019ApJ...871..120E}, the observational value of the circular \textbf{velocity} in the region of the Sun $v_{c}$ ($R$ $=$ 8.122 kpc) = 229.0 km s$^{-1}$
\citep{2019ApJ...871..120E} and the baryonic mass density for to the solar Galactocentric distance $R$ $=$ 8.247 kpc \citet{abuter2020detection} corresponding to the observed baryonic mass density \citep{2017A&A...598A..66P}. We remind that the Galactocentric distance $r$ is related to cylindrical coordinates $R$ and $z$ through the relation $r^{2}$ $=$ $R^{2}$ $+$ $z^{2}$. As for the rotation curve, we put $z$ $=$ 0 which corresponds to the motion in the central plane of the MW.

The total gravitational potential for the baryonic matter of the MW is
\begin{eqnarray}\label{pot-total-new}
\Phi_{BM} \left ( R, z \right ) &=& \sum_{i=1}^{3} \Phi_{i} \left ( R, z \right ) ~,
\end{eqnarray}
where $\Phi_{i} \left ( R, z \right )$, $i$ $=$ 1, 2 and 3 are given by Eqs. (\ref{pot-bulge-new}).
The radial component of the baryonic matter gravitational acceleration is ($z$ $=$ 0):
\begin{eqnarray}\label{g-bar-new}
g_{bar} &=& \left [ \frac{\mbox{d} \Phi_{BM} \left ( R, z \right )}{\mbox{d} R} \right ]_{z=0} ~. 
\end{eqnarray}	

\subsection{Rotation curve of the MW}
Now, we can construct the theoretical rotation curve of the MW and compare the theoretical result with the newest most accurate observational data. 

\subsubsection{Rotation curve of the MW - theory}
On the basis of Eqs. (\ref{obs-r-0}), (\ref{obs-r-1}) and (\ref{g-bar-new}), or,
Eqs. (\ref{obs-Gaugh}) and (\ref{g-bar-new}), we can write {for the rotation speed} 
\begin{eqnarray}\label{rot-curve-theory}
v_{c} \left ( R \right ) &=& \sqrt{R ~\frac{\partial \Phi_{BM}}{\partial R} \left [ 1 - 
	\exp{ \left ( -\sqrt{g_{+}^{-1} \frac{\partial \Phi_{BM}}{\partial R}} \right )} \right ]^{-1} } ~,
\nonumber \\
g_{+} &=& 1.2 \times 10^{-10} ~\mbox{m} ~\mbox{s}^{-2} 
\end{eqnarray}	
and Eqs. (\ref{pot-bulge-new}) and (\ref{pot-total-new}) hold, $z$ $=$ 0. {Explicitly, we have
for the rotation speed at the Galactocentric distance $R$} 
\begin{eqnarray}\label{rot-curve-theory-num-relation} 
v_{c} \left ( R \right ) &=& 143.470221 ~\mbox{km} ~\mbox{s}^{-1} \times
W \sqrt{R_{0} / R} \times 
\nonumber \\
& &   \left [ 1 - \exp{ \left ( -0.82100312 ~W ~R_{0} / R \right )}  \right ]^{-1/2} ~,
\nonumber \\
W^{2} &=& \sum_{j = 1}^{3} ( M_{j}/M_{d} ) [ 1 + ( a_{j} + b_{j})^{2} / R^{2} ]^{-3/2} ~,
\nonumber \\
M_{d} &=& ~3.9440 \times 10^{10} \mbox{M}_{\odot} ~,
\end{eqnarray}
$W$ $>$ 0, $R_0$ $=$ 8.247 $\mbox{kpc}$. Eqs. (\ref{rot-curve-theory}) and (\ref{rot-curve-theory-num-relation}) yield the theoretical rotation curve of the MW.

\subsubsection{Optimal values of $M_{i}$, $a_{i}$ and $b_{i}$, $i$ $=$ 1, 2, 3, }\label{opt_val}
Let us consider, as an approximation to reality, the least square method for rotation speeds. Thus we consider the minimization of the sum 
$\chi$ $\equiv$ $\sum_{j=1}^{N} \left [ v_{cj} (theory) - v_{cj}(observation) \right ]^{2}/ \sigma_{j}^{2}$, where $\sigma_{j}$ denotes error of the $j-$th measurement and the indices $j$ cover the interval of the Galactocentric distances $R$ $\in$ $\langle 5.74, 15.74 \rangle$ kpc \citep{2019ApJ...871..120E}. The theoretical rotation speed 
$ v_{cj} (theory)$ $=$ $v_{c}(R_{j})$ is given by Eqs. (\ref{rot-curve-theory-num-relation}).

The values of the parameters in Eqs. (\ref{pot-bulge-new}) are summarized in Table \ref{Table2}.

\begin{table}[h]
\centering
\def\arraystretch{1.3}
\begin{tabular}{ |c|c|c|c|c| } 
 \hline
 structure & index $i$ & $M_{i}~[10^{10}\mbox{M}_\odot]$ & $a_{i}~[\mbox{kpc}]$ & $b_{i}~[\mbox{kpc}]$\\
 \hline
 thin disc & 1 & 3.3642 & 4.51 & 0.21\\
 thick disc & 2 & 4.1574 & 2.34 & 0.68\\
 bulge & 3 & 0.9071 & 0 & 0.30\\
 \hline
\end{tabular}
\caption{{Values of the parameters for the baryonic matter gravitational potentials of the MW components, see Eqs. (\ref{pot-bulge-new}).
The values are optimized parameters for the model represented by Eqs. (\ref{pot-bulge-new}),
(\ref{pot-total-new}), (\ref{g-bar-new}), (\ref{rot-curve-theory}) and compared with the observational data (Eilers et al. 2019).}}
\label{Table2}
\end{table}


The numerical values summarized in Table \ref{Table2} show that
the model by \citet{2017A&A...598A..66P} overestimates the bulge mass in 20\%. The values summarized in Table \ref{Table2} yield values of the speeds $v_{c}$ ($R$ $=$ 8.122 kpc) = 229.0 km s$^{-1}$, $v_{c}$ ($R$ $=$ 8.247 kpc) = 228.8 km s$^{-1}$ and the value of the derivative of the speed is
$\mbox{d} v_{c} / \mbox{d} R$ ($R$ $=$ 8.247 kpc) $=$ -1.72 km s$^{-1}$ kpc$^{-1}$. 
The results are consistent with the values presented by \citep{2019ApJ...871..120E}.
The result $v_{0}$ $=$ 228.8 $\mbox{km}~ \mbox{s}^{-1}$ is consistent with the value $v_{c} (R=8.25~ \mbox{kpc})$ $=$ (228.9 $\pm$ 0.7) $\mbox{km}~ \mbox{s}^{-1}$ given by \citet{2023ApJ...945....3S}.
The total mass of the baryonic matter of the MW is $M_{BM}$ $=$ $\sum_{i=1}^{3} M_{i}$ $=$ 8.43 $\times$ 10$^{10}$ $\mbox{M}_{\odot}$. 
The error of $M_{BM}$ may be estimated on the basis of Eqs. 
(\ref{rot-curve-theory-num-relation}) and the error of $v_{c}(R_{0})$ $\equiv$ $v_{0}$, $\sigma_{vc0}$ $=$ 0.2 km s$^{-1}$. {The result for the total BM mass of the MW is} 
\begin{equation}\label{MW-mass}
M_{BM} = \sum_{i=1}^{3} M_{i} = ( 8.43 \pm 0.01 ) \times 10^{10} ~\mbox{M}_{\odot} ~.
\end{equation}

If we consider Eqs. (\ref{rot-curve-theory}) and (\ref{rot-curve-theory-num-relation}) with the numerical values summarized in Table \ref{Table2}, then the relation 
$\left [ v_{c} \left ( R \right ) \right ]^{2}/R$ $=$ $g_{+}$ 
and Eqs. (\ref{rot-curve-theory-num-relation})
lead to $R$ $=$ 13.1 $\mbox{kpc}$. The contributions of the BM and the DM components are 
$\left ( \left [ v_{c} \left ( R \right ) \right ]^{2}/R \right)_{BM}$ 
$=$ 0.51 $g_{+}$ and 
$\left ( \left [ v_{c} \left ( R \right ) \right ]^{2}/R \right)_{DM}$ 
$=$ 0.49 $g_{+}$,
$\left ( \left [ v_{c} \left ( R \right ) \right ]^{2}/R \right)_{BM}$ $+$
$\left ( \left [ v_{c} \left ( R \right ) \right ]^{2}/R \right)_{DM}$ $=$ $g_{+}$.

\subsubsection{Baryonic model of the MW}
We have improved baryonic model of \citet{2017A&A...598A..66P}.
More realistic baryonic model of the MW is given by Eqs. 
(\ref{pot-bulge-new}) and the values summarized in Table \ref{Table2}. Also 
Eqs. (\ref{obs-Gaugh}), (\ref{pot-total-new}) and (\ref{g-bar-new}) complement the total gravitational acceleration. 

The agreement between the theoretical and observational results on the rotation curve of the MW is evident. {The improvement represented by 
Eqs. (\ref{pot-bulge-new}) and 
the values summarized in Table \ref{Table2} is presented in Fig. \ref{Fig3}. It is worth to mention that the curve corresponding to Fig. \ref{Fig1} lies below the observational points for Galactocentric distances less than about 8.5 kpc.}

\begin{figure}[h]
\centering
	\includegraphics[width=0.95\linewidth]{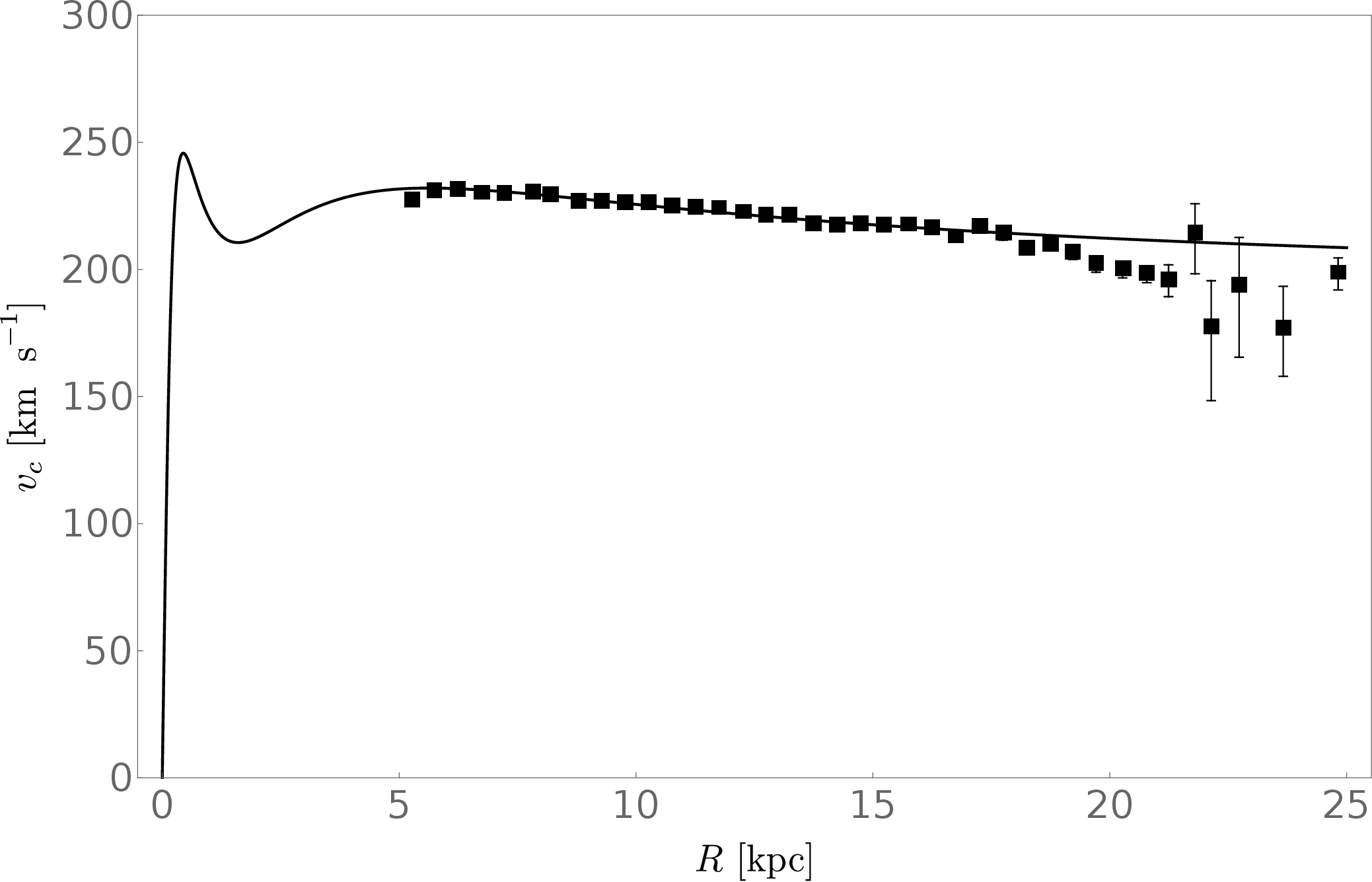}
	\caption{Rotation curve of the Milky Way galaxy. The curve depicts dependence of the circular {velocity} $v_{c}$ of a Galactic body on the Galactocentric distance $R$ of the body.
 The solid curve holds for Eqs. (\ref{pot-bulge-new}), (\ref{pot-total-new}), (\ref{g-bar-new}),
 (\ref{rot-curve-theory}), (\ref{rot-curve-theory-num-relation})
and for the values of the parameters summarized in Table \ref{Table2}. 
The curve fulfills the baryonic Tully-Fisher relation.}\label{curves}
	\label{Fig3}
\end{figure}

%
%
%

\section{Mass-speed relation}\label{Mass-speed-relation}
{We are interested in the rotation speed of a body moving around the center of the MW at large Galactocentric distance.} 

We want to apply Eqs. (\ref{rot-curve-theory}) to large values of the Galactocentric distance $R$, in this section. Eqs. (\ref{pot-bulge-new}), (\ref{pot-total-new}), the values summarized in Table \ref{Table2},
and $z$ $=$ 0 yield {for the total baryonic gravitational potential of the MW}
\begin{eqnarray}\label{large-R-phi}
\Phi_{BM} &=& -~ G M_{BM} / R ~,
\nonumber \\
M_{BM} &=& \sum_{i=1}^{3} M_{i} ~, 
\end{eqnarray}
for large values of $R$. Numerically, {the total mass of the BM of the MW is} $M_{BM}$ $=$ 8.43 $\times$ 10$^{10}$ M$_{\odot}$, see Eq. (\ref{MW-mass}).

Eqs. (\ref{rot-curve-theory}) and (\ref{large-R-phi}) give {for the circular velocity $v_{c}$ at large Galactocentric distance $R$}
\begin{eqnarray}\label{large-R-speed}
v_{c} \left ( R \right ) &=& \sqrt{G M_{BM} R^{-1} \left [ 1 - 
	\exp{ \left ( -\sqrt{G M_{BM} / g_{+}} / R \right ) } \right ]^{-1} } ~,
\nonumber \\
g_{+} &=& 1.2 \times 10^{-10} ~\mbox{m} ~\mbox{s}^{-2} ~.
\end{eqnarray}	
We can define the scale-length $L$ of the MW by the relation
\begin{eqnarray}\label{Mass2}
L &=& \sqrt{G M_{BM}/g_{+}} ~, 
\nonumber \\
M_{BM} &=& \sum_{j=1}^{3} M_j ~,
\end{eqnarray}
\begin{eqnarray}\label{L-value}
L &=& 9.895 ~\mbox{kpc} ~.
\end{eqnarray}
Taking into account that $L / R$ $\ll$ 1, we can reduce Eqs. (\ref{large-R-speed}) into 
\begin{eqnarray}\label{large-R-speed-final}
v_{c} \left ( R \right ) &=& \left ( G g_{+} ~ M_{BM} \right )^{1/4}  ~,
\nonumber \\
g_{+} &=& 1.2 \times 10^{-10} ~\mbox{m} ~\mbox{s}^{-2} ~.
\end{eqnarray}
{The obtained result shows that the circular velocity $v_{c}\left ( R \right )$ does not depend on the value of the large Galactocentric distance $R$, only
dependence on the baryonic mass of the MW $M_{BM}$ exists.}
The last set of equations is equivalent to the set of equations
\begin{eqnarray}\label{large-R-speed-final-T-F}
4 \log{v_{c}} &=& \log{M_{BM}} ~+~ \log{ \left ( G g_{+} \right ) }  ~,
\nonumber \\
g_{+} &=& 1.2 \times 10^{-10} ~\mbox{m} ~\mbox{s}^{-2} ~.
\end{eqnarray}
\citet{2000ApJ...533L..99M} state:
`,The baryonic Tully-Fisher relation appears fundamentally to be a relation between rotation velocity and total baryonic mass of the form $M_{BM}$ $\propto$ $v_{c}^{4}$''. Thus, we
can conclude that Eqs. (\ref{large-R-speed-final-T-F}) correspond to observational data in the form of an empirical relation between the total baryonic mass of a spiral galaxy and its asymptotic rotation speed of the form $M_{BM}$ $\propto$ $v_{c}^{4}$, compare \citet{2000ApJ...533L..99M} and \citet{2016PhRvL.117t1101M}.
As for the original paper by baryonic Tully and Fisher, see \citet{1977A&A....54..661T}.

Considerations presented in this section do not explicitly consider the existence of the DM. The theoretical approach considering explicit existence of the DM is presented in Sec. \ref{sec71} below.

\section{Theoretical results for the Sun}\label{theo-res-sun}
The theoretical approach used in the previous sections and its comparison with the observational data enable us to obtain some important quantities for the region of the Sun in the MW. 

We will use {for the Galactocentric distance of the Sun}
\begin{eqnarray}\label{R-sun}
R_{0} = (8.247 \pm 0.0093) ~\mbox{kpc} ~,
\end{eqnarray}
see \citep{abuter2020detection}. This value is consistent with the value $R_{0}$ $\in$ (8.24, 8.31) $\mbox{kpc}$ presented by \citet{Gaia2021}.

\subsection{Rotation speed}\label{rot-speed}
Eqs. (\ref{pot-bulge-new}), (\ref{pot-total-new}), $z$ $=$ 0, Eqs. (\ref{rot-curve-theory}), (\ref{rot-curve-theory-num-relation})
and the values summarized in Table \ref{Table2} enable us to find circular rotation speed $v_{c} (R_{0})$ $\equiv$ $v_{0}$ for the position of the Sun $R_{0}$ $=$ 8.247 kpc:
\begin{eqnarray}\label{speed-sun}
v_{0} = ( 228.8 \pm 0.2 ) ~\mbox{km}~ \mbox{s}^{-1} ~, 
\end{eqnarray}
where the error of the speed is consistent with the error presented by \citet{2019ApJ...871..120E}. The value 228.8 $\mbox{km}~ \mbox{s}^{-1}$ holds  for the values summarized in Table \ref{Table2}. We can mention that the result 
$v_{0}$ $=$ (228.8 $\pm$ 0.2) $\mbox{km}~ \mbox{s}^{-1}$ is consistent with the value 
$v_{c} (R=8.25~ \mbox{kpc})$ $=$ (228.9 $\pm$ 0.7) $\mbox{km}~ \mbox{s}^{-1}$ given by (\citet{2023ApJ...945....3S}, Table \ref{Table1}).

The value of $v_{c}(R_0)$ $=$ $v_{0}$ is practically insensitive to the inaccuracy of the value $R_0$, $R_0=(8.247 \pm 0.0093)$ kpc.

We can mention that the baryonic model by \citet{2017A&A...598A..66P} leads to $v_{0}$ $=$ 227.4 km s$^{-1}$, if the approach by \citet{2016PhRvL.117t1101M} is used.

The Solar System is orbiting the center of the MW at a speed of  
$( 228.8 \pm 0.2 ) ~\mbox{km}~ \mbox{s}^{-1}$. 

\subsection{Orbital period}\label{orb-per}
The orbital period of the Sun, the period of revolution around the center of the Milky Way galaxy, is
\begin{eqnarray}\label{orbital period}
T_{0} = \frac{2\pi R_{0}}{v_{0}} = ( 221.5 \pm 0.3) \times 10^{6} ~\mbox{yrs} ~,
\end{eqnarray}
if Eqs. (\ref{R-sun}) and (\ref{speed-sun}) are used. 

\subsection{Angular velocity}\label{ang-vel}
The angular {velocity} $\vec{\omega}_{0}$ at the distance $R_{0}$ is 
\begin{eqnarray}\label{angular-speed}
\vec{\omega}_{0} &=& -~ \omega_{0} ~ \hat{z} ~,
\nonumber \\
\omega_{0} &=& \frac{v_{0}}{R_{0}} = (27.74 \pm 0.04) ~\mbox{km} ~\mbox{s}^{-1} ~\mbox{kpc}^{-1}   ~, 
\end{eqnarray}
if Eqs. (\ref{R-sun}) and (\ref{speed-sun}) are used {and the unit vector 
$\hat{z}$ points to the Galactic north pole}.

\subsection{Differentiation of velocity}\label{dif-vel}
We are not interested in $v_{0}$ alone, but also in the differentiation of $v_{c}(R)$, $\left[ \mbox{d}v_{c}(R)/\mbox{d}R \right]_{R=R_{0}}$. Eqs. (\ref{pot-bulge-new}), (\ref{pot-total-new}), $z=0$,  Eqs. (\ref{rot-curve-theory}), 
(\ref{rot-curve-theory-num-relation}) and the values summarized in Table \ref{Table2}, together with Eq. (\ref{speed-sun}) yield
\begin{eqnarray}\label{diff-vel}
\left( \frac{\mbox{d}v_{c}}{\mbox{d}R}\right)_{R=R_0} &=& -~1.72  ~\mbox{km} ~\mbox{s}^{-1} ~\mbox{kpc}^{-1}  
\end{eqnarray}
{for $R_{0}$ $=$ 8.247 $\mbox{kpc}$, see Eq. (\ref{R-sun}).
The result is in good coincidence with Eilers {\it et al.} (2019, Eq. 7), their result is ($-~1.7 \pm 0.1$)  $\mbox{km}~ \mbox{s}^{-1} ~\mbox{kpc}^{-1}$
for $R$ $=$ 8.122 kpc. We can mention that the BM and DM models by \citet{2017A&A...598A..66P} lead to the value $-~0.64$  $\mbox{km}~ \mbox{s}^{-1} ~\mbox{kpc}^{-1}$.}

\subsection{Local rotation curve}\label{loc-rot-cur}
{We can make a local approximation of the rotation speed $v_{c}(R)$ near the Galactocentric distance $R_{0}$.} Let us consider the relation 
\begin{eqnarray}\label{local}
 v_{c}(R) = v_{0} \left( \frac{R_{0}}{R} \right)^{\beta} ~,
\end{eqnarray}
for $R$ close to $R_{0}$. We can take Eq. (\ref{local}) as a definition for the dimensionless parameter $\beta$. Thus 
\begin{eqnarray}\label{local-diff}
\left( \frac{\mbox{d}v_{c}}{\mbox{d}R} \right)_{R=R_{0}} = -~ \beta ~\frac{v_{0}}{R_{0}} ~.
\end{eqnarray}
Comparison between Eqs. (\ref{diff-vel}) and (\ref{local-diff}) gives
\begin{eqnarray}\label{beta}
\beta = 0.062 ~,
\end{eqnarray}
if Eq. (\ref{angular-speed}) is also used. 

The value of $\beta$ is not very sensitive to the value of $v_{0}$. E.g., Eq. (\ref{beta}) holds both for $v_{0} = 227.4 ~\mbox{km}~ \mbox{s}^{-1}$, corresponding to the baryonic model by \citet{2017A&A...598A..66P}, and also for $v_{0} = 228.8 ~\mbox{km}~ \mbox{s}^{-1}$.

The local rotation curve enables us to find some other important quantities. Values of the Oort constants are calculated and discussed in Appendix \ref{ap-oort}, where the results of Secs. \ref{rot-speed}, \ref{orb-per}, \ref{ang-vel}, \ref{dif-vel} and \ref{loc-rot-cur} are also used.

\subsection{Mass density}
Mass density of the baryonic matter $\varrho_{bar}$ can be calculated from the Poisson equation 
\begin{eqnarray}\label{mass-density}
\bigtriangleup \Phi_{BM} &=& 4 ~\pi ~ G~ \varrho_{BM}  ~,
\end{eqnarray}
where $G$ is the Newtonian gravitational constant and
Eqs. (\ref{pot-bulge-new}) and (\ref{pot-total-new}) hold. 
As for the region of the Sun,  $z$ $=$ 0 (approximation) and $R_{0}$ $=$ 8.247 kpc, Eq. (\ref{mass-density}) yields for the mass density $\varrho_{BM}(R_0)$
\begin{eqnarray}\label{mass-density-sun}
\varrho_{BM}(R_0) &=& 8.71 \times 10^{-2} ~\mbox{M}_{\odot}~ \mbox{pc}^{-3} ~. 
\end{eqnarray}
The mass density of the BM at the Galactocentric distance $R_{0}$ is
0.087 $\mbox{M}_{\odot}~ \mbox{pc}^{-3}$ for $z$ $=$ 0, which corresponds to the mass density of the BM in the region of the Sun.


\section{Orbit computations - Vector equation of motion}\label{vec-eq-motion}
We need vector equation of motion of a body moving in the MW, in order to create a model for orbit computations. We have to generalize the radial component of the equation of motion.

Eqs. (\ref{obs-Gaugh}) represent radial component of the equation of motion, for $z=0$. The generalization to 3-D case is the vector equation of motion of the body moving under the action of the gravity of the BM and the DM
\begin{eqnarray}\label{eq-motion-DM}
\vec{\dot{v}} &=& \vec{g}_{BM} + \vec{g}_{DM} ~,
\end{eqnarray}
where the gravitational acceleration of the BM is
\begin{eqnarray}\label{eq-motion-g-bar}
\vec{g}_{BM} &=& -~ \nabla \Phi_{BM} ~,
\end{eqnarray}
see Eqs. (\ref{pot-bulge-new}), (\ref{pot-total-new}) and the values summarized in Table \ref{Table2}.  

Now, we need to find $\vec{g}_{DM}$, the gravitational acceleration generated by the DM of the MW. If we assume spherical distribution of the DM, then we can write for the radial component 
\begin{eqnarray}\label{gbarDM}
 g_{bar} (R) &=& \sum_{j=1}^{3} G M_{j} ~\frac{R}{\left( R^2 + c_{j}^{2}\right)^{3/2}} ~,
 \nonumber \\
 c_k &=& a_k + b_k  ~,~~ k = 1,~2, ~3 ~,
\end{eqnarray}
see Eqs. (\ref{pot-bulge-new}), (\ref{pot-total-new}), 
(\ref{g-bar-new}),
(\ref{rot-curve-theory}), (\ref{rot-curve-theory-num-relation})
and the values summarized in Table \ref{Table2}. 
The second of Eqs.  (\ref{obs-Gaugh}) and Eqs. (\ref{gbarDM}) finally yield for 
the gravitational acceleration of the DM
\begin{eqnarray}\label{vecDM}
\vec{g}_{DM} &=& -~g_{DM}(r) ~\hat{\vec{r}} = -\frac{g_{bar}(r)}{\exp{ \left [ \sqrt{g_{bar}(r)/ g_{+}} \right ]} -1} ~\hat{\vec{r}} ~,
\nonumber \\
 g_{bar} (r) &=& \sum_{j=1}^{3} G M_{j} ~\frac{r}{\left( r^2 + c_{j}^{2}\right)^{3/2}} ~,  ~r \leq r_{DM}  ~,
 \nonumber \\
 c_k &=& a_k + b_k  ~,~~ k = 1,~2, ~3 ~,
\end{eqnarray}
where also Eqs. (\ref{pot-bulge-new}), (\ref{pot-total-new}), (\ref{g-bar-new}),
(\ref{rot-curve-theory}), (\ref{rot-curve-theory-num-relation}) and 
the values summarized in Table \ref{Table2} must be taken into account. We can remind that the DM distribution is spherically symmetric with respect to the center of the MW galaxy, $r$ $=$ $\sqrt{R^2 + z^2}$, $g_{DM}(r) \equiv g_{halo} (r)$ is often used, and, 
$\hat{\vec{r}}$ $\equiv$ $\vec{r}/ r$. Radius of the DM halo is $r_{DM}$. The DM halo of the MW is important in the conventional approach. 

Since $\nabla \times \vec{g}_{DM}=0$, we can write
\begin{eqnarray}\label{nablaG}
\vec{g}_{DM}(r) &=& -\nabla \Phi_{DM}(r) ~, ~r \leq r_{DM} ~,
\nonumber \\
\Phi_{DM}(r) &=& \Phi_{DM}(r_P) - \int_{r_P}^{r}  \vec{g}_{DM} \cdot \mbox{d}\vec{r} 
\nonumber \\
&=& \Phi_{DM}(r_P) + \int_{r_P}^{r}  g_{DM}(r)  ~\mbox{d}r ~,
\end{eqnarray}
where $\Phi_{DM}(r)$ is the DM gravitational potential and $P$ is a point in the space.

The equation of motion of a body moving in the MW is given by Eqs. 
(\ref{eq-motion-DM}), (\ref{eq-motion-g-bar}) and (\ref{vecDM}),
where also  Eqs. (\ref{pot-bulge-new}), (\ref{pot-total-new}), (\ref{g-bar-new}),
(\ref{rot-curve-theory}), (\ref{rot-curve-theory-num-relation}) and 
the values summarized in Table \ref{Table2} hold.

Finally, we define 
\begin{equation}
\label{5final}
g_{DM} (r) = 0 ~, ~r \leq 0.15 ~\mbox{kpc} 
\end{equation}
for the central region of the MW. The value $0.15$ kpc is less than the value $c_3$, see Eqs. (\ref{vecDM}). The value $r=0.15$ kpc corresponds to the negligible value of the mass density $\varrho_{DM}(r)$.

\section{DM - simple consequences}\label{DM-simple-cons}
Several simple consequences of the existence of the DM are treated in this Section.

\subsection{Mass-speed relation}\label{sec71}
{We can write for the rotation speed $v_{c} \left ( R \right )$ at the distance $R$,} on the basis of Eqs. (\ref{eq-motion-DM}), (\ref{eq-motion-g-bar}), (\ref{gbarDM}) and (\ref{vecDM}), {$\left [v_{c} \left ( R \right ) \right ]^2 / R$ $=$ $g_{BM}$ $+$ $g_{DM}$
$=$ $G~M/R^{2}$ $+$ $g_{DM}$, or,}
\begin{eqnarray}\label{eqvDM}
\frac{\left [ v_{c} \left ( R \right ) \right ]^2}{R} &=& g_{DM} ~,
\nonumber \\
 g_{DM} &=& \frac{g_{bar}}{\exp{(\sqrt{g_{bar}/ g_{+}}}) -1} = \sqrt{g_{+} g_{bar}} ~,
 \nonumber \\
 g_{bar} &=& \frac{G M_{BM}}{R^2} ~,
 \nonumber \\
 M_{BM} &=& \sum_{i=1}^{3} M_{i} ~,
\end{eqnarray}
for large values of $R$. Thus 
\begin{eqnarray}\label{VGg}
v_{c} \left ( R \right ) = \left( G g_{+} M_{BM} \right)^{1/4} ~. 
\end{eqnarray}
{The obtained result shows that the rotation speed $v_{c}\left ( R \right )$ does not depend on the value of the  Galactocentric distance $R$, for large value of $R$. Only the dependence on the total baryonic mass $M_{BM}$ of the MW exists.}
One can immediately see, that Eq. (\ref{VGg}) is equivalent to Eq. (\ref{large-R-speed-final}). This subsection shows that the DM explains the observational result $v_{c} \propto {M_{BM}}^{1/4}$.

Our derivation of the relation represented by Eq. (\ref{VGg}) 
helps the physical understanding of the observational result. The action of the DM, the dominant matter of the MW, determines motion of a body at large distance, see the first of Eqs. (\ref{eqvDM}). The action of the DM leads to the relation between the rotation speed $v_{c}$ and the total baryonic mass $M$ of the MW. No information about mass of the DM is contained in Eq. (\ref{VGg}),
the mass-speed relation.

\subsection{Distribution of DM}
The distribution of the DM is given by the mass density of the DM. The mass density $\varrho _{DM} (r) $ can be calculated from the equation
\begin{eqnarray}\label{nablaG}
\nabla \cdot \vec{g}_{DM} = -~4 \pi G \varrho _{DM} ~, 
\end{eqnarray}
where Eqs. (\ref{vecDM}) hold. Considering spherical symmetry, Eq. (\ref{nablaG}) can be rewritten in the form 
\begin{eqnarray}\label{nablaG1}
\frac{1}{r^2} \frac{\mbox{d}}{\mbox{d} r} \left[ r^2 g_{DM} (r) \right]= 4 \pi G \varrho _{DM}(r) ~, 
\end{eqnarray}
or, the mass density is 
\begin{eqnarray}\label{nablaG2}
\varrho_{DM} (r) = \frac{1}{4 \pi G} \left[  \frac{\mbox{d}g_{DM}(r)}{\mbox{d} r} + \frac{2}{r} g_{DM}(r)  \right] ~,
\end{eqnarray}
where
\begin{eqnarray}\label{MDens}
g_{DM}(r) &=& \frac{g_{bar}(r)}{\exp{ \left [ \sqrt{g_{bar}(r)/ g_{+}} \right ]} - 1} ~,
\nonumber \\
 g_{bar}(r) &=& \sum_{j=1}^{3} GM_{j} \frac{r}{\left( r^2 + c_{j}^{2}\right)^{3/2}} ~,  ~r \leq r_{DM} ~,
  \nonumber \\
\frac{\mbox{d}g_{DM}(r)}{\mbox{d} r}  &=& \frac{g' _{bar}(r)}{\exp{ \left [ \sqrt{g_{bar}(r)/ g_{+}} \right ]} -1} \times 
\nonumber \\
&\times& \left\{1 - \frac{\sqrt{g_{bar}(r)/g_{+}}/2}{1-\exp { \left[ -\sqrt{g_{bar}(r)/ g_{+}} \right] } } \right\} ~,
\nonumber \\
g' _{bar}(r) &=& \sum_{j=1}^{3} GM_{j} \frac{1 -3r^2 / (r^2 + c_j^2)}{\left( r^2 + c_{j}^{2}\right)^{3/2}} ~,
\nonumber \\
c_k &=& a_k + b_k  ~,~~ k = 1,~2, ~3 ~,
\end{eqnarray}
where also Eqs. (\ref{pot-bulge-new}), (\ref{pot-total-new}), (\ref{g-bar-new}),
(\ref{rot-curve-theory}), (\ref{rot-curve-theory-num-relation}) and 
the values summarized in Table \ref{Table2} must be taken into account.

Mass of the DM halo $M_{DM}(r)$ follows from Eq. (\ref{nablaG1}): 
\begin{eqnarray}\label{MassDM}
M_{DM}(r) = \frac{1}{G} r^2 g_{DM}(r) ~, ~r \leq r_{DM} ~,
\end{eqnarray}
since $M_{DM}(r) =4 \pi \int_{0}^{r} x^2 \varrho_{DM}(x) ~\mbox{d} x $. 

If $\sqrt{g_{bar}(r)/g_{+}} \ll 1$ and $r^2 \gg c_{1}^{2}$, then Eqs. (\ref{MDens}) and (\ref{MassDM}) give 
\begin{eqnarray}\label{Mass}
M_{DM}(r) = M_{BM} ~\frac{r}{L}  ~,
\end{eqnarray}
where the length scale of the MW $L$ is defined by Eqs. (\ref{Mass2}) and (\ref{L-value}).
Eq. (\ref{Mass}) yields for the mass density $\varrho_{DM} (r)$ at a large distance $r$ 
\begin{eqnarray}\label{rhoDM1}
\varrho_{DM}(r) = \frac{1}{4 \pi} \frac{M_{BM}}{L r^2} ~, ~r \leq r_{DM} ~.
\end{eqnarray}

Eqs. (\ref{5final}) and (\ref{nablaG1}) give 
\begin{equation}
\label{7final}
\varrho_{DM} (r) = 0 ~, ~r \leq 0.15 ~\mbox{kpc} 
\end{equation}
for the central region of the MW. This condition respects decrease of $\varrho_{DM}(r)$ with dereasing $r$. $\varrho_{DM}(r)$ decreases almost to zero and the condition avoids mathematical increase of $\varrho_{DM}(r)$ for values $r \leq 0.15 ~\mbox{kpc}$ when Eqs. (\ref{nablaG2}) and (\ref{MDens}) are used in calculations.

\subsection{Virial radius and virial mass}\label{Virial-radius-mass}
We are interested in the virial radius and the virial mass of the MW. The virial radius corresponds to the
radius of the DM where the mass density decreases to
the value 200 $\varrho_{crit}$, where 
$\varrho_{crit}$ $=$ $3 H_{0}^{2} / ( 8 \pi G )$ and $H_{0}$ is the Hubble constant.

Considering the mass of the baryonic component of the MW, $M_{BM}$ $=$ 8.43 $\times$ 10$^{10}$ $\mbox{M}_{\odot}$ (see Sec. \ref{opt_val} - Eq. \ref{MW-mass}), and using Eqs. (\ref{Mass2}) we obtain $L=9.895 ~\mbox{kpc}$. The critical mass density is
\begin{equation}\label{crit-mass-density}
\varrho_{crit} = \frac{3}{8~\pi} \frac{1}{G}~ H_{0}^{2} ~, 
\end{equation}
where $H_{0}$ is the value of the Hubble constant. The virial radius is obtained by truncating the DM halo at 200 times the critical density. The relation for the virial radius then follows from Eq. (\ref{rhoDM1})
\begin{equation}\label{virial-radius}
r_{vir}=\frac{1}{\sqrt{4\pi}} \frac{1}{\sqrt{200~ \rho_{crit}}} \sqrt{\frac{M_{BM}}{L}} ~.
\end{equation}
The virial mass follows from Eq. (\ref{Mass}), 
\begin{equation}\label{virial-mass}
M_{vir} = M_{BM} ~\frac{r_{vir}}{L} ~,
\end{equation}
{where $M_{BM}$ $=$ 8.43 $\times$ 10$^{10}$ $\mbox{M}_{\odot}$ and $L=9.895 ~\mbox{kpc}$.}

\subsubsection{$H_{0}$ $=$ (67.4 $\pm$ 0.5) $\mbox{km} ~\mbox{s}^{-1} ~\mbox{Mpc}^{-1}$}
We will consider the value of the Hubble constant
$H_{0}$ $=$ (67.4 $\pm$ 0.5) $\mbox{km} ~\mbox{s}^{-1} ~\mbox{Mpc}^{-1}$
following from the ESA's Planck spacecraft data, see \citet{Planck2018}. Using Eqs. (\ref{crit-mass-density}), (\ref{virial-radius}) and (\ref{virial-mass}) we obtain the values 
\begin{eqnarray}\label{virial-mass-H0A}
\varrho_{crit,1} &=& (126.1 \pm 1.9) ~\mbox{M}_{\odot} ~\mbox{kpc}^{-3} ~,
\nonumber \\
r_{vir,1} &=& (164.0 \pm 1.2) ~\mbox{kpc} ~,
\nonumber \\
M_{vir,1} &=& (139.7 \pm 1.0) \times 10^{10} ~\mbox{M}_{\odot} ~,
\end{eqnarray}
for the critical density $\varrho_{crit,1}$, the virial radius $r_{vi1,r}$ and the virial mass $M_{vir,1}$.

\subsubsection{$H_{0}$ $=$ (73.04 $\pm$ 1.04) $\mbox{km} ~\mbox{s}^{-1} ~\mbox{Mpc}^{-1}$}
We will consider the value of the Hubble constant
$H_0$ $=$ (73.04 $\pm$ 1.04) $\mbox{km} ~\mbox{s}^{-1} ~\mbox{Mpc}^{-1}$ based on the standard candles observations, see \citet{RiessHubble}. Using Eqs. (\ref{crit-mass-density}), (\ref{virial-radius}) and (\ref{virial-mass}) we obtain the values 
\begin{eqnarray}\label{virial-mass-H0R}
\varrho_{crit,2} &=& (148.1 \pm 4.2) ~\mbox{M}_{\odot} ~\mbox{kpc}^{-3} ~,
\nonumber \\
r_{vir,2} &=& (151.3 \pm 2.1) ~\mbox{kpc} ~,
\nonumber \\
M_{vir,2} &=& (128.9 \pm 1.8) \times 10^{10} ~\mbox{M}_{\odot} ~,
\end{eqnarray}
for the critical density $\varrho_{crit,2}$, the virial radius $r_{vir,2}$ and the virial mass $M_{vir,2}$.

\subsubsection{Comparison}\label{Comp-MW}
{Although there is a fundamental inconsistency between the two values of the Hubble constant, we can obtain the mass and the radius of the DM halo of the MW. 

The arithmetical average of the numerical results presented by the third equations in Eqs. (\ref{virial-mass-H0A}) and (\ref{virial-mass-H0R}) leads to the mass of the DM halo of the MW
\begin{equation}\label{M_DM}
M_{DM} \doteq  M_{vir} = (134.3 \pm 1.0) \times 10^{10} ~\mbox{M}_{\odot} ~. 
\end{equation}
We can mention that the weighted average of the numerical values presented by the third equations in Eqs. (\ref{virial-mass-H0A}) and (\ref{virial-mass-H0R}), $M_{vir}$ $=$ $\sum_{j=1}^{2} (M_{vir,j}/\sigma_{j}^{2})$ / ($\sum_{i=1}^{2} 1/\sigma_{i}^{2}$) yields $M_{vir}$ $=$ (137.2 $\pm$ 0.9) $\times$ 10$^{10}$ $\mbox{M}_{\odot}$. However, one cannot be sure that the value 
$H_{0}$ $=$ (67.4 $\pm$ 0.5) $\mbox{km} ~\mbox{s}^{-1} ~\mbox{Mpc}^{-1}$ is more significant than the value $H_{0}$ $=$ (73.04 $\pm$ 1.04) $\mbox{km} ~\mbox{s}^{-1} ~\mbox{Mpc}^{-1}$ since we do not correctly understand the evolution of the Universe.

This value is consistent with the value $M_{vir}$
$=$ (1.18 $\pm$ 0.51) $\times$ 10$^{12}$ $\mbox{M}_{\odot}$
presented by \citet{Watkins}. The consistency holds for both values of $H_{0}$. We are using the value from Eq. (22) in \citet{Watkins}, because it accounts for the globular clusters measured by Gaia and Hubble  telescopes, as well as corrections for a past collision of the MW with a dwarf galaxy (see Section 4.4 in \citet{Watkins}).
We can mention that $M_{vir}$ $=$ (7.25 $\pm$ 0.26) $\times$ 10$^{11}$ $\mbox{M}_{\odot}$ according to \citet{2019ApJ...871..120E} using the NFW profile, 
$M_{vir}$ $=$ (6.5 $\pm$ 0.3) $\times$ 10$^{11}$ $\mbox{M}_{\odot}$ according to \citet{2023ApJ...945....3S} using the NFW profile, 
$M_{vir}$ $=$ (1.81 $\pm$ 0.06) $\times$ 10$^{11}$ $\mbox{M}_{\odot}$ according to
\citet{Ou} when using the Einasto profile and $M_{vir}$ $=$ (6.94 $\pm$ 0.12) $\times$ 10$^{11}$ $\mbox{M}_{\odot}$ when using the NFW profile. The dynamical mass, calculated as the sum of the baryonic mass and the DM halo virial mass, is $M_{dyn}$ $=$ (1.99 $\pm$ 0.09) $\times$ 10$^{11}$ $\mbox{M}_{\odot}$ according to \citet{Jiao} using Einasto model for the DM distribution. Low values of the last results imply significant decrease of the DM mass
of the MW in comparison with the results published in previous decades.

We can also find approximate value of the DM radius of the Milky Way. We can use the numerical results presented by the second equations in Eqs. (\ref{virial-mass-H0A}) and (\ref{virial-mass-H0R}). Their arithmetical average leads to (157.7 $\pm$ 1.2) kpc. The weighted average of the numerical values presented by the second equations in Eqs. (\ref{virial-mass-H0A}) and (\ref{virial-mass-H0R}), 
$R_{vir}$ $=$ $\sum_{j=1}^{2} (r_{vir,j}/\sigma_{j}^{2})$ / ($\sum_{i=1}^{2} 1/\sigma_{i}^{2}$) yields 
$R_{vir}$ $=$ (160.9 $\pm$ 1.0 kpc).
Another approach uses the relation $M_{DM}$ $=$ $M R_{DM} / L$ corresponding to Eq. (\ref{virial-mass}) and the result summarized in 
Eq. (\ref{M_DM}), 
$R_{DM}$ $=$ $M_{DM} L / M$, $R_{DM}$ $=$ (157.6 $\pm$ 1.2 kpc). Thus, we can write
\begin{equation}\label{R_DM}
R_{DM} \doteq  R_{vir} = (157.6 \pm 1.2) ~\mbox{kpc} ~,
\end{equation}
approximately. We can mention that $R_{vir}$ $=$ (189.3 $\pm$ 2.2) $\mbox{kpc}$ according to \citet{2019ApJ...871..120E}; $R_{vir}$ $=$ (183 $\pm$ 3) $\mbox{kpc}$ according to \citet{2023ApJ...945....3S};
$R_{vir}$ $=$ (119.35 $\pm$ 1.37) $\mbox{kpc}$ according to
\citet{Ou} when using Einasto profile or (186.81 $\pm$ 1.07) $\mbox{kpc}$ when using NFW profile; $R_{vir}$ $=$ (121.03 $\pm$ 1.80) $\mbox{kpc}$ according to \citet{Jiao}.

\subsection{Discussion on the halo DM mass distribution}
In this subsection the halo DM mass distribution of the MW will be characterized by the ratio between the radius of the DM halo $R_{DM}$ and a scale-length $l_{sca}$ characterizing the DM of the halo.

At first, we can present the result for the model presented by \citet{2017A&A...598A..66P}. The model yields $R_{DM} / l_{sca}$ $\equiv$ $R_{DM}/a_{halo}$ $\doteq$ 7. However, more realistic model given by Eqs. (\ref{eqDM-AS-imp}) yields $R_{DM}/a_{halo}$ $\doteq$ 20. 

Eqs. (\ref{L-value}) and (\ref{R_DM}) lead to  $R_{DM} / L$ $\doteq$ 16.

The model by \citet{2019ApJ...871..120E}  yields
$R_{DM} / l_{sca}$ $\equiv$ $R_{vir} /R_{s}$ $=$ 12.8 $\pm$ 0.3, where $R_{s}$
is the scale radius, the parameter in the NFW profile,
$R_{s}$ $=$ (14.8 $\pm$ 0.4) $\mbox{kpc}$. The ratio $R_{vir} /R_{s}$ is called the concentration parameter.

The model by \citet{2023ApJ...945....3S} (Table 2) yields
$R_{DM} / l_{sca}$ $\equiv$ $R_{vir} /R_{s}$ $=$ 14.5 $\pm$ 0.5. Again, the ratio is the concentration parameter and $R_{s}$ is the parameter in the NFW profile.

The paper by \citet{Ou} considers two models for the DM halo of the MW and the scale radii $r_{s}$ are calculated.  The results are $R_{DM} / r_{s}$ $\doteq$ 31 for the Einasto model and $R_{DM} / r_{s}$ $\doteq$ 35.5 for the NFW profile, if $R_{DM}$ $\doteq$ $R_{vir}$ is accepted.

The model by \citet{Jiao} yields $R_{DM} / l_{sca}$ $\equiv$ 
$R_{DM} / h$ $\doteq$ 10.6 for the DM Einasto model B2.

\subsubsection{Comparison of the published results}
It seems that there are inconsistencies in the results obtained by various authors.

The papers by \citet{2023ApJ...945....3S} and \citet{Ou} consider the same observational data, the same method and the NFW profile, but the results for
$R_{vir} /R_{s}$ are different, 14.5 $\pm$ 0.5 and 35.5. Also the virial masses of the MW are slightly different, $M_{vir}$ $=$ (6.5 $\pm$ 0.3) $\times$ 10$^{11}$ $\mbox{M}_{\odot}$ and $M_{vir}$ $=$ (6.94 $\pm$ 0.12) $\times$ 10$^{11}$ $\mbox{M}_{\odot}$.  The difference in the values of $R_{vir} /R_{s}$ can be understood by the statement by \citet{Ou} that the NFW profile is not suitable for the observational data.

The papers by \citet{Ou} and \citet{Jiao} consider the same observational data, the same method and the Einasto model, but the results for $R_{DM} / l_{sca}$ are 31 and 10.6. However, the masses of the MW are consistent when using the same model, $M_{vir}$ $=$ (1.81 $\pm$ 0.06) $\times$ 10$^{11}$ $\mbox{M}_{\odot}$ and $M_{dyn}$ $=$ (1.99 $\pm$ 0.09) $\times$ 10$^{11}$ $\mbox{M}_{\odot}$.

\subsection{Mass density in the region of the Sun}
The DM mass density in the solar neighborhood can be found from Eqs. (\ref{nablaG2}) and (\ref{MDens}) for $r$ $=$ $R_0$ $=$ 8.247 $\mbox{kpc}$. The result is
\begin{eqnarray}\label{RhoSolDM}
\varrho _{DM} (R_0) = 8.41 \times 10^{-3} ~\mbox{M}_{\odot} ~\mbox{pc}^{-3} ~.
\end{eqnarray}
The found value of the DM mass density $\varrho _{DM} (R_0)$ is one order {of magnitude} smaller than the value of the mass density of the baryonic matter $\varrho _{BM} (R_0)$, compare Eqs. (\ref{mass-density-sun}) and (\ref{RhoSolDM}).

\subsection{Large distance $r$ and $g_{DM}(r)$}
If we consider a large distance $r$, then Eqs. (\ref{MassDM})-(\ref{Mass2}) yield for the gravitational acceleration generated by the DM halo
\begin{eqnarray}\label{gravaccDM}
g_{DM}(r) = \frac{GM_{DM}(r)}{{r^2}} =\frac{G M_{BM}}{Lr} ~, ~r \leq r_{DM} ~,
\end{eqnarray}
where Eqs. (\ref{Mass2}) hold.

The result represented by Eq. (\ref{gravaccDM}) significantly differs from the acceleration generated by the baryonic matter
\begin{eqnarray}\label{gravacc}
g_{bar}(R) = \frac{G M_{BM}}{R^2} ~,
\end{eqnarray}
see Eqs. (\ref{gbarDM}), $z=0$.

Decrease of $g_{bar}(R)$ is more rapid than the decrease of $g_{DM}(R)$ with the increasing distance $R$. 

Radial component of the equation of motion {Eq. (\ref{eq-motion-DM}) for $R$ $\gg$ $L$ leads to}
\begin{eqnarray}\label{radcomp}
\frac{v_{c}^2}{R} = g_{bar}(R) + g_{DM}(R) = \frac{G M_{BM}}{R^2} + \frac{G M_{BM}}{LR} = \frac{G M_{BM}}{LR} ~,
\nonumber \\ 
R \leq r_{DM} ~,
\end{eqnarray}
{where Eqs. (\ref{gravacc}) and (\ref{gravaccDM}) are used}.
This is consistent with Eq. (\ref{VGg}), see also Eqs. (\ref{Mass2}).

\subsection{Other applications}
Other applications of the equation of motion,
Eqs. (\ref{eq-motion-DM}), 
(\ref{eq-motion-g-bar}) and (\ref{vecDM}) are presented in Appendix \ref{appendix-Dm-appl}, where
oscillations perpendicular to the Galactic equatorial plane are discussed.

\section{Discussion}\label{Disc}
In this section we summarize some results for $z$ $=$ 0 for the purpose of comparison with the published results. We concentrate on the rotation curve of the MW and mass densities for the Galactocentric distances of the Sun and for the large Galactocentric distances.

We will discuss also the gravitational acceleration generated by the DM halo of the MW. Various approaches will be considered.

\subsection{Rotation curve of the MW}
Eqs. (\ref{rot-curve-theory-old}) enable to construct the theoretical rotation curve of the MW. The result is in a very good coincidence with the most accurate observational data, see also Fig. \ref{Fig1}. The result is much better than the result obtained with the conventional approach based on the DM halo potential with several free parameters, compare Figs. \ref{Fig1} and \ref{Fig2}.

The data on the rotation curve of the MW enable to improve the model of the baryonic mass distribution. The improved model is 
represented by Eqs.  (\ref{pot-bulge-new}), (\ref{pot-total-new}),
(\ref{rot-curve-theory}) and the values summarized in Table \ref{Table2}. The rotation curve of the MW galaxy is presented in Fig. \ref{Fig3}. 

\subsection{Mass density for the Galactocentric distance of the Sun}\label{mass-den-sun}
The mass density of the baryonic matter is $\varrho_{BM} (R_{0}, z = 0)$ $=$ 0.0871 M$_{\odot}$ pc$^{-3}$ for $R_{0}$ $=$ 8.247 $\mbox{kpc}$, see Eq. (\ref{mass-density-sun}). The value of the mass density of the DM, $\varrho_{DM} (R_{0}, z = 0)$ $=$
8.41 $\times$ 10$^{-3}$ $\mbox{M}_{\odot} ~\mbox{pc}^{-3}$, 
is presented in Eq. (\ref{RhoSolDM}). 
The total mass density in the region of the Sun is
\begin{eqnarray}\label{mass-den-spher-distr-Sun-total}
\varrho_{total}  (R_{0}, z = 0) &=& \varrho_{BM}  (R_{0}, z = 0) + \varrho_{DM}  (R_{0}, z = 0) 
\nonumber \\
&=& 0.0955 ~\mbox{M}_{\odot} ~\mbox{pc}^{-3} ~.  
\end{eqnarray}
The total density in the region of the Sun is $\varrho_{total}$ $=$ (0.095 $\pm$ 0.001) $\mbox{M}_{\odot} ~\mbox{pc}^{-3}$, see Eqs. (\ref{Mden}) and (\ref{mass-den-spher-distr-Sun-total}).

\subsection{Comparison with published results}
\label{comp-w-p-r-density}
Our value of the total density,
(0.095 $\pm$ 0.001) $\mbox{M}_{\odot} ~\mbox{pc}^{-3}$, is in agreement with \citet{McKee}, who give  $\varrho_{total}$ $=$ (0.097 $\pm$ 0.013) $\mbox{M}_{\odot} ~\mbox{pc}^{-3}$, but our 
$\varrho_{DM} (R=8.247~ \mbox{kpc}, z = 0)$ $=$ 8.41 $\times$ 10$^{-3}$ $\mbox{M}_{\odot} ~\mbox{pc}^{-3}$ is lower than their $\varrho_{DM}$ $=$ (0.013 $\pm$ 0.003) $\mbox{M}_{\odot} ~\mbox{pc}^{-3}$. 
The papers \citet{McMillan2011,2017MNRAS.465...76M} present the value 
$\varrho_{DM}$ $=$ (0.011 $\pm$ 0.001) $\mbox{M}_{\odot} ~\mbox{pc}^{-3}$.
Our DM density is in agreement with \citet{Kafle} , who present $\varrho_{DM}$ $=$ (0.0088 $\pm$ 0.0024)~ $\mbox{M}_{\odot} ~\mbox{pc}^{-3}$. The model presented by \citet{2017A&A...598A..66P} leads to
$\varrho_{total}$ $=$ 0.098 $\mbox{M}_{\odot} ~\mbox{pc}^{-3}$ and $\varrho_{DM}$ $=$ 0.011 $\mbox{M}_{\odot} ~\mbox{pc}^{-3}$.

The total mass density is $\varrho_{total} (R_{0})$ $=$ $\varrho_{BM} (R_{0})$ $+$ $\varrho_{DM} (R_{0})$ $\approx$ 0.08 $\mbox{M}_{\odot}$ $\mbox{pc}^{-3}$, calculated from \citet{Jiao}. The local dynamical density (0.076 $\pm$ 0.015) $\mbox{M}_{\odot}$ $\mbox{pc}^{-3}$ has been already presented in the past using data obtained with the HIPPARCOS satellite by \citet{Creze1998}. However, low value of the local dynamical mass density decreases stability of the Oort cloud of comets. The papers dealing with the Oort cloud of comets consider the local density of the Galactic disk in the solar neighbourhood to be 0.10 $\mbox{M}_{\odot}$ $\mbox{pc}^{-3}$ in this century, see, e.g., 
(0.105 $\pm$ 0.005) $\mbox{M}_{\odot}$ $\mbox{pc}^{-3}$ in
\citet{Korchagin2003}, or, 0.10 $\mbox{M}_{\odot}$ $\mbox{pc}^{-3}$ in
\citet{Brasser_Morbidelli}.

\citet{Ou} used the same observational data, the same method and the same Einasto profile for the DM halo as \citet{Jiao}. However, the BM distributions are different in the two papers. The results presented by \citet{Ou} lead to the total mass density is $\varrho_{total} (R_{0})$ $=$ $\varrho_{BM} (R_{0})$ $+$ $\varrho_{DM} (R_{0})$ $=$ 0.124 $\mbox{M}_{\odot}$ $\mbox{pc}^{-3}$. The value is in 50\% higher than the value obtained from the results by \citet{Jiao}.

\subsection{Mass density at large distances}  
We are interested in mass density for large values of $R$ and $z$ $=$ 0. Since $\left ( g_{bar} \right )_{j}$ $=$ $G~M_{j}/R^{2}$, $j$ $=$ 1 to 3, and $g_{bar}$ $=$ $G~M_{BM}/R^{2}$ for large $R$, $M_{BM}$ $=$ $\sum_{j=1}^{3} M_{j}$,
we finally obtain {for the BM and DM mass densities} 
\begin{eqnarray}\label{eq-mass-density-discussion}
\varrho_{BM} \left ( R, z = 0 \right )  &=& \frac{1}{4 \pi} \frac{\sum_{j=1}^{3} M_{j}~ a_{j}/b_{j}}{R^{3}} ~,
\nonumber \\
\varrho_{DM} \left ( R, z = 0 \right ) &=& \frac{1}{4 \pi} \frac{M_{BM}}{L R^{2}} ~, 
\nonumber \\
L &=& \sqrt{G~M_{BM} / g_{+}} ~.
\end{eqnarray}
The result corresponds to the spherical distribution of the DM. 

\subsection{Comparison with published approaches}
The approaches used in Eqs. (\ref{eq-mass-density-discussion}) lead to $\varrho_{DM}$ decreasing with the square of distance. This result yields also, e.g., a spherical pseudo-isothermal halo $\varrho_{DM}$ $=$ $\varrho_{DM0}/ \left ( 1 + r / r_{0} \right )^{2}$, but the fit to the observed rotation curve is poor. Comparison with the second equation of Eqs. (\ref{eq-mass-density-discussion}) for large $r$ yields
$\varrho_{DM0}$ $=$ $M_{BM} / \left ( 4 \pi L r_{0}^{2} \right )$ and the usage of Eq. (\ref{RhoSolDM}) leads to
$\varrho _{DM0}/ \left ( 1 + R_{0} / r_{0} \right )^{2}$ $=$ $8.41 \times 10^{-3} ~\mbox{M}_{\odot} ~\mbox{pc}^{-3}$. Numerically, $\varrho_{DM0}$ $=$
1.273 $\mbox{M}_{\odot} ~\mbox{pc}^{-3}$ and
$r_{0}$ $=$ 0.730 $\mbox{kpc}$. Another example of the spherical distribution of the DM in the MW is given by Eq. (\ref{eqDM-Poul}) leading to $\varrho_{DM} (r)$ $=$ $(4 \pi)^{-1}$ $M_{halo}/(a_{halo} r^{2})$ for large distance $r$ instead of $\varrho_{DM} (r)$ $=$ $(4 \pi)^{-1}$ $M_{BM}/(L r^{2})$, compare with the second of 
Eqs. (\ref{eq-mass-density-discussion}) and see also Secs. \ref{sec-DM-P}, \ref{Conv-DM-halo}, \ref{Imp-DM-halo} and \ref{com-DM} for discussion.

\subsection{Acceleration generated by the DM - spherical mass distribution}
Gravitational acceleration generated by the DM halo of the MW is represented by Eqs. (\ref{vecDM}). The Taylor expansion leads to
\begin{eqnarray}\label{vecDM-disc}
\vec{g}_{DM} &=& -~g_{DM}(r) ~\hat{\vec{r}} ~,
\nonumber \\
g_{DM}(r) &=& \frac{G M_{BM}}{L r} \left [ 1 - \frac{1}{2} \frac{L}{r} 
+ \frac{1}{6} \left ( \frac{L}{r} \right )^{2}
-\frac{3}{4} \sum_{j=1}^{3} \frac{M_{j}}{M} ~ \left ( \frac{c_{j}}{r} \right )^{2} \right ] ~,
\nonumber \\
M_{BM} &=& \sum_{j=1}^{3} M_{j} ~,
\nonumber \\
c_k &=& a_k + b_k  ~,~~ k = 1,~2, ~3 ~,
\end{eqnarray}
if higher powers of $1/r$ are neglected. The values of the parameters in Eqs. (\ref{vecDM-disc}) are summarized in Table \ref{Table2} and Eqs. (\ref{Mass2}) and (\ref{L-value}) hold. We can remind that $M$ is the total mass of the BM of the MW.

The DM model of the MW halo presented by \citet{2017A&A...598A..66P},
see also Eqs. (\ref{eqDM-Poul}) in Sec. \ref{sec-DM-P} leads to
\begin{eqnarray}\label{eqDM-Poul-dis}
\vec{g}_{DM} &=& -~g_{DM}(r) ~\hat{\vec{r}} ~,
\nonumber \\
g_{DM}(r) &=& \frac{G~M_{halo}}{a_{halo}~r} \left [ 1 + 
1.02 \left ( \frac{a_{halo}}{r} \right )^{1.02} \right ] ~,
\nonumber \\
M_{halo} &=& 1.392 ~\times 10^{11} ~\mbox{M}_{\odot} ~,
\nonumber \\
a_{halo} &=& 14 ~\mbox{kpc} ~,
\end{eqnarray}
if higher powers of $1/r$ are neglected.

The improvement of the DM model by \citet{2017A&A...598A..66P} considers also the approach by \citet{Allen1991}, see Sec. \ref{fixed-model}.
The DM model of the MW halo leads to
\begin{eqnarray}\label{grav-acc-eq-DM-disc}
\vec{g}_{DM} &=& -~g_{DM}(r) ~\hat{\vec{r}} ~,
\nonumber \\
g_{DM} \left ( r \right )  &=& 
\frac{G M_{BM}}{L r} ~\left [ 1 - \left ( \frac{7.86 ~\mbox{kpc}}{r} \right )^{1.02} \right ] 
\end{eqnarray}
for large values of the galactocentric distance $r$,
see also Eqs. (\ref{grav-acc-eq-DM}), (\ref{third-req-fin}) and
(\ref{three-req-a-halo}). 

The DM model of the MW halo presented by \citet{Barros2016} leads to
\begin{eqnarray}\label{vecDM-disc-Barros}
\vec{g}_{DM} &=& -~g_{DM}(r) ~\hat{\vec{r}} ~,
\nonumber \\
g_{DM}(r) &=& \frac{G M_{BM}}{L r} \left [ 1 - \left ( \frac{r_{h}}{r} \right )^{2} \right ] ~,
\nonumber \\
r_h &=& 1.009 ~R_{0} = 8.32 ~\mbox{kpc} ~,
\end{eqnarray}
if higher powers of $1/r$ are neglected, see also Sec. \ref{Comp-DM-models}.

We can summarize that our approach represented by Eqs. (\ref{vecDM}), see also Eqs. (\ref{vecDM-disc}), 
contains both the first, the second and higher orders in $1/r$, 
\begin{eqnarray}\label{vecDM-disc-summary}
g_{DM}(r) &=& \frac{G M_{BM}}{L r} \left ( 1 - \frac{4.95~\mbox{kpc}}{r} 
- \frac{0.56~ \mbox{kpc}^{2}}{r^{2}} \right ) ~,
\end{eqnarray}
while the published approaches lead to the results when besides the first order in $1/r$ exist only orders higher than $(1/r)^{2}$, see Eqs. 
(\ref{eqDM-Poul-dis}), (\ref{grav-acc-eq-DM-disc}) and
(\ref{vecDM-disc-Barros}). The dominant term $g_{DM}(r)$ $\propto$ $1/r$
may not even exist. Considerations in recent papers \citep{2023ApJ...945....3S}, \citep{Ou} and \citep{Jiao} are equivalent to $g_{DM}(r)$ $\propto$ $(\ln{r}) /r^{2}$
if higher orders in $1/r$ are neglected, or, in the case of an Einasto profile $g_{DM}(r) \propto 1/r^{2} - \Gamma(3/\alpha;(r/r_{s})^{\alpha})/r^{2}$.

\section{Conclusion}\label{Conc}
The equation of motion of a body (massless point) moving in the MW is given by Eqs. (\ref{eq-motion-DM}), (\ref{eq-motion-g-bar}), (\ref{gbarDM}) and (\ref{vecDM}), where also Eqs. (\ref{pot-bulge-new}), (\ref{pot-total-new}) and the values summarized in Table \ref{Table2} hold. The equation of motion considers spherical distribution of the DM in the MW. The gravitational model is entirely analytical and provides continuous derivatives at all points. 

The paper presents important results on the values of the Oort constants. The results are obtained on the basis of: (i) Eqs. (\ref{obs-r-0})-(\ref{obs-r-1}), and, (ii) gravitational potentials for the baryonic matter components of MW, see Eqs. (\ref{pot-bulge-new}), (\ref{pot-total-new}), (\ref{g-bar-new}) and the values summarized in Table \ref{Table2}. Papers by \citet{2019ApJ...871..120E}, \citet{abuter2020detection}, \citet{2017A&A...598A..66P} and \citet{2016PhRvL.117t1101M} are also utilized. 

The great advantage of our approach is that we use, at the beginning, recently published baryonic model of the MW 
\citep{2017A&A...598A..66P}. We have made only a slight improvement of the model, see 
 Eqs. (\ref{pot-bulge-new}), (\ref{pot-total-new}), (\ref{g-bar-new}), (\ref{rot-curve-theory}) and the values summarized in Table \ref{Table2}. {The empirical results of the observations of the SPARC extragalactic galaxies by \citet{2016PhRvL.117t1101M}, represented by Eqs. (\ref{obs-r-0}), (\ref{obs-r-1}) and (\ref{obs-Gaugh}),
are added to the baryonic model and generalized to the vector form represented by Eq. (\ref{eq-motion-DM}).}

The results are compared with the results on the rotation curve data, see \citet{2019ApJ...871..120E}. Fitting a newer rotation curve \citet{2019ApJ...871..120E} using an older baryonic model \citet{2017A&A...598A..66P} and an interpolating function \citet{2016PhRvL.117t1101M} is a fundamentally different approach from fitting a measured rotation curve with new BM and DM models.
 
The rotation curve obtained by \citet{2019ApJ...871..120E} is confirmed by newer observational data and newer results presented by \citet{2023ApJ...945....3S}, \citet{Ou} and by \citet{Jiao}. However, \citet{2023ApJ...945....3S}, 
\citet{Ou} and \citet{Jiao} present relevant decrease of the rotation curve for Galactocentric distances above 15 kpc. The decrease is not consistent with the flat rotation curves, see also Eqs. (\ref{large-R-speed-final}) or (\ref{VGg})
(`,the baryonic Tully-Fisher relation'', according to \citet{2000ApJ...533L..99M} or \citet{2016PhRvL.117t1101M}), holding for spiral galaxies in the Universe.
Our theoretical approaches are consistent with Eqs. (\ref{large-R-speed-final}) or (\ref{VGg}), see Secs. \ref{Mass-speed-relation} and \ref{sec71}.
Moreover, the papers by \citet{2023ApJ...945....3S}, \citet{Ou} and by \citet{Jiao} suggests various models of the Galactic mass distribution
and some of the models lead to inconsistent values of the mass density in the region of the Sun, the set of models do not enable to improve the accuracy of the Oort constants. It may be that either the data or the data analyses by \citet{2023ApJ...945....3S}, \citet{Ou} and \citet{Jiao} contain some systematic errors or invalid assumptions, compare e.g., \citet{Chae}.

Our theoretical rotation curve, including the value of the Galactocentric rotation speed of the Sun, is consistent with observational results. Values of the Oort constants $A$ and $B$ are found, see Eqs. (\ref{AB-num2}). Baryonic mass density in the region of the Sun is given by Eq. (\ref{mass-density-sun}).
 Values of several important quantities characterizing Solar System in the MW are summarized in the Table \ref{tab:my_label1}: the Galactocentric distance $R_0$, rotation speed $v_0$, the period of revolution around the center of the Milky Way $T_{0}$, baryonic mass density $\varrho_{BM}(R_0)$, DM mass density $\varrho_{DM}(R_0)$, total mass density $\varrho(R_0)$ and the Oort constants $A$ and $B$. Our results are determined more precisely than previously published results, see, e.g., Secs. \ref{Discussion on-the-Oort-constants}, \ref{Sec-Oort-ap}, \ref{Sec-sum-Oort} and \ref{comp-w-p-r-density}. 

The model of the three component Galactic discs is also developed, see Appendix \ref{3-components-disc}. The model yields values of the relevant quantities consistent with the model of the two component Galactic discs presented in the main part of the paper.
 
Values of several important quantities characterizing the MW are summarized in the Table \ref{tab:my_label2}: $M$ - mass of the baryonic matter of the MW, $M_{DM}$ - mass of the DM of the MW, $R_{DM}$ - radius of the DM of the MW, $M_{vir,1}$ and $r_{vir,1}$ - virial mass and virial radius of the MW for $H_{0}$ $=$ (67.4 $\pm$ 0.5) $\mbox{km} ~\mbox{s}^{-1} ~\mbox{Mpc}^{-1}$, $M_{vir,2}$ and $r_{vir,2}$  - virial mass and virial radius of the MW for $H_{0}$ $=$ (73.04 $\pm$ 1.04) $\mbox{km} ~\mbox{s}^{-1} ~\mbox{Mpc}^{-1}$, see Sec. \ref{Virial-radius-mass}. 

It is shown that the corrected model by \citet{2017A&A...598A..66P} and \citet{Allen1991} can be understood as a rough approximation to our more realistic model (see Secs. \ref{sec-DM-AS-P} and \ref{C-simple-ideas}). Moreover, improvement of the DM model by \citet{2017A&A...598A..66P} and \citet{Allen1991}, see Eqs. (\ref{eqDM-AS-app}), (\ref{three-req-a-halo}), (\ref{three-req-M-halo}) and (\ref{third-req-res12-sum-R-DM}), leads to consequences for motion of a body in the MW which are consistent with the new MW model presented in this paper (the values of the quantities presented in Table \ref{tab:my_label1} hold).

The viability of the model created in this paper is confirmed by additional, currently the newest, observational data. E.g., Sec. \ref{opt_val} presents results found for the baryonic matter mass distribution and the found value of the circular rotation speed $v_{0}$ $=$ (228.8 $\pm$ 0.2) km/s is consistent with the value (228.9 $\pm$ 0.7) km/s based on the observational data by \citet{2023ApJ...945....3S}. Similarly, the virial mass of the MW, based on the model and the newest values of the Hubble constant, $M_{vir}$ $=$ (134.3 $\pm$ 1.0) $\times$ 10$^{10}$ $\mbox{M}_{\odot}$, 
is consistent with recently published value $M_{vir}$ $=$ (1.18 $\pm$ 0.51) $\times$ 10$^{12}$ $\mbox{M}_{\odot}$ \citep{Watkins}, see Sec. \ref{Comp-MW}. 
We do not consider older observational data, because of lower precision and accuracy.

 \begin{table}\label{tab-char}
    \caption{Galactic characteristics for the Solar System. \\ $R_0$ - Galactocentric distance of the Sun, $v_{0}$ - circular velocity at $R_0$, $T_0$ - orbital period at $R_0$, $T_{osc}$ - period of oscillations perpendicular to the Galactic equator at $R_0$, $\varrho_{BM}$ - mass density of baryonic matter at $R_0$, $\varrho_{DM}$ - mass density of dark matter at $R_0$, $\varrho$ - total mass density at $R_0$, $A$, $B$ - Oort's constants at $R_0$}
        \label{tab:my_label1}
    \centering
    \setlength{\tabcolsep}{1pt}
    \begin{tabular}{lrcll} \hline
        Quantity & \multicolumn{3}{c}{Value} & Unit \\\hline
        $R_0$ & $8247$ & $\pm$ & $9$ & pc \\
        $v_{0}$ & $228.8$ & $\pm$ & $0.2$ & km s$^{-1}$ \\
        $T_{0}$ & $221.5$ & $\pm$ & $0.3$ &  Myrs \\
        $T_{osc}$ & $84.7$ & $\pm$ & $0.5$ & Myrs \\
        $\varrho_{BM}$ & $8.7$ & $\pm$ & $0.1$ & 10$^{-2}$ M$_{\odot}$ pc$^{-3}$ \\
        $\varrho_{DM}$ & $0.84$ & $\pm$ & $0.01$ & 10$^{-2}$ M$_{\odot}$ pc$^{-3}$ \\
        $\varrho$ & $9.5$ & $\pm$ & $0.1$ & 10$^{-2}$ M$_{\odot}$ pc$^{-3}$ \\
        $A$ & $14.73$ & $\pm$ & $0.03$ & km s$^{-1}$ kpc$^{-1}$ \\
        $B$ & $-13.01$ & $\pm$ & $0.03$ & km s$^{-1}$ kpc$^{-1}$ \\
         \hline
    \end{tabular}
\end{table}

 \begin{table}\label{tab-MW}
    \caption{Characteristics of the Milky Way. \\ 
    $M_{BM}$ - mass of the baryonic matter of the MW, 
    $M_{DM}$ - mass of the dark matter of the MW,
    $R_{DM}$ - radius of the dark matter of the MW,
    $M_{vir,1}$ and $r_{vir,1}$ - virial mass and virial radius
    of the MW for $H_{0}$ $=$ (67.4 $\pm$ 0.5) $\mbox{km} ~\mbox{s}^{-1} ~\mbox{Mpc}^{-1}$,
    $M_{vir,2}$ and $r_{vir,2}$ - virial mass and virial radius
    of the MW for $H_{0}$ $=$ (73.04 $\pm$ 1.04) $\mbox{km} ~\mbox{s}^{-1} ~\mbox{Mpc}^{-1}$. }
        \label{tab:my_label2}
    \centering
    \setlength{\tabcolsep}{1pt}
    \begin{tabular}{lrcll} \hline
    Quantity & \multicolumn{3}{c}{Value} & Unit \\\hline
    $M_{BM}$  & 8.43 & $\pm$ & 0.01 & 10$^{10}$ $\mbox{M}_{\odot}$ \\
    $M_{DM}$ & 134.3 & $\pm$ & 1.0 & 10$^{10}$ $\mbox{M}_{\odot}$ \\
    $R_{DM}$ & 157.6 & $\pm$ & 1.2 & kpc \\
    $M_{vir,1}$ & 139.7 & $\pm$ & 1.0 & 10$^{10}$ $\mbox{M}_{\odot}$ \\
    $r_{vir,1}$ & 164.0 & $\pm$ & 1.2 & kpc \\
    $M_{vir,2}$ & 128.9 & $\pm$ & 1.8 & 10$^{10}$ $\mbox{M}_{\odot}$ \\
    $r_{vir,2}$ & 151.3 & $\pm$ & 2.1 & kpc \\
         \hline
    \end{tabular}
\end{table}

\section*{Acknowledgement} This work was supported by the Scientific Grant Agency VEGA, Slovak Republic, No. 1/0761/21 and by Comenius University Grants, No. UK/346/2023 and No. UK/3216/2024.

\bibliography{biblio}{}

\begin{thebibliography}{}
\expandafter\ifx\csname natexlab\endcsname\relax\def\natexlab#1{#1}\fi
\providecommand{\url}[1]{\href{#1}{#1}}
\providecommand{\dodoi}[1]{doi:~\href{http://doi.org/#1}{\nolinkurl{#1}}}
\providecommand{\doeprint}[1]{\href{http://ascl.net/#1}{\nolinkurl{http://ascl.net/#1}}}
\providecommand{\doarXiv}[1]{\href{https://arxiv.org/abs/#1}{\nolinkurl{https://arxiv.org/abs/#1}}}

\bibitem[{Abuter {et~al.}(2021)Abuter, Amorim, Bauböck, {et~al.}}]{Gaia2021}
Abuter, R., Amorim, A., Bauböck, J.~P., {et~al.} 2021, Astronomy and
  Astrophysics, 647, 17

\bibitem[{Abuter {et~al.}(2020)Abuter, Amorim, Baub{\"o}ck, Berger, Bonnet,
  Brandner, Cardoso, Cl{\'e}net, de~Zeeuw, Dexter,
  {et~al.}}]{abuter2020detection}
Abuter, R., Amorim, A., Baub{\"o}ck, M., {et~al.} 2020, Astronomy and
  Astrophysics, 636, L5

\bibitem[{Aghanim {et~al.}(2020)Aghanim, Akrami, \& Ashdown}]{Planck2018}
Aghanim, N., Akrami, Y., \& Ashdown, M. 2020, Astronomy and Astrophysics, 641,
  67

\bibitem[{{Allen} \& {Santillan}(1991)}]{Allen1991}
{Allen}, C., \& {Santillan}, A. 1991, \rmxaa, 22, 255

\bibitem[{{Barros} {et~al.}(2016){Barros}, {L{\'e}pine}, \&
  {Dias}}]{Barros2016}
{Barros}, D.~A., {L{\'e}pine}, J.~R.~D., \& {Dias}, W.~S. 2016, \aap, 593,
  A108, \dodoi{10.1051/0004-6361/201527535}

\bibitem[{{Bensby} {et~al.}(2011){Bensby}, {Alves-Brito}, {Oey}, {Yong}, \&
  {Mel{\'e}ndez}}]{Bensby2011}
{Bensby}, T., {Alves-Brito}, A., {Oey}, M.~S., {Yong}, D., \& {Mel{\'e}ndez},
  J. 2011, \apjl, 735, L46, \dodoi{10.1088/2041-8205/735/2/L46}

\bibitem[{{Bovy}(2017)}]{bovy17}
{Bovy}, J. 2017, Monthly Notices of the Royal Astronomical Society, 468, L63,
  \dodoi{10.1093/mnrasl/slx027}

\bibitem[{{Bovy} {et~al.}(2012){Bovy}, {Rix}, {Liu}, {Hogg}, {Beers}, \&
  {Lee}}]{Bovy2012}
{Bovy}, J., {Rix}, H.-W., {Liu}, C., {et~al.} 2012, \apj, 753, 148,
  \dodoi{10.1088/0004-637X/753/2/148}

\bibitem[{{Bovy} {et~al.}(2016){Bovy}, {Rix}, {Schlafly}, {Nidever},
  {Holtzman}, {Shetrone}, \& {Beers}}]{Bovy2016}
{Bovy}, J., {Rix}, H.-W., {Schlafly}, E.~F., {et~al.} 2016, \apj, 823, 30,
  \dodoi{10.3847/0004-637X/823/1/30}

\bibitem[{{Brasser} \& {Morbidelli}(2013)}]{Brasser_Morbidelli}
{Brasser}, R., \& {Morbidelli}, A. 2013, \icarus, 225, 40,
  \dodoi{10.1016/j.icarus.2013.03.012}

\bibitem[{{Chae}(2023)}]{Chae}
{Chae}, K.-H. 2023, \apj, 952, 128, \dodoi{10.3847/1538-4357/ace101}

\bibitem[{{Creze} {et~al.}(1998){Creze}, {Chereul}, {Bienayme}, \&
  {Pichon}}]{Creze1998}
{Creze}, M., {Chereul}, E., {Bienayme}, O., \& {Pichon}, C. 1998, \aap, 329,
  920, \dodoi{10.48550/arXiv.astro-ph/9709022}

\bibitem[{Dehnen \& Binney(1998)}]{dehnen1998local}
Dehnen, W., \& Binney, J.~J. 1998, Monthly Notices of the Royal Astronomical
  Society, 298, 387

\bibitem[{{Eilers} {et~al.}(2019){Eilers}, {Hogg}, {Rix}, \&
  {Ness}}]{2019ApJ...871..120E}
{Eilers}, A.-C., {Hogg}, D.~W., {Rix}, H.-W., \& {Ness}, M.~K. 2019, The
  Astrophysical Journal, 871, 120, \dodoi{10.3847/1538-4357/aaf648}

\bibitem[{{Famaey} \& {McGaugh}(2012)}]{Famaey2012}
{Famaey}, B., \& {McGaugh}, S.~S. 2012, Living Reviews in Relativity, 15, 10,
  \dodoi{10.12942/lrr-2012-10}

\bibitem[{Feast \& Whitelock(1997)}]{feast1997galactic}
Feast, M., \& Whitelock, P. 1997, Monthly Notices of the Royal Astronomical
  Society, 291, 683

\bibitem[{{Halle} {et~al.}(2018){Halle}, {Di Matteo}, {Haywood}, \&
  {Combes}}]{Halle2018}
{Halle}, A., {Di Matteo}, P., {Haywood}, M., \& {Combes}, F. 2018, \aap, 616,
  A86, \dodoi{10.1051/0004-6361/201832603}

\bibitem[{{Irrgang} {et~al.}(2013){Irrgang}, {Wilcox}, {Tucker}, \&
  {Schiefelbein}}]{Irrgang2013}
{Irrgang}, A., {Wilcox}, B., {Tucker}, E., \& {Schiefelbein}, L. 2013, \aap,
  549, A137, \dodoi{10.1051/0004-6361/201220540}

\bibitem[{{Jiao} {et~al.}(2021){Jiao}, {Hammer}, {Wang}, \&
  {Yang}}]{2021A&A...654A..25J}
{Jiao}, Y., {Hammer}, F., {Wang}, J.~L., \& {Yang}, Y.~B. 2021, \aap, 654, A25,
  \dodoi{10.1051/0004-6361/202141058}

\bibitem[{{Jiao} {et~al.}(2023){Jiao}, {Hammer}, {Wang, Haifeng}, {Wang,
  Jianling}, {Amram, Philippe}, {Chemin, Laurent}, \& {Yang, Yanbin}}]{Jiao}
{Jiao}, Y., {Hammer}, F., {Wang, Haifeng}, {et~al.} 2023, Astronomy and
  Astrophysics, 678, A208, \dodoi{10.1051/0004-6361/202347513}

\bibitem[{Kafle {et~al.}(2014)Kafle, Sharma, Lewis, \& Bland-Hawthorn}]{Kafle}
Kafle, P.~R., Sharma, S., Lewis, G.~F., \& Bland-Hawthorn, J. 2014, The
  Astrophysical Journal, 794, 17

\bibitem[{{Karim} \& {Mamajek}(2017)}]{2017MNRAS.465..472K}
{Karim}, T., \& {Mamajek}, E.~E. 2017, Monthly Notices of the Royal
  Astronomical Society, 465, 472, \dodoi{10.1093/mnras/stw2772}

\bibitem[{{Karimova} \& {Pavlovskaya}(1974)}]{karimova1973}
{Karimova}, D.~K., \& {Pavlovskaya}, E.~D. 1974, Soviet Astronomy, 17, 470

\bibitem[{{Kerr} \& {Lynden-Bell}(1986)}]{kerr}
{Kerr}, F.~J., \& {Lynden-Bell}, D. 1986, Monthly Notices of the Royal
  Astronomical Society, 221, 1023, \dodoi{10.1093/mnras/221.4.1023}

\bibitem[{{Korchagin} {et~al.}(2003){Korchagin}, {Girard}, {Borkova},
  {Dinescu}, \& {van Altena}}]{Korchagin2003}
{Korchagin}, V.~I., {Girard}, T.~M., {Borkova}, T.~V., {Dinescu}, D.~I., \&
  {van Altena}, W.~F. 2003, \aj, 126, 2896, \dodoi{10.1086/379138}

\bibitem[{{Kulikovskij}(1985)}]{kulik}
{Kulikovskij}, P.~G. 1985, {Stellar \uppercase{A}stronomy} (Nauka, Moscow, 2nd
  edition, 272pp)

\bibitem[{{Levison} {et~al.}(2001){Levison}, {Dones}, \&
  {Duncan}}]{Levison2001}
{Levison}, H.~F., {Dones}, L., \& {Duncan}, M.~J. 2001, \aj, 121, 2253,
  \dodoi{10.1086/319943}

\bibitem[{McGaugh {et~al.}(2016)McGaugh, Lelli, \&
  Schombert}]{2016PhRvL.117t1101M}
McGaugh, S.~S., Lelli, F., \& Schombert, J.~M. 2016, Physical Review Letters,
  117, 201101

\bibitem[{McGaugh {et~al.}(2000)McGaugh, Schombert, Bothun, \&
  De~Blok}]{2000ApJ...533L..99M}
McGaugh, S.~S., Schombert, J.~M., Bothun, G.~D., \& De~Blok, W. 2000, The
  Astrophysical Journal Letters, 533, L99

\bibitem[{McKee {et~al.}(2015)McKee, Parravano, \& Hollenbach}]{McKee}
McKee, C.~F., Parravano, A., \& Hollenbach, D.~J. 2015, The Astrophysical
  Journal, 814, 13

\bibitem[{{McMillan}(2011)}]{McMillan2011}
{McMillan}, P.~J. 2011, \mnras, 414, 2446,
  \dodoi{10.1111/j.1365-2966.2011.18564.x}

\bibitem[{{McMillan}(2017)}]{2017MNRAS.465...76M}
---. 2017, Monthly Notices of the Royal Astronomical Society, 465, 76,
  \dodoi{10.1093/mnras/stw2759}

\bibitem[{{Miyamoto} \& {Nagai}(1975)}]{Miyamoto}
{Miyamoto}, M., \& {Nagai}, R. 1975, \pasj, 27, 533

\bibitem[{Moore \& Rees(2014)}]{2014pmdb.book.....M}
Moore, P., \& Rees, R. 2014, Patrick Moore's data book of astronomy (Cambridge
  University Press, Cambridge)

\bibitem[{{Necib} \& {Lin}(2022)}]{2022ApJ...926..189N}
{Necib}, L., \& {Lin}, T. 2022, \apj, 926, 189,
  \dodoi{10.3847/1538-4357/ac4244}

\bibitem[{Ou {et~al.}(2023)Ou, Eilers, Necib, \& Frebel}]{Ou}
Ou, X., Eilers, A.-C., Necib, L., \& Frebel, A. 2023, arXiv preprint
  arXiv:2303.12838

\bibitem[{Pouliasis {et~al.}(2017)Pouliasis, Di~Matteo, \&
  Haywood}]{2017A&A...598A..66P}
Pouliasis, E., Di~Matteo, P., \& Haywood, M. 2017, Astronomy and Astrophysics,
  598, A66

\bibitem[{Riess {et~al.}(2022)Riess, Yuan, Macri, {et~al.}}]{RiessHubble}
Riess, A.~G., Yuan, W., Macri, M.~M., {et~al.} 2022, The Astrophysical Journal
  Letters, 934, L7

\bibitem[{{Smith} {et~al.}(2015){Smith}, {Flynn}, {Candlish}, {Fellhauer}, \&
  {Gibson}}]{Smith2015}
{Smith}, R., {Flynn}, C., {Candlish}, G.~N., {Fellhauer}, M., \& {Gibson},
  B.~K. 2015, \mnras, 448, 2934, \dodoi{10.1093/mnras/stv228}

\bibitem[{{Snaith} {et~al.}(2015){Snaith}, {Haywood}, {Di Matteo}, {Lehnert},
  {Combes}, {Katz}, \& {G{\'o}mez}}]{Snaith2015}
{Snaith}, O., {Haywood}, M., {Di Matteo}, P., {et~al.} 2015, \aap, 578, A87,
  \dodoi{10.1051/0004-6361/201424281}

\bibitem[{{Snaith} {et~al.}(2014){Snaith}, {Haywood}, {Di Matteo}, {Lehnert},
  {Combes}, {Katz}, \& {G{\'o}mez}}]{Snaith2014}
{Snaith}, O.~N., {Haywood}, M., {Di Matteo}, P., {et~al.} 2014, \apjl, 781,
  L31, \dodoi{10.1088/2041-8205/781/2/L31}

\bibitem[{{Sylos Labini} {et~al.}(2023){Sylos Labini}, {Chrob{\'a}kov{\'a}},
  {Capuzzo-Dolcetta}, \& {L{\'o}pez-Corredoira}}]{2023ApJ...945....3S}
{Sylos Labini}, F., {Chrob{\'a}kov{\'a}}, {\v{Z}}., {Capuzzo-Dolcetta}, R., \&
  {L{\'o}pez-Corredoira}, M. 2023, \apj, 945, 3,
  \dodoi{10.3847/1538-4357/acb92c}

\bibitem[{Tully \& Fisher(1977)}]{1977A&A....54..661T}
Tully, R.~B., \& Fisher, J.~R. 1977, Astronomy and Astrophysics, 54, 661

\bibitem[{Watkins {et~al.}(2019)Watkins, van~der Marel, Sohn, \&
  Evans}]{Watkins}
Watkins, L.~L., van~der Marel, L.~P., Sohn, S.~T., \& Evans, N.~W. 2019, The
  Astrophysical Journal, 873, 13

\end{thebibliography}
\bibliographystyle{aasjournal}

\begin{appendix}

\section{Oort constants - new results and comparison with the published results}\label{ap-oort}

This appendix deals with some theoretical results for the Sun. Results obtained in the main text will be applied to Oort constants.

\subsection{Oort constants}
The first and the second Oort constants are defined by the relations
\begin{eqnarray}\label{oort}
A &=& +\frac{1}{2} \left[ \frac{v_{0}}{R_0} - \left( \frac{\mbox{d}v_{c}}{\mbox{d}R}\right)_{R=R_0} \right]  ~,
\nonumber \\
B &=& -\frac{1}{2} \left[ \frac{v_{0}}{R_0} + \left( \frac{\mbox{d}v_{c}}{\mbox{d}R}\right)_{R=R_0} \right] ~.
\end{eqnarray}
Numerically, on the basis of Eqs. (\ref{angular-speed}) and (\ref{diff-vel})
\begin{eqnarray}\label{AB-num}
A &=& +14.73 ~\mbox{km} ~\mbox{s}^{-1} ~\mbox{kpc}^{-1} ~,
\nonumber \\
B &=& -13.01 ~\mbox{km} ~\mbox{s}^{-1} ~\mbox{kpc}^{-1} ~.
\end{eqnarray}

\subsection{Discussion on the Oort constants}\label{Discussion on-the-Oort-constants}
Eqs. (\ref{local-diff}) and (\ref{oort}) yield {for the Oort constants $A$ and $B$}
\begin{eqnarray}\label{AB-const}
A &=& \frac{\beta +1}{2} \frac{v_{0}}{R_0} ~,
\nonumber \\
B &=& \frac{\beta - 1}{2} \frac{v_{0}}{R_0} ~.
\end{eqnarray}
The value of $\beta$ is much less sensitive to inaccuracies in $v_{0}$ and $\left( \mbox{d}v_{c} / \mbox{d}R \right)_{R=R_{0}}$ than the values of $A$ and $B$.
The value $v_{0} = 228.8 ~\mbox{km}~ \mbox{s}^{-1}$ (Eq. \ref{speed-sun}), together with Eqs. (\ref{R-sun}), (\ref{beta}) and (\ref{AB-const}), leads to 
\begin{eqnarray}\label{AB-num2}
A &=& (+14.73 \pm 0.03) ~\mbox{km} ~\mbox{s}^{-1} ~\mbox{kpc}^{-1} ~,
\nonumber \\
B &=& (-13.01 \pm 0.03) ~\mbox{km} ~\mbox{s}^{-1} ~\mbox{kpc}^{-1} ~,
\end{eqnarray}
where the errors are calculated from Eqs. (\ref{R-sun}), (\ref{speed-sun}) and (\ref{AB-const}), also (\ref{local-diff-num}) below can be used.

The values of $A$, $B$ and $\beta$ are obtained from the rotation curve of the Milky Way. The results summarized in Eqs. (\ref{AB-num2}) practically do not depend on the baryon mass model of the MW. 
The values of $\beta$, $v_{0}$ and $R_0$ are known to a high precision and Eqs. (\ref{AB-const}) lead to very small relative errors of $A$ and $B$, e.g. $\sigma_{A}/A$ $=$ 0.0244 $\sigma_{\beta}/\beta$ and $\sigma_{B}/B$ $=$ $-$0.0256 $\sigma_{\beta}/\beta$, if the errors in $R_{0}$ and $v_{0}$ are neglected. Therefore, any improvement of the baryon mass model of the MW will not change the values summarized in Eqs. (\ref{AB-num2}). 
The values improve the results found in previous papers, e.g.,
\citet{bovy17}, \citet{feast1997galactic}, \citet{kerr}, or, the IAU recommended values $ A = +15 ~\mbox{km} ~\mbox{s}^{-1} ~\mbox{kpc}^{-1} ~, ~B = -10 ~\mbox{km} ~\mbox{s}^{-1} ~\mbox{kpc}^{-1}$. We can mention that
the conventional approach uses spatial variations of the velocity field of local stars. We have determined the values of $A$ and $B$ from the rotation curve of the MW.

Approximate error of $\beta$ is 0.001, in reality less than 0.001. Any improvement of the baryonic model will not change the value of $\beta$. 
 
\subsection{Oort constants and older approaches}\label{Sec-Oort-ap}
We have calculated values of the Oort constants, see Eqs. (\ref{AB-num2}).
As for the older approach, we used the data obtained by \citet{karimova1973}, \citep[see also][p. 98]{kulik} for the variable stars of the type $\delta$~Cephei. While \citet{karimova1973} and \citet{kulik} used numerical differentiation of 
the observational results of the angular \textbf{velocities}
$\omega (R_{i})$, $i \in \{ 1, 2, 3 ... \}$, we use analytical fit $\omega = \omega (R)$ and analytical differentiation $\mbox{d}\omega(R) / \mbox{d}R$. Our results are much more accurate. We obtain $-B=A=\omega_{0}/2=(14.6 \pm 0.4) ~\mbox{km} ~\mbox{s}^{-1} ~\mbox{kpc}^{-1}$ and $\beta = 0 \pm 0.02$, where also Eqs. (\ref{AB-const}) are used, $\beta = (A+B)/(A-B)$ and $\sigma_{\beta} = 2 \sqrt{B^2 \sigma_{A}^{2} + A^2 \sigma_{B}^{2}  }/(A-B)^{2}$. The found values of $A$ and $B$ seem to be roughly consistent with the results collected in Eqs. (\ref{beta}) and (\ref{AB-num}) or Eqs. (\ref{AB-const}) and (\ref{AB-num2}). 

Eqs. (\ref{beta}), (\ref{AB-const}) and (\ref{AB-num2}) lead to 
\begin{equation}\label{beta2}
\beta = 0.062 \pm 0.001 ~.
\end{equation}
 
As for the conventional approach, much less accurate results are the IAU recommended values $ A = +15 ~\mbox{km} ~\mbox{s}^{-1} ~\mbox{kpc}^{-1} ~, ~B = -10 ~\mbox{km} ~\mbox{s}^{-1} ~\mbox{kpc}^{-1}$ leading to $\beta = 0.20$. The values $A = (14.4 \pm 1.2 )  ~\mbox{km} ~\mbox{s}^{-1} ~\mbox{kpc}^{-1} ~, ~B = (-12.0 \pm 2.8) ~\mbox{km} ~\mbox{s}^{-1} ~\mbox{kpc}^{-1}$ \citep{kerr} are not accurate, although they admit $\beta$ close to zero, $\beta = 0.09 \pm 0.12$. \citet{feast1997galactic} present 
$A = (+14.82 \pm 0.84) ~\mbox{km} ~\mbox{s}^{-1} ~\mbox{kpc}^{-1}$, $B = (-12.37 \pm 0.64 )  ~\mbox{km} ~\mbox{s}^{-1} ~\mbox{kpc}^{-1}$ yielding $\beta=0.09 \pm 0.04$. We can mention also the results based on the main-sequence stars from the Gaia DR1 Tycho-Gaia Astrometric Solution \citep{bovy17}, $A = (15.3 \pm 0.4 )  ~\mbox{km} ~\mbox{s}^{-1} ~\mbox{kpc}^{-1}$ and $B = (-11.9 \pm 0.4) ~\mbox{km} ~\mbox{s}^{-1} ~\mbox{kpc}^{-1}$ leading to $\beta = 0.125 \pm 0.021$. 

Potentials given by Eqs. (\ref{pot-bulge}) and (\ref{eqDM-Poul}) give the rotation speed $v_{c} \left ( R_0 \right )$ $=$ 243.9 km s$^{-1}$ and $\left ( \mbox{d}v_{c} /  \mbox{d}R \right  )_{R =R_{0}}$ $=$ $-$ 0.644 km $s^{-1}$ kpc$^{-1}$, for $R_0$ $=$ 8.247 kpc. The values of the Oort constants are $A$ $=$ 15.11 km $s^{-1}$ kpc$^{-1}$ and $B$ $=$ $-$14.47 km $s^{-1}$ kpc$^{-1}$. The values of $A$ and $B$ are not consistent with Eqs. (\ref{AB-num2}). The value of $\beta$ $=$ 0.0218 also differs from the value in Eq. (\ref{beta2}).

The IAU recommended values $ A = +15 ~\mbox{km} ~\mbox{s}^{-1} ~\mbox{kpc}^{-1} ~, ~B = -10 ~\mbox{km} ~\mbox{s}^{-1} ~\mbox{kpc}^{-1}$ corresponds to $\beta = 0.2$.

Eqs. (\ref{pot-bulge})-(\ref{g-bar-rad}) and 
(\ref{rot-curve-theory-old}) lead to the values 
$A$ $=$ (14.14 $\pm$ 0.04) km s$^{-1}$ kpc$^{-1}$, 
$B$ $=$ (-13.47 $\pm$ 0.04) km s$^{-1}$ kpc$^{-1}$. However, Eqs. (\ref{pot-bulge})-(\ref{g-bar-rad}) and (\ref{rot-curve-theory-old}) are not consistent with 
Eq. (\ref{speed-sun}).

Finally, we can present the values which follow from
Eilers et al. (2019, Eq. 7):
$A$ $=$ (14.95 $\pm$ 0.08) km s$^{-1}$ kpc$^{-1}$, 
$B$ $=$ (-13.25 $\pm$ 0.08) km s$^{-1}$ kpc$^{-1}$, $\beta$ $=$ 0.060 $\pm$ 0.004, and, $v_{c}(R=8.247 \mbox{kpc})$ $=$ (228.8 $\pm$ 0.2) km s$^{-1}$ (Eq. 7 from \citet{2019ApJ...871..120E} is used). The value of $\beta$ $=$ 0.060 $\pm$ 0.004 is consistent with the value given in Eq. (\ref{beta2}), 
the value of $(\mbox{d}v_{c}/\mbox{d}R)_{R=R_0}$ $=$ $-$ 1.7 $~\mbox{km} ~\mbox{s}^{-1} ~\mbox{kpc}^{-1}$ is also consistent with the values presented in Eqs. (\ref{diff-vel}). 

\subsection{Summary on the Oort constants}\label{Sec-sum-Oort} 
Several of the published values of the Oort constants $A$, $B$ and our values are collected in Table \ref{tab:my_label}, three other results are derived and discussed in Sec. \ref{Sec-Oort-ap} and they are also added to Table \ref{tab:my_label}. Their results are: \\
i) $A = -B = (14.6 \pm 0.4) ~\mbox{km} ~\mbox{s}^{-1} ~\mbox{kpc}^{-1}$, based on \citet{karimova1973},\\
ii) $A$ $=$ (14.95 $\pm$ 0.08) km s$^{-1}$ kpc$^{-1}$, $B$ $=$ (-13.25 $\pm$ 0.08) km s$^{-1}$ kpc$^{-1}$, based on \citet{2019ApJ...871..120E}, \\
iii) $A$ $=$ 15.11 km $s^{-1}$ kpc$^{-1}$, $B$ $=$ $-$14.47 km $s^{-1}$ kpc$^{-1}$, 
based on \citet{2017A&A...598A..66P} and \\ 
iv) $A$ $=$ 13.45 km $s^{-1}$ kpc$^{-1}$, $B$ $=$ $-$13.50 km $s^{-1}$ kpc$^{-1}$, 
based on corrected \citet{2017A&A...598A..66P} using \citet{Allen1991}, see Eqs. (\ref{pot-bulge}), Table \ref{Table1}
and Eqs. (\ref{eqDM-AS}). \\ 
Any improvement of the baryon mass model must be consistent with the values presented in Eqs. (\ref{AB-num2}).

The values of  $A$ and  $B$ found in this paper, see first line in Table \ref{tab:my_label}, are presented to a high precision. The values are independent on potential improvements of the baryon mass model of the MW. The approach based on Eqs. (\ref{obs-r-0}) and (\ref{obs-r-1}), or, (\ref{obs-Gaugh}) and (\ref{pot-bulge-new}), (\ref{pot-total-new}), (\ref{g-bar-new}), (\ref{rot-curve-theory}), (\ref{rot-curve-theory-num-relation}) and the values summarized in Table \ref{Table2}
yields much better results than other approaches.

\begin{table}

    \caption{Values of the Oort constants $A$ and $B$ (in km s$^{-1}$ kpc$^{-1}$). The resulting values from this paper can be compared with other published values or values following from published papers.}
        \label{tab:my_label}
    \centering
    \begin{tabular}{lcc} \hline
        source & $A$  & $B$    \\
        \hline
        \hline
        & & \\
        this paper & $14.73 \pm 0.03$ & $-13.01 \pm 0.03$ \\
        & & \\
        \hline
        & & \\
        Bovy (2017) & $15.30\pm 0.40$ & $-11.90\pm0.40$ \\
         Feast \& Whitelock (1997) & $14.82\pm 0.84$ & $-12.37 \pm 0.64$ \\
         Kerr \& Lynden-Bell (1986) & $14.40\pm 1.20$ & $-12.00\pm2.80$ \\
         IAU recommended values & $15$ & $-10$ \\
         & & \\
         \hline
         & & \\
         Eilers et al. (2019)  & $14.95 \pm 0.08$ & $-13.25 \pm 0.08$ \\
         Pouliasis et al. (2017) & $15.11$ & $-14.47$ \\
         fixed Pouliasis et al. (2017) & $13.45$ & $-13.50$ \\
         Karimova \& & & \\
         Pavlovskaya (1974) & $14.6 \pm 0.4$ &
         $-14.6 \pm 0.4$ \\
         Appendix C (this paper) & 14.68 & -13.06 \\
         & & \\
    \hline
    \end{tabular}
\end{table}


\subsection{Accuracy of $(\mbox{d}v / \mbox{d} R)_{R=R_{0}}$}
The value of $\beta$ is 0.062, both for $\delta$ $=$ 0 and
$\delta$ $=$ 0.15 within the accuracy 0.00024. Thus, Eq. (\ref{beta2}) can be considered as a very good determination of the dimensionless quantity $\beta$ offering the value and the error of $(\mbox{d}v_{c} / \mbox{d} R)_{R=R_{0}}$. 

If we use Eq. (\ref{local-diff}) and Eq. (\ref{beta2}) with
Eqs. (\ref{R-sun})-(\ref{speed-sun}), then
\begin{eqnarray}\label{local-diff-num}
\left ( \frac{\mbox{d}v_{c}}{\mbox{d}R} \right )_{R=R_{0}} = \left ( -~ 1.72 \pm 0.03 \right ) ~\mbox{km} ~\mbox{s}^{-1} ~\mbox{kpc}^{-1} ~.
\end{eqnarray}
{The result is consistent with the value presented in Eq. (\ref{diff-vel}).
The error is smaller than the error $\pm$ 0.1 $\mbox{km} ~\mbox{s}^{-1} ~\mbox{kpc}^{-1}$ given by \citet[Eq. 7]{2019ApJ...871..120E}. The continuous curve in Fig. \ref{Fig3} is smooth and it does not exhibit any fluctuation. This can be an explanation of the small error in Eq. (\ref{local-diff-num}).}

\section{DM - application of the equation of motion}\label{appendix-Dm-appl}
In this appendix, we will deal with applications of Eqs. (\ref{eq-motion-DM}), (\ref{eq-motion-g-bar}) and (\ref{vecDM}). Oscillations normal to the central equatorial plane, the Galactic midplane $z=0$, will be treated. The Sun and a test particle far away from the center of the MW will be considered. 
Spherically symmetric mass distribution of the DM will be taken into account.

\subsection{Sun - motion along $z$-axis}
We can deal with up and down motion of the Sun through the Galactic disc. Two approaches will be presented.

\subsubsection{Approach 1 - Poisson equation}
Let us start with {the Poisson equation}
\begin{eqnarray}\label{poisson}
\bigtriangleup \Phi = 4 \pi G \varrho ~,
\end{eqnarray}
where 
\begin{eqnarray}\label{phirho}
\Phi (R,z) &=& \Phi_{BM}(R,z) + \Phi_{DM} (R,z) ~,
\nonumber \\
\varrho(R,z) &=& \varrho_{BM}(R,z) +\varrho_{DM}(R,z) 
\end{eqnarray} 
are gravitational potentials and mass densities at the Galactocentric distance $R$ and
at the distance $|z|$ from the Galactic equatorial plane, all the quantities are even functions of the coordinate $z$. Both the baryonic matter and the dark matter components are considered. We remind that $\Phi_{DM} (r)$ $=$
$\Phi_{DM} (\sqrt{R^{2} + z^{2}})$ $\equiv$ $\Phi_{DM} (R,z)$. The Poisson equation reads
\begin{eqnarray}\label{derivePhi}
\frac{1}{R}\frac{\partial}{\partial R} \left( R\frac{\partial \Phi}{\partial R} \right) + \frac{\partial^2 \Phi}{\partial z^2} = 4 \pi G \varrho ~.
\end{eqnarray}
Since {the radial component of the equation of motion Eq. (\ref{eq-motion-DM}) leads to} $v^2 / R = \partial \Phi / \partial R$, we can write {for the Galactocentric distance $R_{0}$}
\begin{eqnarray}\label{derivePhi2}
2\frac{v_0}{R_0} \left(\frac{\mbox{d}v}{\mbox{d}R} \right)_{R=R_0} + \frac{\partial}{\partial z} \left( -\ddot{z} \right) = 4 \pi G \varrho(R_0) ~,
\end{eqnarray}
where the total mass density in the region of the Sun $\varrho (R_0) \equiv \varrho (R_0,z=0)$ is 
\begin{eqnarray}\label{Mden}
\varrho(R_0) = \varrho_{BM} (R_0) + \varrho_{DM} (R_0) = 0.0954 ~\mbox{M}_{\odot} ~\mbox{pc}^{-3} ~,
\end{eqnarray}
see also Eqs. (\ref{mass-density-sun}) and (\ref{RhoSolDM}).
Eqs. (\ref{derivePhi2}) and (\ref{oort}) yield
\begin{eqnarray}\label{zddotted}
\ddot{z} &=& -~ \omega^2 ~z ~, 
\nonumber \\
\omega^2 &=& 4 \pi G \varrho(R_0) + 2\left( A^2 -B^2 \right)
\end{eqnarray}
and Eqs. (\ref{AB-num2}) and (\ref{Mden}) hold. Numerically
\begin{eqnarray}\label{omegaBM}
\omega = 2.350 \times 10^{-15} ~\mbox{s}^{-1} ~.
\end{eqnarray}

Solution of Eqs.(\ref{zddotted}) is
\begin{eqnarray}\label{zddottedsol}
z &=& A_{DM} \cos \left( \omega t + \phi \right) 
\nonumber \\
&=& \left( A_{DM} \cos \phi \right) \cos\left( \omega t \right) - \left( A_{DM} \sin \phi \right) \sin \left( \omega t \right) ~,
\end{eqnarray}
where $A_{DM}$ is the amplitude of the oscillatory motion,
and, the initial conditions are
\begin{eqnarray}\label{initconSol}
z (t=0) &\equiv& z_0 = (17.40 \pm 1.90) ~\mbox{pc} ~,
\nonumber \\
\dot{z}(t=0) &\equiv& \dot{z}_0 = (+7.17 \pm 0.38) ~\mbox{km} ~\mbox{s}^{-1} ~,
\end{eqnarray}
see \citet{dehnen1998local}, \citet{2017MNRAS.465..472K}. The present position of the Sun is $17.40$ pc above the Galactic midplane and the current motion of the Sun is vertically upwards. 
Eqs. (\ref{zddottedsol}) and (\ref{initconSol}) give 
\begin{eqnarray}\label{zddottedsol2}
 z_0 &=& A_{DM} ~ \cos \phi ~,
\nonumber \\
\dot{z}_0 &=& -~ \omega ~A_{DM} ~\sin \phi ~,
\end{eqnarray}
Thus
\begin{eqnarray}\label{zddottedsol3}
A_{DM} &=& \sqrt{\left(z_0 \right)^2 +\left( \dot{z}_0 / \omega \right)^2 } ~,
\nonumber \\
\cos \phi &=& z_0 / A_{DM} ~,
\nonumber \\
\sin \phi &=& -\dot{z}_0 / (\omega A_{DM}) ~. 
\end{eqnarray}
Numerically,
\begin{eqnarray}\label{zddottedsol4}
A_{DM} &=& (100.40 \pm 5.20) ~\mbox{pc} ~,
\nonumber \\
\phi &=& 280.0^{\circ} \pm 1.2^{\circ}  ~,
\nonumber \\
T_{osc}(R_0) &=& \frac{2 \pi}{\omega} = ( 84.7  \pm 0.5) \times 10 ^{6} ~\mbox{yrs} ~.
\end{eqnarray}

The Sun intersects the Galactic equatorial plane at times $t = (1/2 - \phi/180 +k)T_{osc}/2$, where $k \in Z$. Numerically, we can obtain $t=-87.07 \times 10^{6} ~\mbox{yrs}$ for $k=-1$,  $t=-44.71 \times 10^{6} ~\mbox{yrs}$ for $k=0$ and  $t=-2.35\times 10^{6} ~\mbox{yrs}$ for $k=+1$. The error in $t$ is $\sigma_t = (\sigma_\phi/180)  T_{osc} /2 = (1.2/180)  T_{osc} /2 =  0.28 \times 10^{6}  ~\mbox{yrs}$. 

The found value $84.7$ Myrs is consistent with the value $83.3$ Myrs of \citet[p. 5]{2014pmdb.book.....M} within $1.7$\%.

\subsubsection{Approach 2 - Equation of motion}\label{7.1.222}

In this subsection, we will consider another approach to derive the motion of the Sun in the z-axis. The approach considers direct integration of equation of motion for the case of simultaneous action of the baryonic matter and the dark matter. The equation of motion is given by Eq. (\ref{eq-motion-DM}) with Eqs. (\ref{eq-motion-g-bar}) and (\ref{vecDM}).

The expression caused by the baryonic matter and its z-component is given as
\begin{equation}\label{SZ2-1}
(\vec{g}_{BM})_{z} = -~ \frac{\partial \Phi_{BM} (R, z)}{\partial z} ~.
\end{equation}

The expression for the $z$-component of the acceleration caused by the dark matter is given as
\begin{equation}\label{SZ2-2}
(\vec{g}_{DM})_{z} = -~ g_{DM} (r) ~\frac{z}{r} ~,
\end{equation}
which reduces, for $|z| \ll R$, to
\begin{equation}\label{SZ2-3}
(\vec{g}_{DM})_{z} = -~ g_{DM} (R) ~\frac{z}{R} ~.
\end{equation}

Using these accelerations {in the $z$-component of the equation of motion Eq. (\ref{eq-motion-DM}), $\ddot{z}$ $=$ $(\vec{g}_{BM})_{z}$ $+$ $(\vec{g}_{DM})_{z}$}, we obtain 
\begin{equation}\label{SZ2-4}
\ddot{z} = -~ \omega^{2} ~z ~,
\end{equation}
where the angular frequency of small oscillations along the $z$-axis, $\omega$, is
\begin{equation}\label{SZ2-5}
\omega^{2} = \omega_{BM}^{2} + \omega_{DM}^{2} ~.
\end{equation}
Here, for $R=R_{0}$,
\begin{eqnarray}\label{SZ2-6}
\omega_{BM}^{2} &=& \frac{G M_{b}}{\left ( R_{0}^{2} + b_{b}^{2} \right )^{3/2}} +
\sum_{j = 1}^{2}  \frac{G M_{j} \left ( 1 + a_{j}/b_{j} \right )}{\left ( R_{0}^{2} + c_{j}^{2}
\right )^{3/2}}
\end{eqnarray}
and
\begin{eqnarray}\label{SZ2-7}
\omega_{DM}^{2} &=& g_{DM} (R_{0}) / R_{0} ~,
\end{eqnarray}
see also Eqs. (\ref{vecDM}).



Finally, the angular frequency $\omega$ is
\begin{equation}\label{SZ2-10}
\omega = \sqrt{\omega_{BM}^{2} + \omega_{DM}^{2}}  
\end{equation}
and the resulting oscillating period {for the Galactocentric distance $R_{0}$}  
\begin{eqnarray}\label{SZ2-10}
T_{osc} (R_{0})  &=& \frac{2 \pi}{\omega} = 84.7 \times 10^{6} ~\mbox{yrs}  
\end{eqnarray}
corresponds to Eqs. (\ref{zddottedsol4}).


\subsubsection{Approach 3 - Equation of motion}\label{7.1.222}
This subsection does not consider linearized equation of motion corresponding to Eqs. (\ref{SZ2-4}),
(\ref{SZ2-5}), (\ref{SZ2-6}) and (\ref{SZ2-7}). We will numerically solve Eqs. (\ref{SZ2-1}) and 
(\ref{SZ2-3}).

The equation of motion reads
\begin{eqnarray}\label{Mi-1}
\ddot{z} &=& -~ \frac{\partial \Phi_{BM} (R, z)}{\partial z} - g_{DM} (r) ~\frac{z}{r} ~,
\end{eqnarray}
where the first term is given by Eq. (\ref{pot-total-new}) and the second term by Eq. (\ref{MDens}). The numerical solution of the nonlinearized equation of motion 
Eq. (\ref{Mi-1}) yields
\begin{eqnarray}\label{osc-non-lin-my}
A_{DM} &=& 99.9 ~\mbox{pc} ~,
\nonumber \\
T_{osc}(R_0) &=& 85.0 \times 10 ^{6} ~\mbox{yrs} ~.
\end{eqnarray}
The results summarized in Eqs. (\ref{osc-non-lin-my}) are consistent
with the results based on the linearization of the equation of motion,
compare Eqs. (\ref{osc-non-lin-my}) and Eqs. (\ref{zddottedsol4}).

\subsection{Revolution and oscillation periods}
Finally, we can compare the orbital period, the period of the revolution, $T_{rev}(R_0)=T_0$ and oscillation period $T_{osc}(R_0)$, see Eqs. (\ref{orbital period}) and (\ref{zddottedsol4}). We obtain
\begin{eqnarray}\label{7_2_110}
\left[ \frac{T_{rev}(R_0)}{T_{osc}(R_0)} \right]_{DM}= \frac{T_0}{T_{osc}(R_0)} = 2.61 \pm 0.02 ~.
\end{eqnarray}

\subsection{Test particle - motion along $z$-axis}
Let us consider that the test particle is far away from the center of the MW. 

\subsubsection{Conventional DM halo}\label{sec731}
As the conventional DM halo we can take the one described by the gravitational potential $\tilde{\Phi}_{halo}(r)$ represented by Eq. (\ref{eqDM-Poul}). The case of large distance $R$ gives
\begin{eqnarray}\label{7_2_11}
\ddot{z} &=& -~\frac{\partial \tilde{\Phi}_{halo}(r)}{\partial z} = -~\frac{G M_{halo}}{a_{halo} R^2} ~z ~,
\end{eqnarray}
if $|z| \ll R$.

\subsubsection{Improved DM halo - Equation of motion}\label{sec732}
The conventional DM halo represented by Eqs. (\ref{eqDM-Poul}) is not consistent with the rotation curve of the MW galaxy. The consistency is achieved by the improved DM model given by Eqs. (\ref{vecDM}).

Eqs. (\ref{eq-motion-DM}) and (\ref{vecDM}) yield {for the $z$-component of the acceleration of the test particle at large Galactocentric distance $R$}
\begin{eqnarray}\label{7_2_111}
\ddot{z} &=& -~\frac{G M_{BM}}{\exp (L/R )-1} \frac{z}{R^3}  = -~\frac{G M_{BM}}{L R^2} ~z ~,
\nonumber \\
M_{BM} &=& \sum_{j=1}^{3} M_j ~,
\nonumber \\
L &=& \sqrt{G M_{BM}/g_{+}} ~,
\end{eqnarray}
if $|z| \ll R$.

Solution of Eqs. (\ref{7_2_111}) 
\begin{eqnarray}\label{7_2_1111}
\ddot{z} &=& -~ \omega_{DM}^{2} ~z ~,
\nonumber \\
\omega_{DM}^{2} &=& \frac{G M_{BM}}{LR^2} 
\end{eqnarray}
is an oscillatory motion with the period $T_{DM-osc}$,
\begin{eqnarray}\label{7_2_111111}
T_{DM-osc}(R) = \frac{2 \pi}{\omega_{DM}} = \frac{2 \pi R}{\sqrt{G M_{BM}/L}} ~.
\end{eqnarray}

\subsubsection{Improved DM halo - Poisson equation}
Another approach to the acceleration $\ddot{z}$ is based on the Poisson equation. 

We can write {for the gravitational acceleration acting on the test particle at large Galactocentric distance $r$}
\begin{eqnarray}\label{eqgDM}
\vec{g}_{DM} = -~ \frac{G M_{BM}}{Lr} ~\hat{\vec{r}} ~.
\end{eqnarray}
This enables us to find the mass density $\varrho_{DM}$ 
\begin{eqnarray}\label{massDensity1}
\nabla \cdot \vec{g}_{DM} &=& -~ 4 \pi G \varrho_{DM} ~,
\nonumber \\
\frac{1}{r^2}\frac{\mbox{d} }{\mbox{d} r} \left[ r^2 \left( - \frac{G M_{BM}}{Lr} \right)    \right] &=& -~ 4 \pi G \varrho_{DM} ~.
\end{eqnarray}
Thus
\begin{eqnarray}\label{eqgGMLr}
 \frac{G M_{BM}}{Lr^2} =  4 \pi G \varrho_{DM} ~.
\end{eqnarray}

Now we are interested in the motion in the $z$-direction. We {can write, 
on the basis of the Poisson equation $\nabla \cdot \vec{g}_{DM}$ $=$ $-~ 4 \pi G \varrho_{DM}$ and $\vec{g}_{DM}$ $=$ $-~ \nabla \Phi_{DM}$, in cylindrical coordinates}
\begin{eqnarray}\label{eqgGMLrZ}
\frac{1}{R}\frac{\partial }{\partial R} \left( R \frac{\partial \Phi_{DM}}{\partial R}\right) +  \left(  \frac{\partial^2 \Phi_{DM}}{\partial z^2}\right) &=& 4 \pi G \varrho_{DM} ~,
\nonumber \\
\frac{\partial \Phi_{DM}}{\partial R} &=& \frac{G M_{BM}}{LR} ~.
\end{eqnarray}
Since $\ddot{z}$ $=$ $-~\partial \Phi_{DM}/\partial z$, we obtain 
\begin{eqnarray}\label{zdotdot1}
\ddot{z} = -~ 4 \pi G \varrho_{DM} ~z 
\end{eqnarray}
and, using Eq. (\ref{eqgGMLr}), finally
\begin{eqnarray}\label{zdotdot2}
\ddot{z} = -~\frac{G M_{BM}}{LR^2} ~z ~.
\end{eqnarray}
The found result is consistent with Eqs. (\ref{7_2_1111}).

\subsubsection{Comparison}\label{sec733}
We can compare the results given by Eqs. (\ref{7_2_11}) and (\ref{7_2_111}). 

Eq. (\ref{7_2_11}) contains two halo characteristics, the mass $M_{halo}$ and the length $a_{halo}$. However, Eqs. (\ref{7_2_111}) contain mass $M_{BM}$ of the baryonic matter and a linear characteristics $L$ of the baryonic matter, $M_{halo}$ $\dot{=}$ 1.4 $M_{BM}$, $a_{halo}$ $\dot{=}$ 1.4 $L$. While $M_{halo}$ and $a_{halo}$ were taken as free parameters for the DM halo, Eqs. (\ref{7_2_111}) contain no free parameter for the DM halo. Eqs. (\ref{7_2_111}) contain information solely on the baryonic matter.

If the radius of the DM halo is $r_{DM}$, then Eqs. (\ref{Mass})-(\ref{Mass2}) lead to the gravitational potential 
\begin{eqnarray}\label{largefi}
\Phi (r) = -~ \frac{GM r_{DM}/L}{r} ~, ~r \geq r_{DM} ~.
\end{eqnarray}
Eq. (\ref{largefi}) differs from {the conventional $\tilde{\Phi} (r)$ $=$ $-~G M_{halo} /r$, see Eq. (\ref{eqDM-Poul})}. The total mass of the DM halo is $M r_{DM}/L$, not $M_{halo}$.

\subsection{Body - revolution in the plane $z=0$}
The body moves around the center of the MW in the plane $z=0$ far away from the Galactic center. Radial components of Eqs. (\ref{large-R-speed}), (\ref{eq-motion-DM}) and (\ref{vecDM}) are relevant for large value of $r=R$.

\subsubsection{Conventional DM halo}\label{Conv-DM-halo}
On the basis of Eq. (\ref{eqDM-Poul}) we can write {for the radial component of the equation of motion}
\begin{eqnarray}\label{7_2_1}
\frac{v_{c}^2}{R} = \frac{\partial \tilde{\Phi}_{halo}}{ \partial R} = \frac{G M_{halo}}{a_{halo} R} ~,
\end{eqnarray}
where $M_{halo} = 1.392 \times 10^{11}$ M$_{\odot}$ and $a_{halo}=14$ kpc. The terms $1/R^2$ are neglected.

\subsubsection{Improved DM halo}\label{Imp-DM-halo}
On the basis of Eqs. (\ref{eq-motion-DM}), (\ref{eq-motion-g-bar}) and (\ref{vecDM}) we obtain
{for the radial component of the equation of motion}
\begin{eqnarray}\label{7_2_2}
\frac{v_{c}^2}{R} &=& \frac{G M_{DM}}{R^2} = g_{DM}(R) = \sqrt{g_{bar} g_{+}} = \frac{G M_{BM}}{LR} ~,
\nonumber \\
L&=& \sqrt{G M_{BM}/ g_{+}} ~,
\end{eqnarray}
where $M_{BM}=8.43\times 10^{10}$ M$_{\odot}$ and $L=9899.3$ pc. 

Eqs. (\ref{7_2_2}) is consistent with Eq. (\ref{gravaccDM}) and also Eq. (\ref{largefi}), for $r=R=r_{DM}$ since $[v_{c}^2/R]_{R=r_{DM}} = [\mbox{d}\Phi (r) / \mbox{d}r ]_{r=r_{DM}}$.

\subsubsection{Comparison}\label{com-DM}
We can compare the results given by Eqs.(\ref{7_2_1}) and (\ref{7_2_2}). Eq. (\ref{7_2_1}) contains $M_{halo}$ and $a_{halo}$. Eqs. (\ref{7_2_2}) contain only baryonic matter characteristics, the mass $M_{BM}$ and the scale length $L$. The situation is analogous to the results obtained in Sections \ref{sec731} and \ref{sec732}, see discussion in Section \ref{sec733}.

\subsection{Revolution and oscillation periods}
If the object is far away from the Galactic center, then we can write for the period of the revolution
\begin{eqnarray}\label{7_2_3}
T_{DM-rev} = \frac{2\pi R}{v_{c}} = \frac{2 \pi R}{\sqrt{G M_{BM} / L}} ~,
\end{eqnarray}
{see Eqs. (\ref{7_2_2}). The DM action is relevant for large Galactocentric distances.}

The period of oscillation is given by Eq.  (\ref{7_2_111111}).

Finally, we can compare Eqs. (\ref{7_2_111111}) and (\ref{7_2_3}). The result is 
\begin{eqnarray}\label{7_2_31}
\frac{T_{DM-rev}(R)}{T_{DM-osc}(R)} =  1 ~.
\end{eqnarray}

The result given in Eq. (\ref{7_2_31}) is easily understood, since the motion occurs in a circular orbit in a spherically symmetric potential $\Phi_{DM}(R)$.

\subsection{Comparison for $R_0$ and large $R$}
We can summarize the results found for the region of the Sun and for objects far away from the Galactic center. The results are represented by Eqs. (\ref{7_2_110}) and (\ref{7_2_31})
\begin{eqnarray}\label{RR0comp}
\left[ \frac{T_{rev}(R_0)}{T_{osc}(R_0)} \right]_{DM} &=& \frac{13}{5} ~,
\nonumber \\
\left[\frac{T_{rev}(R)}{T_{osc}(R)}\right]_{DM}  &=&  \frac{T_{DM-rev}(R)}{T_{DM-osc}(R)} = 1 ~, ~ R \gg R_0 ~.
\end{eqnarray}

\section{Improved model by  \citet{2017A&A...598A..66P} and \citet{Allen1991}}\label{appendix-fixed-model}
In this appendix, we will deal with the model presented by \citet{2017A&A...598A..66P}. 
The visible matter model, the BM model, will be considered to be identical to the BM model by \citet{2017A&A...598A..66P}. However, the model of the DM will be considered in the form given by \citet{Allen1991} with free parameters $M_{halo}$, $a_{halo}$ and $R_{DM}$ and their values will be found. 

\subsection{Summary of the model}
The circular velocity $v_{c}$ obtained from Eqs. (\ref{pot-bulge})-(\ref{g-bar-rad}) and (\ref{eqDM-AS})-(\ref{g-DM-rad-AS}) is
\begin{eqnarray}\label{30-AS}
v_{c}^{2} &=& \sum_{j=1}^{3} v_{cj}^{2} + v_{cDM-AS}^{2} ~,
\end{eqnarray}
where Eqs. (\ref{31}) hold for the baryonic matter and
\begin{eqnarray}\label{eqDM-AS-app}
\tilde{\Phi}_{halo}(r) &=& -~ \frac{G M_{halo}}{r}  
\frac{\left ( r/a_{halo} \right )^{2.02}}{1 + \left ( r/a_{halo} \right )^{1.02}}
- \frac{G M_{halo}}{1.02 a_{halo}} \times
\nonumber \\
& & \left [ \frac{-1.02}{1 + ( \xi / a_{halo} )^{1.02}} + \ln \left ( 1 + 
\left ( \frac{\xi}{a_{halo}} \right )^{1.02} \right ) \right ]_{\xi = r}^{R_{DM}}  ,
\end{eqnarray}
together with
\begin{eqnarray}\label{32-ap}
v_{cDM-AS}^{2} &=&  \left ( G M_{halo} / a_{halo} \right ) \zeta / \left ( 1 + \zeta \right ) ~,
\nonumber \\
\zeta &=& \left ( R/a_{halo} \right )^{1.02} 
\end{eqnarray}
hold for the DM.  Sec. \ref{sec-DM-AS-P} shows that the values of $M_{halo}$, $a_{halo}$ and $R_{DM}$ used by \citet{2017A&A...598A..66P} and \citet{Allen1991} do not match the observational data.

The DM model represented by Eq. (\ref{eqDM-AS-app}) contains three free parameters, 
$M_{halo}$, $a_{halo}$ and $R_{DM}$. To be the DM model complete, we have to determine the values of the three parameters. 

\subsection{Specification of the DM model - simple ideas}\label{C-simple-ideas}
The model given by  Eq. (\ref{eqDM-AS-app}) leads to the gravitational acceleration
\begin{eqnarray}\label{grav-acc-eq-DM}
g_{DM} \left ( r \right )  &=& \frac{G~ M_{DM} \left ( r \right )}{r^{2}} =
\frac{G~M_{halo}}{a_{halo}} ~\frac{1/r}{1 + \left ( a_{halo} / r \right )^{1.02}} 
\nonumber \\
&\doteq& \frac{G~M_{halo}}{a_{halo}~r} ~\left [ 1 - \left ( \frac{a_{halo}}{r} \right )^{1.02} \right ] 
\end{eqnarray}
for $r$ $\le$ $R_{DM}$ and large values of the galactocentric distance $r$. 

The third of Eqs. (\ref{obs-Gaugh}) yields for large values of 
the galactocentric distance $r$
\begin{eqnarray}\label{grav-acc-eq-DM-my}
g_{DM} \left ( r \right )  &=& \frac{G M_{BM} / r^{2}}{\exp{\left ( L/r
\right )} - 1} \doteq \frac{G M_{BM}}{L~r} \left (  1 - \frac{1}{2} \frac{L}{r} \right ) 
\end{eqnarray}
for $r$ $\le$ $R_{DM}$. We remind that $L$ $=$ $\sqrt{G M_{BM}/g_{+}}$ is the characteristic scale-length of the BM of the MW.

Comparison of Eqs. (\ref{grav-acc-eq-DM}) and (\ref{grav-acc-eq-DM-my}) immediately yields for the approximation $r$ $=$ $R_{DM}$
the following results
\begin{eqnarray}\label{grav-acc-eq-DM-my-comp}
\frac{M_{halo}}{a_{halo}} &=& \frac{M_{BM}}{L} ~,
\nonumber \\
\left ( \frac{a_{halo}}{R_{DM}} \right )^{1.02} &\approx& 
\frac{1}{2} \frac{L}{R_{DM}} ~.
\end{eqnarray}

Comparison of Eqs. (\ref{grav-acc-eq-DM}) and (\ref{grav-acc-eq-DM-my}) shows that the approach used by \citet{2017A&A...598A..66P} and \citet{Allen1991} is an approximation to the third of Eqs. (\ref{obs-Gaugh}) and the value 1.02 is purely a numerical factor found from fit to observational data without any physical meaning. The relations in Eqs. (\ref{grav-acc-eq-DM-my-comp}) suggest that $M_{halo}$ is less than the total BM mass of the MW 
and $a_{halo}$ is less than the BM scale-length of the MW.
Thus, $M_{halo}$ and $a_{halo}$ are not physical characteristics of the DM of the MW. Detailed values of $M_{halo}$, $a_{halo}$ and $R_{DM}$ will be found in the following subsection Appendix \ref{spec-DM-num-res}.

\subsection{Specification of the DM model - numerical results}
\label{spec-DM-num-res}
We have to determine the values of $M_{halo}$, $a_{halo}$ and $R_{DM}$. 

We need to fulfill three requirements. We will take into account three observational results.

\subsubsection{First requirement}
The first requirement will consider the critical mass density, see Eq. (\ref{crit-mass-density}),
and the observational values of the Hubble constant,
$H_{0}$ $=$ (67.4 $\pm$ 0.5) $\mbox{km} ~\mbox{s}^{-1} ~\mbox{Mpc}^{-1}$
following from the ESA's Planck spacecraft data, see \citet{Planck2018}, and,
$H_0$ $=$ (73.04 $\pm$ 1.04) $\mbox{km} ~\mbox{s}^{-1} ~\mbox{Mpc}^{-1}$ based on the standard candles observations, see \citet{RiessHubble}. The virial radius $r_{vir}$ is obtained by truncating the DM halo at 200 times the critical density. We will use 
$R_{DM}$ $\doteq$ $r_{vir}$.

The critical mass density is 
\begin{eqnarray}\label{crit-mass-den-H0A}
\varrho_{crit,1} &=& (126.1 \pm 1.9) ~\mbox{M}_{\odot} ~\mbox{kpc}^{-3} 
\end{eqnarray}
for $H_{0}$ $=$ (67.4 $\pm$ 0.5) $\mbox{km} ~\mbox{s}^{-1} ~\mbox{Mpc}^{-1}$, and, the critical density is 
\begin{eqnarray}\label{crit-mass-den-H0R}
\varrho_{crit,2} &=& (148.1 \pm 4.2) ~\mbox{M}_{\odot} ~\mbox{kpc}^{-3} 
\end{eqnarray}
for $H_{0}$ $=$ (73.04 $\pm$ 1.04) $\mbox{km} ~\mbox{s}^{-1} ~\mbox{Mpc}^{-1}$. 

The condition for the $R_{DM}$ will be considered in the form
\begin{eqnarray}\label{R-DM-crit-mass-den-H0A}
\varrho_{halo} (R_{DM,j}) &=& 200~ \varrho_{crit,j} ~,~~ j = 1,~ 2~, 
\end{eqnarray}
where the mass density $\varrho_{halo} (R_{DM})$ follows from Eq. (\ref{eqDM-AS-app}), 
\begin{eqnarray}\label{mass-den-halo-eqDM-AS-app}
\varrho_{halo} \left ( r \right ) &=& \frac{1}{4 \pi} ~\frac{M_{halo}}{a_{halo}^{3}}
\left ( \frac{r}{a_{halo}} \right )^{-0.98} 
\frac{2.02 + \left ( r/a_{halo} \right )^{1.02}}{\left [ 1 + \left ( r/a_{halo} \right )^{1.02} \right ]^{2}} ~.
\end{eqnarray}

\subsubsection{Second requirement}
The second requirement is based on the 
observational data on circular rotation speeds - the speed (229.0 $\pm$ 0.2) km s$^{-1}$ holds for the galactocentric distance $R$ $=$ 8.122 kpc according to
\citet{2019ApJ...871..120E}. Thus, we require 
\begin{eqnarray}\label{sec-req-speed}
v_{c} \left ( R=8.122 ~\mbox{kpc} \right ) &=& 222.9 ~\mbox{km} ~ \mbox{s}^{-1} ~.
\end{eqnarray}

\subsubsection{Third requirement}
The third requirement is based on Eq. (\ref{VGg}), the baryonic Tully-Fisher relation. We require for the circular rotation speed at large galactocentric distance $R$ 
\begin{eqnarray}\label{third-req-speed}
v_{c} \left ( R \right ) &=& \left ( G g_{+} M_{BM} \right )^{1/4} ~, 
\nonumber \\
M_{BM} &=& \sum_{j=1}^{3} M_{j} ~
\end{eqnarray}
where the values of $M_{j}$, $j$ $=$ 1,2 and 3 are summarized in  Table \ref{Table2}. 

Eqs. (\ref{30-AS}), (\ref{eqDM-AS-app}), (\ref{32-ap}) and (\ref{third-req-speed}) yield
\begin{eqnarray}\label{third-req-fin}
\frac{M_{halo}}{a_{halo}} &=& \left ( \frac{g_{+}}{G}~ \sum_{j=1}^{3} M_{j} \right )^{1/2} ~,
\end{eqnarray}
see also Table \ref{Table2}.

\subsubsection{Numerical results}
The three requirements finally lead to the numerical results summarized in this section.

At first, Eq. (\ref{third-req-fin}) enables to find the ration $M_{halo}/a_{halo}$. Eqs. (\ref{31}),
the values in Table \ref{Table1}. Eqs. (\ref{30-AS}), (\ref{32-ap}) and (\ref{sec-req-speed}) finally yield
$M_{halo}$ and $a_{halo}$. Finally, Eqs. (\ref{crit-mass-den-H0A}), (\ref{crit-mass-den-H0R}),
(\ref{R-DM-crit-mass-den-H0A}) and (\ref{mass-den-halo-eqDM-AS-app}) give $r_{vir,j}$ $\doteq$ $R_{DM, j}$,
$j$ $=$ 1 and 2.

Finally, the numerical values of $a_{halo}$ and $M_{halo}$ are
\begin{eqnarray}\label{three-req-a-halo}
a_{halo} &=& 7.863 ~\mbox{kpc}  ~,
\end{eqnarray}
\begin{eqnarray}\label{three-req-M-halo}
M_{halo} &=&  6.696 \times 10^{10} ~\mbox{M}_{\odot} ~.
\end{eqnarray}

The condition represented by Eq. (\ref{crit-mass-den-H0A}) leads to  
\begin{eqnarray}\label{third-req-res1}
R_{DM,1} &=& 163.83 ~\mbox{kpc} ~,
\nonumber \\
M_{DM,1} &=& 133.48 \times 10^{10} ~\mbox{M}_{\odot} ~,
\end{eqnarray}
since
\begin{eqnarray}\label{third-req-M-DM}
M_{DM} &=& M_{halo} ~\frac{\left ( R_{DM}/a_{halo} \right )^{2.02}}{1 + \left ( R_{DM}/a_{halo} \right )^{1.02}} ~.
\end{eqnarray}
Similarly, the condition represented by Eq. (\ref{crit-mass-den-H0R}) leads to
\begin{eqnarray}\label{third-req-res2}
R_{DM,2} &=& 151.16 ~\mbox{kpc} ~,
\nonumber \\
M_{DM,2} &=& 122.70 \times 10^{10} ~\mbox{M}_{\odot} ~.
\end{eqnarray}

Eqs. (\ref{third-req-res1}) and (\ref{third-req-res2}) can be summarized as 
\begin{eqnarray}\label{third-req-res12-sum-R-DM}
R_{DM} &=& \left ( 157.5 \pm 6.4 \right ) ~\mbox{kpc} 
\end{eqnarray}
and
\begin{eqnarray}\label{third-req-res12-sum-M-DM}
M_{DM} &=& \left ( 128.1 \pm 5.4 \right ) \times 10^{10} ~\mbox{M}_{\odot} ~.
\end{eqnarray}
The results summarized in Eqs. (\ref{third-req-res12-sum-R-DM}) and (\ref{third-req-res12-sum-M-DM})
are consistent with Eq. (\ref{third-req-M-DM}).

\subsection{Consequences}
The numerical results for the DM halo model represented by Eq. (\ref{eqDM-AS-app}) are given by 
Eqs. (\ref{three-req-a-halo}),  (\ref{three-req-M-halo}) and (\ref{third-req-res12-sum-R-DM}). 

The found value of $R_{DM}$ is consistent with the value presented by Eq. (\ref{R_DM}). Similarly,
the value of $M_{DM}$ given by Eq. (\ref{third-req-res12-sum-M-DM}) is consistent with the value presented by
Eq. (\ref{M_DM}).  

However, the numerical results of $a_{halo}$, $M_{halo}$ and $R_{DM}$ 
significantly differ from the numerical values used by \citet{2017A&A...598A..66P} and \citet{Allen1991}, compare
Eqs. (\ref{three-req-a-halo}), (\ref{three-req-M-halo}),  (\ref{third-req-res12-sum-R-DM}), and 
Eqs. (\ref{eqDM-AS}). The value of $M_{halo}$ used by \citet{2017A&A...598A..66P} is more than two-times higher than the value consistent with the observations on rotation curve of the MW. Also the real value of $a_{halo}$ is approximately half the value of the value used by \citet{2017A&A...598A..66P}. The value of the DM radius of the MW is in more than 50 \% higher than the value given by  \citet{2017A&A...598A..66P} or \citet{Allen1991}.

The value of $a_{halo}$ is much less than the radius of the DM halo, $a_{halo}$ $\ll$ $R_{DM}$. Thus, Eq. (\ref{third-req-M-DM}) can be simplified to the form
\begin{eqnarray}\label{third-req-M-DM-simple}
M_{DM} &=& \frac{M_{halo}}{a_{halo}}~ R_{DM} ~,
\end{eqnarray}
which is similar to Eq. (\ref{Mass}), $M_{DM}(R_{DM})$ $=$ $(M_{BM}/L) R_{DM}$, where $L$ is the scale-length of the visible matter of the MW, see Eqs. (\ref{Mass2})-(\ref{L-value}). Really, the relation 
$M_{halo} / a_{halo}$ $=$ $M_{BM}/L$ holds. The value of $a_{halo}$ is less than the distance of the Sun from the center of the MW, $a_{halo}$ $<$ $R_{0}$ $<$ $L$.

The relation $M_{DM}(R_{DM})$ $=$ $(M_{BM}/L) R_{DM}$ relates mass of the DM halo of the MW to the total mass of the BM of the MW and the scale-length of the BM of the MW. The radius of the DM halo is much greater than $a_{halo}$ and thus, $a_{halo}$ cannot be considered as a scale-length of the DM halo of the MW. Similarly, 
$M_{halo}$ cannot be considered as a characteristic mass of the DM halo of the MW. Thus, the model of the DM halo of the MW represented by Eq. (\ref{eqDM-AS-app}) cannot be considered as a physical model containing a characteristic mass $M_{halo}$ and a scale-length $a_{halo}$ of the DM halo. The model of the DM halo is a mathematical fit to some observational data and $M_{halo}$ and $a_{halo}$ are some free parameters which do not give any information about mass of the DM halo and a radius or the scale-length of the DM halo.

\subsection{Consequences - rotation speed, Oort constants}
Eqs. (\ref{eqDM-AS-app}), (\ref{three-req-a-halo}), (\ref{three-req-M-halo}) and (\ref{third-req-res12-sum-R-DM}) enable us to calculate some other important quantities characterizing our galaxy, the MW.

The circular rotation speed $v_{c} (R)$ can be calculated from Eqs. (\ref{30-AS}), (\ref{31}) and (\ref{32-ap}). The speed $v_{c} (R_{0})$ for the position of the Sun $R_{0}$ $=$ 8.247 kpc is 
\begin{eqnarray}\label{cons-1}
v_{c}  \left ( R_{0} \right ) = 228.8  ~\mbox{km}~ \mbox{s}^{-1} ~.
\end{eqnarray}
The result is coincidental with Eq. (\ref{speed-sun}).

The derivative of the circular rotation speed is
\begin{eqnarray}\label{cons-2}
\left( \frac{\mbox{d}v_{c}}{\mbox{d}R}\right)_{R=R_0} &=& -~1.63 ~\mbox{km}~ \mbox{s}^{-1} ~ \mbox{kpc}^{-1}  ~.
\end{eqnarray}
The result is slightly higher than the result represented by Eq. (\ref{diff-vel}), 
$v_{c}'$ $=$ $-$~1.72  $\mbox{km} ~\mbox{s}^{-1} ~\mbox{kpc}^{-1}$. 

The results represented by Eqs. (\ref{cons-1})-(\ref{cons-1}) enable us to obtain the Oort constants. 
The values of the Oort constants are 
\begin{eqnarray}\label{cons-3}
A  &=& 14.68 ~\mbox{km} ~\mbox{s}^{-1} ~\mbox{kpc}^{-1} ~,
\nonumber \\
B  &=& -~13.06 ~\mbox{km} ~\mbox{s}^{-1} ~\mbox{kpc}^{-1} ~. 
\end{eqnarray}
The obtained results are slightly different from the results given by Eqs. (\ref{AB-num2}). 

\subsection{Consequence - motion of the Sun along $z$-axis}\label{appendix-Dm-appl-z}
We can deal with up and down motion of the Sun through the Galactic disc. The BM of the MW is represented by Eqs. (\ref{pot-bulge}), (\ref{g-bar-rad}) and the values summarized in Table \ref{Table1}. 
Equation $\dot{\vec{v}}$ $=$ $-~\nabla \tilde{\Phi}_{BM}$ $-$ $\nabla \tilde{\Phi}_{halo}$, (\ref{eqDM-AS-app}), (\ref{three-req-a-halo}), (\ref{three-req-M-halo}) and 
(\ref{third-req-res12-sum-R-DM}) enable us to calculate oscillation of the Sun normal to the central equatorial plane, the Galactic midplane $z=0$.  

We remind that the gravitational acceleration of the DM is
\begin{eqnarray}\label{eq-motion-g-DM-C}
\vec{g}_{DM} &=& -~ \nabla \tilde{\Phi}_{halo} ~.
\end{eqnarray}

The equation of motion reads
\begin{eqnarray}\label{Mi-1-C}
\ddot{z} &=& 
-~ \frac{\partial \tilde{\Phi}_{BM} (R, z)}{\partial z} 
- \frac{\partial \tilde{\Phi}_{halo} (R, z)}{\partial z} ~,
\end{eqnarray}
with the initial conditions given by Eqs. (\ref{initconSol}).
We remind that $\tilde{\Phi}_{halo} (R, z)$ $=$ 
$\tilde{\Phi}_{halo} (r)$, $r$ $=$ $\sqrt{R^{2} + z^{2}}$ .
The numerical solution of the equation of motion 
Eq. (\ref{Mi-1-C}) yields
\begin{eqnarray}\label{z-osc-C}
A_{DM} &=& 102.8 ~\mbox{pc} ~,
\nonumber \\
T_{osc}(R_0) &=& 87.4 \times 10 ^{6} ~\mbox{yrs} ~,
\end{eqnarray}
 where $A_{DM}$ is the amplitude of the oscillatory motion
 and $T_{osc}(R_0)$ is the period of the oscillatory motion.
The results summarized in Eqs. (\ref{z-osc-C}) are less than in 3\%
higher than the values presented in Eqs. (\ref{osc-non-lin-my}). 

\subsection{Consequence - mass density of the DM component}\label{appendix-mass-density}
The DM halo model is represented by Eqs. (\ref{eqDM-AS-app}),  
(\ref{three-req-a-halo}),  (\ref{three-req-M-halo}) and (\ref{third-req-res12-sum-R-DM}). The model enables us to find DM mass density.
The result for the region of the Sun reads
\begin{eqnarray}\label{DM-density-R0-app-C}
\varrho_{DM} \left ( R_{0} \right ) &=& 7.64 \times 10^{-3}  ~\mbox{M}_{\odot} ~\mbox{pc}^{-3} ~,
\nonumber \\
R_{0} &=& 8.247 ~\mbox{kpc} ~. 
\end{eqnarray}
This value is in 9\% less than the value presented in Eq. (\ref{RhoSolDM}). The total mass density is
\begin{eqnarray}\label{total-density-R0-app-C}
\varrho_{total} \left ( R_{0} \right ) &=& 
\varrho_{BM} \left ( R_{0} \right ) + \varrho_{DM} \left ( R_{0} \right ) 
\nonumber \\
&=& 9.47 \times 10^{-2}  ~\mbox{M}_{\odot} ~\mbox{pc}^{-3} ~,
\end{eqnarray}
see also Eq. (\ref{mass-density-sun}). The total mass density is consistent with
the value $\varrho_{total}$ $=$ (0.095 $\pm$ 0.001) $\mbox{M}_{\odot} ~\mbox{pc}^{-3}$, see Eqs. (\ref{mass-den-spher-distr-Sun-total}) and Sec.
\ref{mass-den-sun}. 

\subsection{Corollary}\label{col-ap-C}
The gravitational potential $\tilde{\Phi}_{halo}(r)$ describes distribution of the DM and the motion of a star moving in the MW,
$\bigtriangleup \tilde{\Phi}_{halo}$ $=$ $4 \pi G \varrho_{DM}$
and $\dot{\vec{v}}$ $=$ $-~\nabla \tilde{\Phi}_{bar}$ $-$ 
$\nabla \tilde{\Phi}_{halo}$. The potential contains three characteristic quantities of the DM, 
$M_{halo}$ $=$ 1.392 $\times$ $10^{11} ~\mbox{M}_{\odot}$,
$a_{halo}$ $=$ 14 $\mbox{kpc}$ and $R_{DM}$ $=$ 100 $\mbox{kpc}$,
according to \citet{2017A&A...598A..66P}. However, the values of the quantities, including the ratio $R_{DM}/a_{halo}$ $\doteq$ 7,
are not consistent with observations. The observational data lead to
$M_{halo}$ $=$ 6.695 $\times$ $10^{10} ~\mbox{M}_{\odot}$,
$a_{halo}$ $=$ 7.863 $\mbox{kpc}$ and $R_{DM}$ $\doteq$ 157.5 $\mbox{kpc}$. Moreover, $M_{halo} / a_{halo}$ $=$ $M_{BM}/L$ $=$ 
$0.97 ~{\tilde{M}}_{BM}/\tilde{L}$, where $\tilde{L}$ $=$ 
$\sqrt{G {\tilde{M}}_{BM}/g_{+}}$ is the scale-length of the BM of the MW.
Simply, the total mass of the BM of the MW determines the ratio of the two characteristics of the DM of the MW. And, 
$a_{halo}$ $<$ $R_{0}$ $<$ $\tilde{L}$ $=$ 10.2 $\mbox{kpc}$, $M_{halo}$ $<$ $\tilde{M_{BM}}$ $=$ 8.9552 $\times$ 10$^{10}$ $\mbox{M}_{\odot}$ and $R_{DM}/a_{halo}$ $=$ 20 $\pm$ 1. These results significantly differ from the published values by \citet{2017A&A...598A..66P}. Very small values of $a_{halo}$ and $M_{halo}$ indicate that 
 $\tilde{\Phi}_{halo}(r)$ is only a mathematical function fitting the observational data, $a_{halo}$ and $M_{halo}$ are mathematical parameters, not physical characteristics of the DM of the MW.

The observational rotation curve of the MW is mathematically fitted by the function $\tilde{\Phi}_{halo}(r)$ with three free parameters 
$M_{halo}$, $a_{halo}$ and $R_{DM}$, see Eqs. (\ref{eqDM-Poul})  and (\ref{eqDM-AS}), according to \citet{Allen1991} 
and \citet{2017A&A...598A..66P}. The obtained numerical values for $v (R_{0})$, the Oort constants $A$, $B$, the amplitude and period of the oscillatory motion of the Sun confirm correctness of our approach developed in Secs. \ref{basis-eq}, \ref{sec-DM-G}, 
\ref{BM-improved}, \ref{Mass-speed-relation}, 
\ref{theo-res-sun}, \ref{DM-simple-cons} and 
Appendix \ref{ap-oort}. It is important to stress that our approach introduces no free parameter characterizing DM of the MW. The approach with $\tilde{\Phi}_{halo}(r)$ used by \citet{Allen1991} 
and \citet{2017A&A...598A..66P} is a mathematical approximation to our approach.

\section{Three component Galactic disc}\label{3-components-disc}
This appendix shows that a more component MW can be used to model motions of objects in our galaxy, the Milky Way.

We consider the BM model represented by our bulge and three component Galactic disc, a thin disc, an intermediate disk and a thick disc, in accordance with 
\citet{Halle2018}. 

We take into account the following forms of the baryonic potentials (Miyamoto-Nagai)
\begin{eqnarray}\label{pot-bulge-new-D}
\Phi_{i} (R,z) &=& -~ \frac{G M_{i}}{\sqrt{R^{2} + \left ( a_{i} + \sqrt{z^{2} + b_{i}^{2}} \right )^{2} }} ~, ~i~=~1, ...,~4
\end{eqnarray}
and the values of the parameters are summarized in Table \ref{TableD}.

Mass of the bulge corresponds to Table \ref{Table2}. Masses of the discs are calculated, their ratios correspond to the ratios presented by 
\citet{Halle2018} and $M_{BM}$ $=$ $\sum_{i=1}^{4} M_{i}$ $=$ 8.43 $\times$ 10$^{10}$ $\mbox{M}_{\odot}$, compare Eq. (\ref{MW-mass}). 
Eqs. (\ref{large-R-speed-final}) and (\ref{VGg}) are fulfilled. 
Moreover, we require
$v_{c} (R_{0})$ $=$ 228.8 $\mbox{km}~ \mbox{s}^{-1}$ and $\varrho_{bar}(R_0, z=0)$ $=$ 0.087 $\mbox{M}_{\odot}$ $\mbox{pc}^{-3}$, see Eqs. (\ref{speed-sun}), (\ref{mass-density-sun}). As for the bulge, we consider the values consistent with the values presented in Table \ref{Table2}, $a_{4}$ $=$ 0, $b_{4}$ $=$ 0.3 $\mbox{kpc}$. As for the length-scales of the discs, we consider $a_{1}$ $=$ $a R_{0}$, 
$b_{1}$ $=$ $b R_{0}$, $a_{2}$ $=$ $(a/2) R_{0}$, 
$b_{2}$ $=$ $2 b R_{0}$, $a_{3}$ $=$ $(a/2) R_{0}$, $b_{3}$ $=$ 
$3 b R_{0}$, compare with \citet{Halle2018}. The requirements $v_{c} (R_{0})$ $=$ 228.8 $\mbox{km}~ \mbox{s}^{-1}$, $\varrho_{bar}(R_0)$ $=$ 0.087 $\mbox{M}_{\odot}$ $\mbox{pc}^{-3}$ lead to $a$ $=$ 0.5415 and $b$ $=$ 0.0295. The relevant values for the bulge and the discs are summarized in Table \ref{TableD}.

\begin{figure}[h]
\centering
	\includegraphics[width=0.95\linewidth]{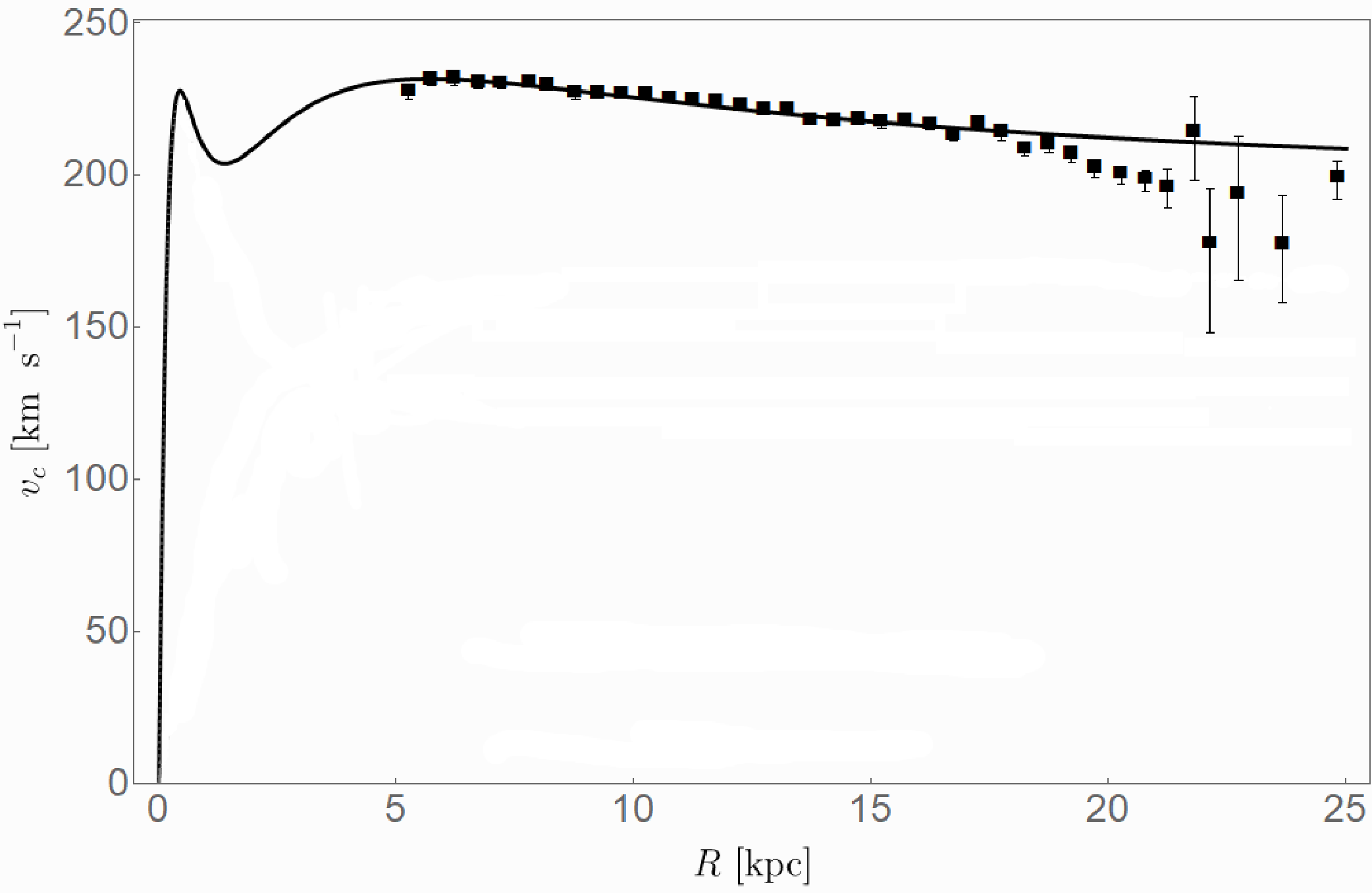}
 \caption{Rotation curve of the Milky Way galaxy for three component disc.
The curve depicts dependence of the circular {velocity} $v_{c}$ of a Galactic body on the Galactocentric distance $R$ of the body. The solid curve holds for Eqs. (\ref{pot-bulge-new-D}), (\ref{eq-motion-DM-D}), 
(\ref{eq-motion-g-bar-D}) and (\ref{vecDM-D})
and for the values of the parameters summarized in Table \ref{TableD}. 
The rotation curve holds for spherical mass distributions of the DM. The curve fulfills the baryonic Tully-Fisher relation.}\label{curvesD}
	\label{FigD}
\end{figure}



\begin{table}[h]
\centering
\def\arraystretch{1.3}
\begin{tabular}{ |c|c|c|c|c| } 
 \hline
 structure & index $i$ & $M_{i}~[10^{10}\mbox{M}_\odot]$ & $a_{i}~[\mbox{kpc}]$ & $b_{i}~[\mbox{kpc}]$\\
 \hline
 thin disc & 1 & 3.8352 & 4.4658 & 0.24329 \\
 intermediate & 2 & 2.2127 & 2.2329 & 0.48657 \\
 disc &  &  &  &  \\
 thick disc & 3 & 1.4750 & 2.2329 & 0.72986 \\
 bulge & 4 & 0.9071 & 0 & 0.30 \\
 \hline
\end{tabular}
\caption{Values of the parameters for the baryonic matter gravitational potentials of the MW components, see Eqs. (\ref{pot-bulge-new-D}).}
\label{TableD}
\end{table}

The vector equation of motion of the body moving under the action of the gravity of the BM and the DM
\begin{eqnarray}\label{eq-motion-DM-D}
\vec{\dot{v}} &=& \vec{g}_{BM} + \vec{g}_{DM} ~,
\end{eqnarray}
where the gravitational acceleration of the BM is
\begin{eqnarray}\label{eq-motion-g-bar-D}
\vec{g}_{BM} &=& -~ \nabla \Phi_{BM} ~,~~
\Phi_{bar} = \sum_{j=1}^{4} \Phi_{j} ~,
\end{eqnarray}
see Eqs. (\ref{pot-bulge-new-D}) and the values summarized in Table \ref{TableD}, compare Eqs. (\ref{eq-motion-DM}) and (\ref{eq-motion-g-bar}).
The gravitational acceleration $\vec{g}_{DM}$ generated by the DM of the MW is 
\begin{eqnarray}\label{vecDM-D}
\vec{g}_{DM} &=& -~g_{DM}(r) ~\hat{\vec{r}} = -\frac{g_{bar}(r)}{\exp{ \left [ \sqrt{g_{bar}(r)/ g_{+}} \right ]} -1} ~\hat{\vec{r}} ~,
\nonumber \\
 g_{bar} (r) &=& \sum_{j=1}^{4} G M_{j} ~\frac{r}{\left( r^2 + c_{j}^{2}\right)^{3/2}} ~,  ~r \leq r_{DM}  ~,
 \nonumber \\
 c_k &=& a_k + b_k  ~,~~ k = 1,~2, ~3, ~4~,
\end{eqnarray}
compare with Eqs. (\ref{vecDM}). We remind that the DM distribution is spherically symmetric with respect to the center of the MW galaxy, $r$ $=$ $\sqrt{R^2 + z^2}$, $g_{DM}(r) \equiv g_{halo} (r)$ is often used, and, 
$\hat{\vec{r}}$ $\equiv$ $\vec{r}/ r$. Radius of the DM halo is $r_{DM}$. The DM halo of the MW is important in the conventional approach. Finally, Eq. (\ref{5final}) can be used.

\subsection{Application: Oort constants}
The first and the second Oort constants are defined by the relations
\begin{eqnarray}\label{oort-D}
A &=& +\frac{1}{2} \left[ \frac{v_{0}}{R_0} - \left( \frac{\mbox{d}v_{c}}{\mbox{d}R}\right)_{R=R_0} \right]  ~,
\nonumber \\
B &=& -\frac{1}{2} \left[ \frac{v_{0}}{R_0} + \left( \frac{\mbox{d}v_{c}}{\mbox{d}R}\right)_{R=R_0} \right] ~.
\end{eqnarray}

We can use Eqs. (\ref{angular-speed}) and  
\begin{eqnarray}\label{diff-vel-D}
\left( \frac{\mbox{d}v_{c}}{\mbox{d}R}\right)_{R=R_0} &=& -~1.74  ~\mbox{km} ~\mbox{s}^{-1} ~\mbox{kpc}^{-1}  
\end{eqnarray}
for $R_{0}$ $=$ 8.247 $\mbox{kpc}$, see Eq. (\ref{R-sun}). We have used Eqs. (\ref{pot-bulge-new-D}), (\ref{eq-motion-DM-D}), 
(\ref{eq-motion-g-bar-D}) and (\ref{vecDM-D}) together with the values summarized in Table \ref{TableD}.
More exact value is given in Eq. (\ref{diff-vel}). The new value presented in Eq. (\ref{diff-vel-D}) is consistent with the value given in  Eq. (\ref{diff-vel}). 

Finally, the numerical values of the Oort constants are
\begin{eqnarray}\label{AB-num-D}
A &=& +14.74 ~\mbox{km} ~\mbox{s}^{-1} ~\mbox{kpc}^{-1} ~,
\nonumber \\
B &=& -13.00 ~\mbox{km} ~\mbox{s}^{-1} ~\mbox{kpc}^{-1} ~,
\end{eqnarray} 
if Eqs. (\ref{angular-speed}) and (\ref{diff-vel-D}) are used for $R_{0}$ $=$ 8.247 $\mbox{kpc}$, see Eq. (\ref{R-sun}). The found result is consistent with Eqs. 
(\ref{AB-num2}).

\end{appendix}

\end{document}